\documentclass{aastex63}
\usepackage[utf8]{inputenc}
\usepackage{hyperref}

\accepted{Astronomical Journal}
\usepackage{xcolor}

\newcommand\kms{\ifmmode{\rm km\thinspace s^{-1}}\else km\thinspace s$^{-1}$\fi}
\newcommand\ticstar{TIC\,1729}

\begin{document}

\title{TIC 172900988: A Transiting Circumbinary Planet Detected in One Sector of TESS Data}

\correspondingauthor{Veselin Kostov}
\email{veselin.b.kostov@nasa.gov}

\author[0000-0001-9786-1031]{Veselin~B.~Kostov}
\affiliation{NASA Goddard Space Flight Center, 8800 Greenbelt Road, Greenbelt, MD 20771, USA}
\affiliation{SETI Institute, 189 Bernardo Ave, Suite 200, Mountain View, CA 94043, USA}
\affiliation{GSFC Sellers Exoplanet Environments Collaboration}
\author[0000-0003-0501-2636]{Brian P. Powell}
\affiliation{NASA Goddard Space Flight Center, 8800 Greenbelt Road, Greenbelt, MD 20771, USA}
\author[0000-0001-9647-2886]{Jerome A. Orosz}
\affil{Department of Astronomy, San Diego State University, 5500 Campanile Drive, San Diego, CA 92182, USA}
\author[0000-0003-2381-5301]{William F. Welsh}
\affil{Department of Astronomy, San Diego State University, 5500 Campanile Drive, San Diego, CA 92182, USA}
\author[0000-0001-9662-3496]{William Cochran}
\affil{Center for Planetary Systems Habitability and McDonald Observatory, The University of Texas, Austin, Texas USA 78712}
\affil{McDonald Observatory, The University of Texas at Austin, 
Austin, TX 78712-0259, USA}
\author[0000-0001-6588-9574]{Karen A.\ Collins}
\affiliation{Center for Astrophysics $\vert$ Harvard \& Smithsonian, 60 Garden St, Cambridge, MA, 02138, USA}
\author{Michael Endl}
\affil{McDonald Observatory, The University of Texas at Austin, 
Austin, TX 78712-0259, USA}
\author{Coel Hellier}
\affil{Astrophysics Group, Keele University, Staffordshire, ST5 5BG, United Kingdom}
\author[0000-0001-9911-7388]{David W.\ Latham}
\affiliation{Center for Astrophysics $\vert$ Harvard \& Smithsonian, 60 Garden St, Cambridge, MA, 02138, USA}
\author{Phillip MacQueen}
\affil{McDonald Observatory, The University of Texas at Austin, 
Austin, TX 78712-0259, USA}
\author[0000-0002-3827-8417]{Joshua Pepper}
\affil{Department of Physics, Lehigh University, 413 Deming Lewis Lab,
16 Memorial Drive East
Bethlehem, PA  18015, USA}
\author[0000-0002-9644-8330]{Billy Quarles}
\affil{Center for Relativistic Astrophysics, School of Physics, Georgia Institute of Technology, 770 State St NW, Atlanta, GA 30313, USA}
\affil{Department of Physics, Astronomy, Geosciences and Engineering Technology, Valdosta State University, GA, 31698}
\author[0000-0001-8102-3033]{Lalitha Sairam}
\affil{School of Physics \& Astronomy, University of Birmingham, Edgbaston, Birmingham B15 2TT, United Kingdom}
\author[0000-0002-5286-0251]{Guillermo Torres}
\affiliation{Center for Astrophysics $\vert$ Harvard \& Smithsonian, 60 Garden St, Cambridge, MA, 02138, USA}
\author{Robert F. Wilson}
\affil{Department of Astronomy, University of Virginia, 
530 McCormick Road, Charlottesville, VA, 22904, USA}
\author{Serge Bergeron}
\affil{UCLAN, Casselman, Ontario}
\author{Pat Boyce}
\affil{Boyce Research Initiatives and Education Foundation, 3540 Carleton Street, San Diego, CA, 92106}
\author{Allyson Bieryla}
\affiliation{Center for Astrophysics $\vert$ Harvard \& Smithsonian, 60 Garden St, Cambridge, MA, 02138, USA}
\author{Robert Buchheim}
\affil{Lost Gold Observatory, Society for Astronomical Sciences, Gold Canyon, AZ 85118, USA}
\author{Caleb Ben Christiansen}
\affil{Department of Astronomy, San Diego State University, 5500 Campanile Drive, San Diego, CA 92182, USA}
\author[0000-0002-5741-3047]{David R.~Ciardi}
\affil{NASA Exoplanet Science Institute - Caltech-IPAC, 1200 E California Ave, MS100-22 Pasadena, CA 91125 USA}
\author[0000-0003-2781-3207]{Kevin I.\ Collins}
\affiliation{George Mason University, 4400 University Drive, Fairfax, VA, 22030 USA}
\author[0000-0003-2239-0567]{Dennis M.\ Conti}
\affiliation{American Association of Variable Star Observers, 49 Bay State Road, Cambridge, MA 02138, USA}
\author{Scott Dixon}
\affil{San Diego Astronomy Association, P.O. Box 23215, San Diego, CA 92193-3215}
\author{Pere Guerra}
\affiliation{Observatori Astronòmic Albanyà, Camí de Bassegoda S/N, Albanyà 17733, Girona, Spain}
\author{Nader Haghighipour}
\affiliation{Planetary Science Institute, 1700 East Fort Lowell, Tucson, AZ 85719, USA}
\affil{ Institute for Astronomy, University of Hawaii-Manoa, 2680 Woodlawn Dr, Honolulu, HI 96822}
\author{Jeffrey Herman}
\affil{San Diego Astronomy Association, P.O. Box 23215, San Diego, CA 92193-3215}
\author[0000-0002-9867-7938]{Eric G. Hintz}
\affiliation{Department of Physics and Astronomy, Brigham Young University, Provo, UT 84602, USA}
\author{Ward S. Howard}
\affil{Department of Astronomy, University of North Carolina at Chapel Hill,
Philips Hall, CB \#3255, 120 E. Cameron Avenue, Chapel Hill, NC 27599, USA}
%
%
\author[0000-0002-4625-7333]{Eric L.\ N.\ Jensen}
\affiliation{Department of Physics \& Astronomy, Swarthmore College, Swarthmore PA 19081, USA}
\author[0000-0003-0497-2651]{John F.\ Kielkopf}
\affiliation{Department of Physics and Astronomy, University of Louisville, Louisville, KY 40292, USA}
\author[0000-0002-0493-1342]{Ethan Kruse}
\affiliation{NASA Goddard Space Flight Center, 8800 Greenbelt Road, Greenbelt, MD 20771, USA}
\author{Nicholas M. Law}
\affil{Department of Astronomy, University of North Carolina at Chapel Hill, Philips Hall, CB \#3255, 120 E. Cameron Avenue, Chapel Hill, NC 27599, USA}
\author[0000-0002-7595-6360]{David Martin}
\affil{Department of Astronomy, The Ohio State University, 4055 McPherson Laboratory, Columbus, OH 43210, USA}
\affil{NASA Sagan Fellow}
\author{Pierre F.~L.~Maxted}
\affiliation{Keele University, Staffordshire, ST5 5BG}
\author[0000-0001-7516-8308]{Benjamin~T.~Montet}
\affil{School of Physics, University of New South Wales, Sydney NSW 2052, Australia}
\affiliation{UNSW Data Science Hub, University of New South Wales, Sydney, NSW 2052, Australia}
%
\author{Felipe Murgas}
\affiliation{Instituto de Astrof\'isica de Canarias (IAC), E-38205 La Laguna, Tenerife, Spain}
\affiliation{Departamento de Astrof\'isica, Universidad de La Laguna (ULL), E-38206 La Laguna, Tenerife, Spain}
\author{Matt Nelson}
\affil{Department of Astronomy, University of Virginia, 
530 McCormick Road, Charlottesville, VA, 22904, USA}
\author[0000-0001-8472-2219]{Greg Olmschenk}
\affiliation{NASA Goddard Space Flight Center, 8800 Greenbelt Road, Greenbelt, MD 20771, USA}
\affiliation{Universities Space Research Association, 7178 Columbia Gateway Drive, Columbia, MD 21046}
\author{Sebastian Otero}
\affil{American Association of Variable Star Observers}
\author{Robert Quimby}
\affil{Department of Astronomy, San Diego State University, 5500 Campanile Drive, San Diego, CA 92182, USA}
\author{Michael Richmond}
\affil{School of Physics and Astronomy, Rochester Institute of Technology, RIT Observatory}
\author[0000-0001-8227-1020]{Richard P. Schwarz}
\affiliation{Patashnick Voorheesville Observatory, Voorheesville, NY 12186, USA}
\author[0000-0002-1836-3120]{Avi Shporer}
\affil{Department of Physics and Kavli Institute for Astrophysics and Space Research, Massachusetts Institute of Technology, Cambridge, MA 02139, USA}
\author[0000-0002-3481-9052]{Keivan G.\ Stassun}
\affil{Department of Physics \& Astronomy, Vanderbilt University, 6301 Stevenson Center Ln., Nashville, TN 37235, USA}
\author{Denise C.\ Stephens} 
\affiliation{Brigham Young University, N486 ESC, Provo, UT, 84602, USA}
\author[0000-0002-5510-8751]{Amaury H. M. J. Triaud}
\affil{School of Physics \& Astronomy, University of Birmingham, Edgbaston, Birmingham B15 2TT, United Kingdom}
\author{Joe Ulowetz}
\affil{Center for Backyard Astrophysics, 855 Fair Ln, Northbrook, IL 60062}
\author{Bradley S. Walter}
\affil{Central Texas Astronomical Society, Paul and Jane Meyer Observatory, Waco TX}
\author{Edward Wiley}
\affil{Stellar Skies Observatories, AAVSO, 125 Mountain Creek Pass, Georgetown, TX 78633}
\author{David Wood}
\affil{San Diego Astronomy Association, P.O. Box 23215, San Diego, CA 92193-3215}
\author{Mitchell Yenawine}
\affil{Department of Astronomy, San Diego State University, 5500 Campanile Drive, San Diego, CA 92182, USA}
\author{Eric Agol}
\affil{Department of Astronomy, University of Washington}
\author[0000-0001-7139-2724]{Thomas~Barclay}
\affiliation{NASA Goddard Space Flight Center, 8800 Greenbelt Road, Greenbelt, MD 20771, USA}
\affiliation{University of Maryland, Baltimore County, 1000 Hilltop Cir,
Baltimore, MD 21250, USA}
\author[0000-0002-9539-4203]{Thomas G. Beatty}
\affiliation{Department of Astronomy and Steward Observatory, University of Arizona, Tucson, AZ 85721}
\author[0000-0001-8388-8399]{Isabelle Boisse}
\affil{Aix Marseille Univ CNRS CNES LAM Marseille France}
\author{Douglas A. Caldwell}
\affil{SETI Institute, 189 Bernardo Ave, Suite 200, Mountain View, CA 94043, USA}
\author[0000-0002-8035-4778]{Jessie Christiansen}
\affil{NASA Exoplanet Science Institute - Caltech-IPAC, 1200 E California Ave, MS100-22 Pasadena, CA 91125 USA}
\author[0000-0001-8020-7121]{Knicole~D.~Col\'{o}n}
\affiliation{NASA Goddard Space Flight Center, 8800 Greenbelt Road, Greenbelt, MD 20771, USA}
\author{Magali Deleuil}
\affil{Aix Marseille Univ CNRS CNES LAM Marseille France}
\author{Laurance Doyle}
\affiliation{SETI Institute, 189 Bernardo Ave, Suite 200, Mountain View, CA 94043, USA}
%
%
\author[0000-0002-9113-7162]{Michael~Fausnaugh}
\affiliation{Department of Physics and Kavli Institute for Astrophysics and Space Research, Massachusetts Institute of Technology, Cambridge, MA 02139, USA}
\author{G{\' a}bor~F{\H u}r{\' e}sz}
\affiliation{Department of Physics and Kavli Institute for Astrophysics and Space Research, Massachusetts Institute of Technology, Cambridge, MA 02139, USA}
\author[0000-0002-0388-8004]{Emily A. Gilbert}
\affiliation{NASA Goddard Space Flight Center, 8800 Greenbelt Road, Greenbelt, MD 20771, USA}
\affiliation{Department of Astronomy and Astrophysics, University of
Chicago, 5640 S. Ellis Ave, Chicago, IL 60637, USA}
\affiliation{The Adler Planetarium, 1300 South Lakeshore Drive, Chicago, IL 60605, USA}
\affiliation{GSFC Sellers Exoplanet Environments Collaboration}
\affiliation{University of Maryland, Baltimore County, 1000 Hilltop Cir,
Baltimore, MD 21250, USA}
\author{Guillaume Hébrard}
\affil{Institut d'Astrophysique de Paris, UMR7095 CNRS, Université Pierre \& Marie Curie, 98bis boulevard Arago, 75014 Paris, France}
\author[0000-0001-5160-4486]{David J. James}
\affiliation{ASTRAVEO LLC, PO Box 1668, MA 01931}
\author[0000-0002-4715-9460]{Jon Jenkins}
\affil{NASA Ames Research Center, Moffett Field, CA, 94035, USA}
\author{Stephen R. Kane}
\affiliation{Department of Earth and Planetary Sciences,
University of California, Riverside,
900 University Avenue,
Riverside, CA 92521, USA}
\author{Richard C. Kidwell Jr.}
\affil{Mikulski Archive for Space Telescopes, Space Telescope Science Institute}
\author{Ravi Kopparapu}
\affil{NASA Goddard Space Flight Center, 8800 Greenbelt Road, Greenbelt, MD 20771, USA}
\author{Gongjie Li}
\affil{Center for Relativistic Astrophysics, School of Physics, Georgia Institute of Technology, 770 State St NW, Atlanta, GA 30313, USA}
\author[0000-0001-6513-1659]{Jack J. Lissauer}
\affil{Space Science and Astrobiology Division,
MS 245-3, NASA Ames Research Center, Moffett Field, CA 94035, USA}
\author[0000-0003-2527-1598]{Michael B. Lund}
\affil{NASA Exoplanet Science Institute, Caltech/IPAC 1200 E. California Ave, Pasadena, CA 91125, USA}
\author{Steve R. Majewski}
\affil{Department of Astronomy, University of Virginia, 530 McCormick Road, Charlottesville, VA, 22904, USA}
\author{Tsevi Mazeh}
\affiliation{Department of Astronomy and Astrophysics, Tel Aviv University, 69978 Tel Aviv, Israel}
\author[0000-0002-8964-8377]{Samuel N. Quinn}
\affiliation{Center for Astrophysics $\vert$ Harvard \& Smithsonian, 60 Garden St, Cambridge, MA, 02138, USA}
\author{Elisa Quintana}
\affiliation{NASA Goddard Space Flight Center, 8800 Greenbelt Road, Greenbelt, MD 20771, USA}
\author{George Ricker}
\affil{Department of Physics and Kavli Institute for Astrophysics and Space Research, Massachusetts Institute of Technology, Cambridge, MA 02139, USA}
\author[0000-0001-8812-0565]{Joseph E. Rodriguez}
\affil{Department of Physics and Astronomy, Michigan State University, East Lansing, MI 48824, USA}
\author{Jason Rowe}
\affiliation{Department of Physics and Astronomy, Bishop's University, 2600 College Street
Sherbrooke, QC, Canada, J1M 1Z7}
\author[0000-0002-3586-1316]{Alexander Santerne}
\affil{Aix Marseille Univ CNRS CNES LAM Marseille France}
\author{Joshua Schlieder}
\affiliation{NASA Goddard Space Flight Center, 8800 Greenbelt Road, Greenbelt, MD 20771, USA}
\author[0000-0002-6892-6948]{Sara Seager}
\affiliation{Department of Physics and Kavli Institute for Astrophysics and Space Research, Massachusetts Institute of Technology, Cambridge, MA 02139, USA}
\affiliation{Department of Earth, Atmospheric and Planetary Sciences, Massachusetts Institute of Technology, Cambridge, MA 02139, USA}
\affiliation{Department of Aeronautics and Astronautics, MIT, 77 Massachusetts Avenue, Cambridge, MA 02139, USA}
\author[0000-0002-7608-8905]{Matthew R. Standing}
\affil{School of Physics \& Astronomy, University of Birmingham, Edgbaston, Birmingham B15 2TT, United Kingdom}
\author[0000-0002-5951-8328]{Daniel J. Stevens}
\altaffiliation{Eberly Fellow}
\affiliation{Center for Exoplanets and Habitable Worlds, The Pennsylvania State University, 525 Davey Lab, University Park, PA 16802, USA}
\affiliation{Department of Astronomy \& Astrophysics, The Pennsylvania State University, 525 Davey Lab, University Park, PA 16802, USA}
\author{Eric B. Ting}
\affil{NASA Ames Research Center, Moffett Field, CA 94035, USA}
\author{Roland Vanderspek}
\affil{Department of Physics and Kavli Institute for Astrophysics and Space Research, Massachusetts Institute of Technology, Cambridge, MA 02139, USA}
\author[0000-0002-4265-047X]{Joshua N. Winn}
\affil{Department of Astrophysical Sciences, Princeton University, Princeton, NJ 08544, USA}

\date{\today}

\begin{abstract}
We report the first discovery of a transiting circumbinary planet detected from a single sector of TESS data. During Sector 21, the planet TIC 172900988b transited the primary star and then 5 days later it transited the secondary star. The binary is itself eclipsing, with a period of $P\approx 19.7$ days and an eccentricity of $e\approx  0.45$. Archival data from ASAS-SN, Evryscope, KELT, and SuperWASP reveal a prominent apsidal motion of the binary orbit, caused by the dynamical interactions between the binary and the planet. A comprehensive photodynamical analysis of the TESS, archival and follow-up data yields stellar masses and radii of $M_1=1.2384\pm 0.0007\,M_{\odot}$ and $R_1=1.3827\pm 0.0016\,R_{\odot}$ for the primary and $M_2=1.2019\pm 0.0007\,M_{\odot}$ and $R_2=1.3124\pm 0.0012\,R_{\odot}$ for the secondary. The radius of the planet is $R_3=11.25\pm0.44\,R_{\oplus}$ ($1.004\pm0.039 R_{\rm Jup}$). The planet's mass and orbital properties are not uniquely determined---there are six solutions with nearly equal likelihood. Specifically, we find that the planet's mass is in the range of $824 \lesssim M_3\lesssim 981\,M_{\oplus}$ ($2.65 \lesssim M_3 \lesssim 3.09 M_{\rm Jup}$), its orbital period could be 188.8, 190.4, 194.0, 199.0, 200.4, or 204.1 days, and the eccentricity is between 0.02 and 0.09. At a V=10.141 mag, the system is accessible for high-resolution spectroscopic observations, e.g.~Rossiter-McLaughlin effect and transit spectroscopy. 
\end{abstract}



\section{Introduction}\label{sec:intro}

Long before ${\emph Kepler}$'s discovery of transiting circumbinary planets (CBPs, Doyle et al. 2011, Welsh \& Orosz 2018), Schneider \& Chevreton (1990) discussed an unusual observational signature such planets would have -- the occurrence of multiple transits during one conjunction. This effect is caused by the planet transiting one or both stars of the host eclipsing binary (EB) several times over the course of a fraction of its orbital period. The configuration of such transits depends on the relative sky-projected velocities of the CBP and the stars being transited. Importantly, the orbital period of the CBP can be estimated from such transits provided the host system is a double-lined spectroscopic binary, as argued by Schneider \& Chevtron (1990) and Kostov et al. (2020b), and demonstrated in Kostov et al. (2016). In this methodology, the radial velocities of both the primary and secondary stars are required to calculate the sky positions of the transited star(s) during the observed transits. The situation is similar to that of triply-eclipsing stellar triples that produce multiple sets of eclipses and occultations (e.g., Marsh et al. 2014, Borkovits et al. 2020). 

Several groups attempted to detect transits of circumbinary planets (CBP) from the ground before the turn of the century  -- single-conjunction or otherwise -- but were ultimately hampered by the limited time sampling (e.g. Doyle et al. 1996, Deeg et al. 1997). Fortunately, thanks to its long dwell time and high photometric precision, NASA's ${\emph Kepler}$ mission enabled the discovery of a $\sim$dozen transiting CBPs and also demonstrated that the occurrence of pairs of transits during one conjunction is common. Four of the eleven known ${\emph Kepler}$ CBP systems exhibit such transits---Kepler-16, -34, -35, and -1647 (see Kostov et al. 2020b for more details). 

Kostov et al. (2020a) recently found a transiting CBP (TOI-1338b) using data from the Transiting Exoplanet Survey Satellite (TESS, Ricker et al.\ 2015), highlighting its detection potential. By virtue of residing in the southern continuous viewing zone of TESS\footnote{The continuous viewing zones are regions of the sky near the ecliptic poles which are covered by multiple TESS sectors.}, TOI-1338b was observed in 12 consecutive sectors (about 336 days). This provided sufficient observational coverage for this CBP to be at conjunction three times and produce one transit across the primary star at each conjunction. In this regard, the discovery process of TESS' first CBP was identical to that used for Kepler's CBPs ---the orbital period of TOI-1338b was well-constrained from the TESS photometry alone prior to the detailed photodynamical modeling, and the analysis of the system followed a familiar path (Welsh et al. 2016, Kostov et al. 2020a). 

Most of the sky, however, is continuously covered by only a single sector of TESS observations (about 27.4 days). Compared to the orbital periods of Kepler's CBPs with a minimum of 49.5 days and median of 175 days (Welsh \& Orosz 2018, Martin 2018, Socia et al. 2020), the observing duration of one TESS sector is too short to allow for the detection of more than one conjunction. Fortunately, the duration is more than sufficient to detect a CBP producing two (or more) transits during a single conjunction as these occur over a small fraction of the planet's orbital period (Kostov et al. 2020b). For the four {\it Kepler} CBPs exhibiting pairs of transits during one conjunction, the separation between these ``1-2 punch'' events was at most several days --- well within the 27.4 day span of a TESS sector. As demonstrated by Kostov et al. (2020b), measuring the times of such transits, when supplemented with radial velocity data for both stars in the host binary system, yields an estimate of the orbital period
of the planet with an uncertainty of $\lesssim10\%$. A remarkable result given that it is based on observational data covering a small fraction of the planet's orbit. Perhaps even more importantly, finding such transits in TESS EB data is the only pathway towards detecting a significant number of transiting CBPs for years to come (Kostov et al. (2020b)). These detections will enable improved population studies of a large number of these interesting worlds and provide deeper understanding of the formation of circumbinary planets. We note that CBPs, i.e. P-type planets (Dvorak 1982), represent a small fraction of exoplanets discovered in binary star systems. The majority of exoplanets with binary star hosts are in S-type configuration (e.g. Haghighipour 2010, Ciardi et al. 2015, Furlan et al. 2017,  Matson et al. 2018, Mugrauer \& Michel 2020, Ziegler et al. 2021, Howell et al. 2021, and reference therein) and a comprehensive understanding of planet formation in binary star systems depends on the analysis of both populations.

Here we report the discovery of the first TESS CBP using the multiple-transits-in-one-conjunction technique. This paper is organized as follows. In Section \ref{sec:detection}, we describe the detection and preliminary analysis of the system in the TESS data, as well as the archival data and new observations of the system. Section \ref{sec:ELC} presents the comprehensive photometric-dynamical analysis of the system. We discuss the results in Section \ref{sec:discussion} and draw our conclusions in Section \ref{sec:conclusions}.

\section{Detection}
\label{sec:detection}

Finding transiting planets orbiting around binary stars is much more difficult than around single stars. The transits are shallower (due to the constant `third-light' dilution from the binary companion), noisier (due to starspots and stellar activity from two stars), and can be blended with the stellar eclipses\footnote{We note that the first two complications are also present when searching for transiting planets in S-type configurations}. This difficulty is greatly compounded when the observations cover a single conjunction and, even if multiple transits are detected as in the system presented here, they are neither periodic, nor have the same depth and duration (e.g.~Kostov et al. 2014, Windemuth et al. 2019, Martin \& Fabrycky 2021). The transit times and shapes depend on the orientation and motion of the binary stars and of the CBP at the observed times. The complexity of such transits is both a curse and a blessing: it is much more difficult to find them, but once identified and confirmed as real events rather than instrumental artifacts, and shown to be originating from the same source as the eclipses, there are very few viable false positive scenarios that can explain the morphology of the transit(s) (Kostov et al. 2020a). When combined with additional observations and modeling, the information that can be extracted from such transits is richer than what can be obtained from a single transit of a single-star planet as these transits enable direct estimation of the planet's period for double-lined spectroscopic binary stars, regardless of the stellar radius or impact parameters.

As part of our TESS GI Cycle 2 and 3 programs, V.K. is leading a group that is performing a search for transiting CBPs using targets from the GSFC TESS EB Catalog (Kruse et al. in prep). The catalog is generated by a neural network classifier and is based on the long-cadence {\texttt{eleanor}} light curves (Feinstein et al.\ 2019) we created locally from the TESS Full Frame Images (FFIs). The calibrated FFIs were produced by the Science Procesing Operations Center (SPOC) and NASA Ames Research Center (Jenkins et al. 2016). The neural network, a one dimensional adaptation of ResNet (He et al.\ 2015) modified for light curve vector input, was trained to identify eclipse-like features, without any requirement for periodicity or consistency in the shape of the light curves. While this approach does allow for eclipse-shaped noise to be classified as an EB, it also provides for the identification of single-eclipse EBs. The lack of a requirement for multiple eclipses maximizes the utility of the GSFC TESS EB Catalog for a CBP search in that it allows for a greater number of candidates to be selected for further examination.

Emphasizing the need for single-eclipse detection, transit features are most readily identifiable on a light curve with a flat baseline, are found predominantly in detached binaries (in contrast to the variable baselines of semi-detached or contact binaries). Long-period detached binaries are ideal in that the flat baseline is present for a longer duration. Transit-like events are found through visual examination of the baseline from stellar systems identified in the GSFC TESS EB Catalog.  While nearly all of these events have been determined through extensive vetting to be false alarms, there have been several which cannot be ruled out as false positives.  TIC 172900988 exhibited two such transit-like events which are on-target and proved to be caused by a $\sim Jupiter$-sized object on a circumbinary orbit.

\subsection{TESS data}

\begin{figure}[h]
    \centering

    \includegraphics[width=0.99\textwidth]{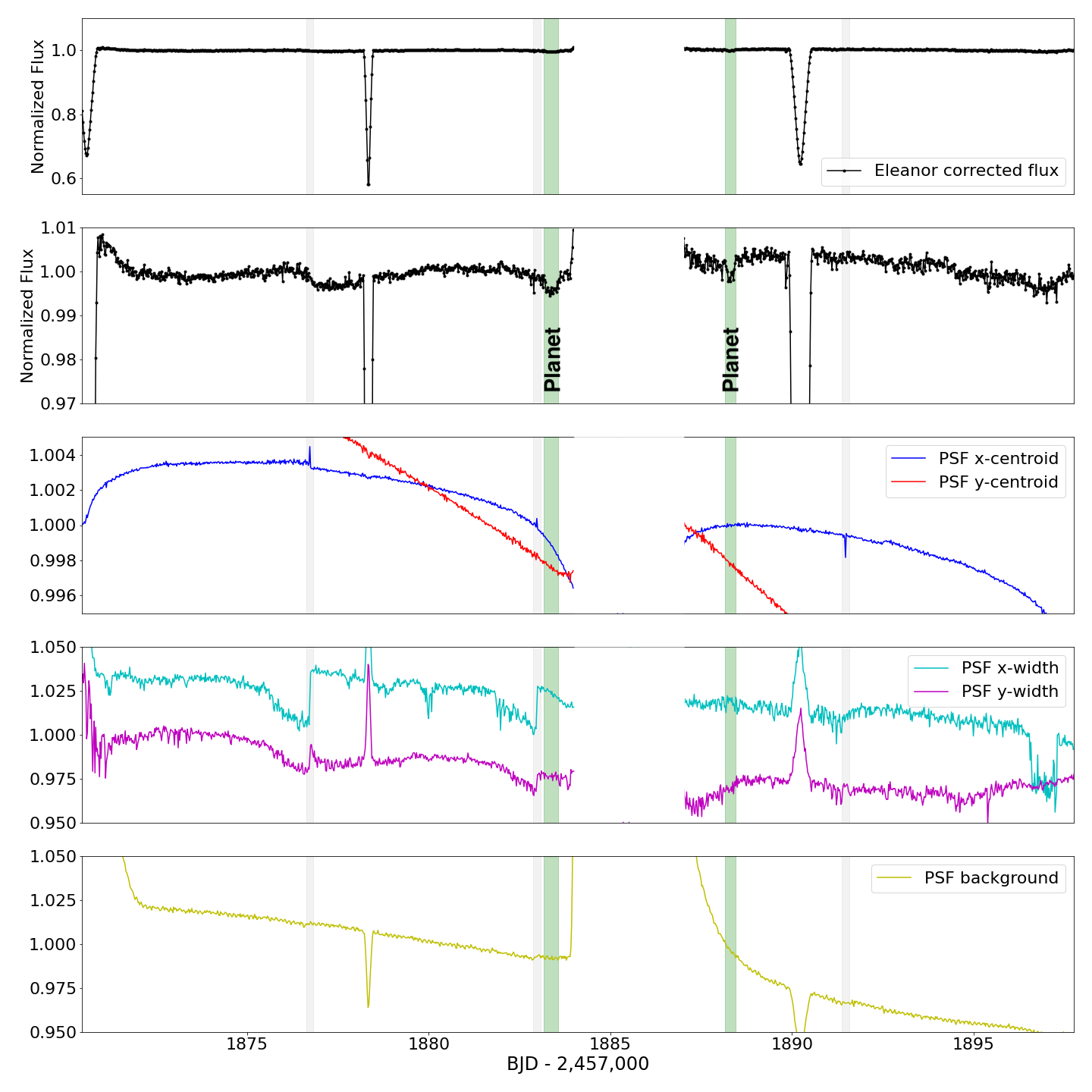}
    \caption{First panel from top: \texttt{eleanor's} Corrected Flux light curve of \ticstar\ for Sector 21. The primary eclipse is at day 1878.3, the two secondary eclipses are at days 1870.57 and 1890.23. The CBP transits are highlighted in green. Momentum dump events are highlighted in light grey. Second panel: Same as above but zoomed in to showcase the two CBP transits. Third to fifth panels: Diagnostic tests to rule out photocenter shift or systematic artifacts associated with the CBP transits. Third panel: normalized $x$- and $y$-photocenters as a function of time; Fourth panel: $x$- and $y$-widths of the PSF used for light curve extraction; Fifth panel: same as above but for the background. There are no significant photocenter shifts or systematic artifacts associated with the CBP transits, confirming them as real signals. 
    }
    \label{fig:lc_psf_}
\end{figure}

TESS observed TIC 172900988 (hereafter \ticstar) during Sector 21
(2020 January 21 through 2020 February 18). Preliminary inspection of the light curve revealed one primary and two secondary eclipses (the
first of which is only partially covered) with depths of $\approx 40\%$ and $\approx 35\%$ respectively, and indicated an EB orbital period of $\approx 19.7$ days with a significantly non-zero eccentricity. The light curve exhibits two additional transit-like events at days 1883.37 (BJD $-$ 2,457,000) and 1888.30 (just before and right after the downlink data gap), with depths of $\approx 3000$ ppm (0.3\%), and durations of $\approx 0.4$ days and $\approx 0.27$ days respectively. Figure \ref{fig:lc_psf_} shows the FFI {\texttt{eleanor}} light curve, both full-scale to present the stellar eclipses and also zoomed-in to highlight the CBP transits, as well as the diagnostic tests we used to rule out artifacts associated with the latter. The variations in the durations of the transits, combined with the lack of significant photocenter shift or systematic artifacts during the two events rule out false positive scenarios due to a background EB or instrumental artifacts, and indicate a circumbinary body.

\begin{figure*}
    \centering
    \includegraphics[width=0.91\linewidth,angle=0]{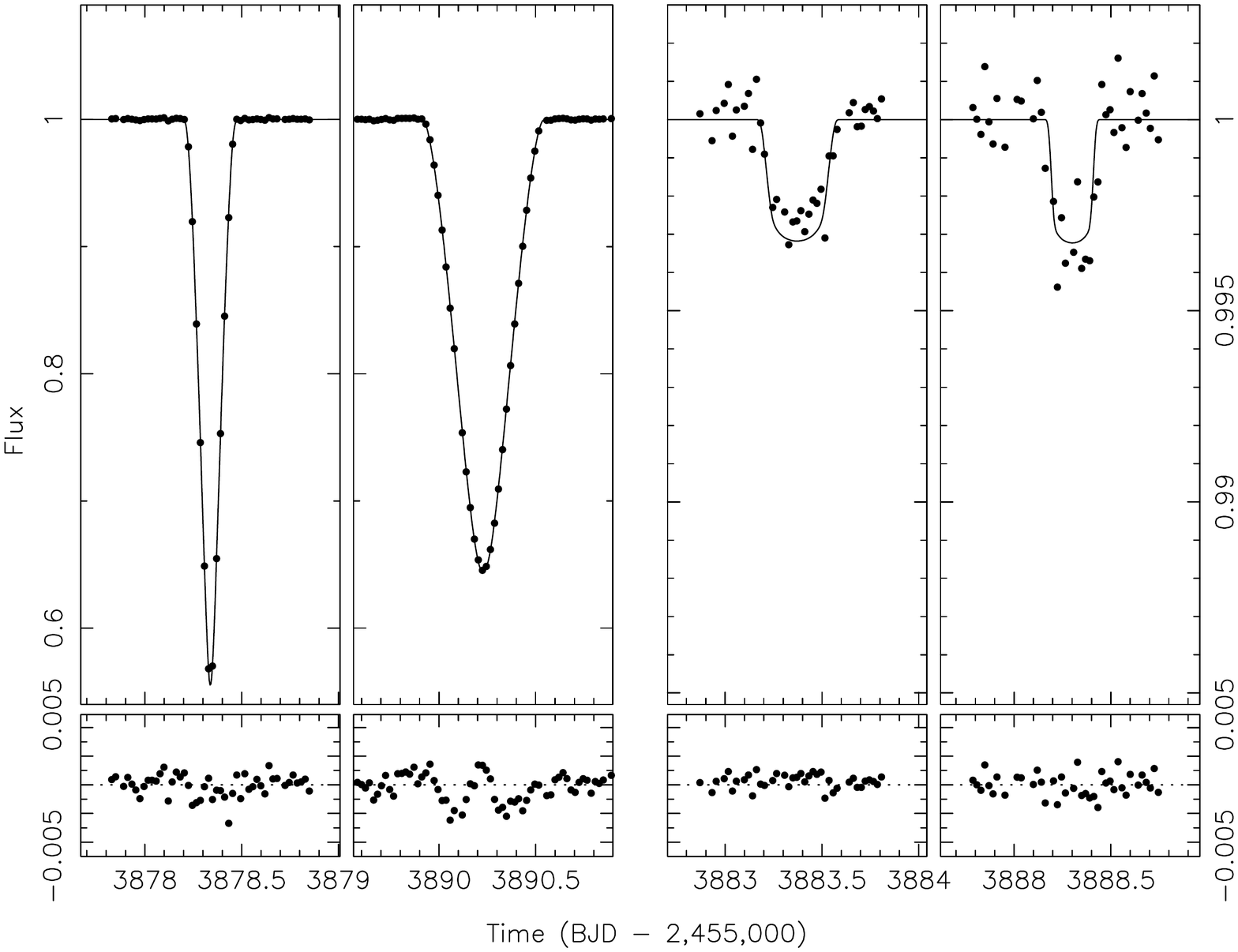}
    \includegraphics[width=0.9\linewidth,angle=0]{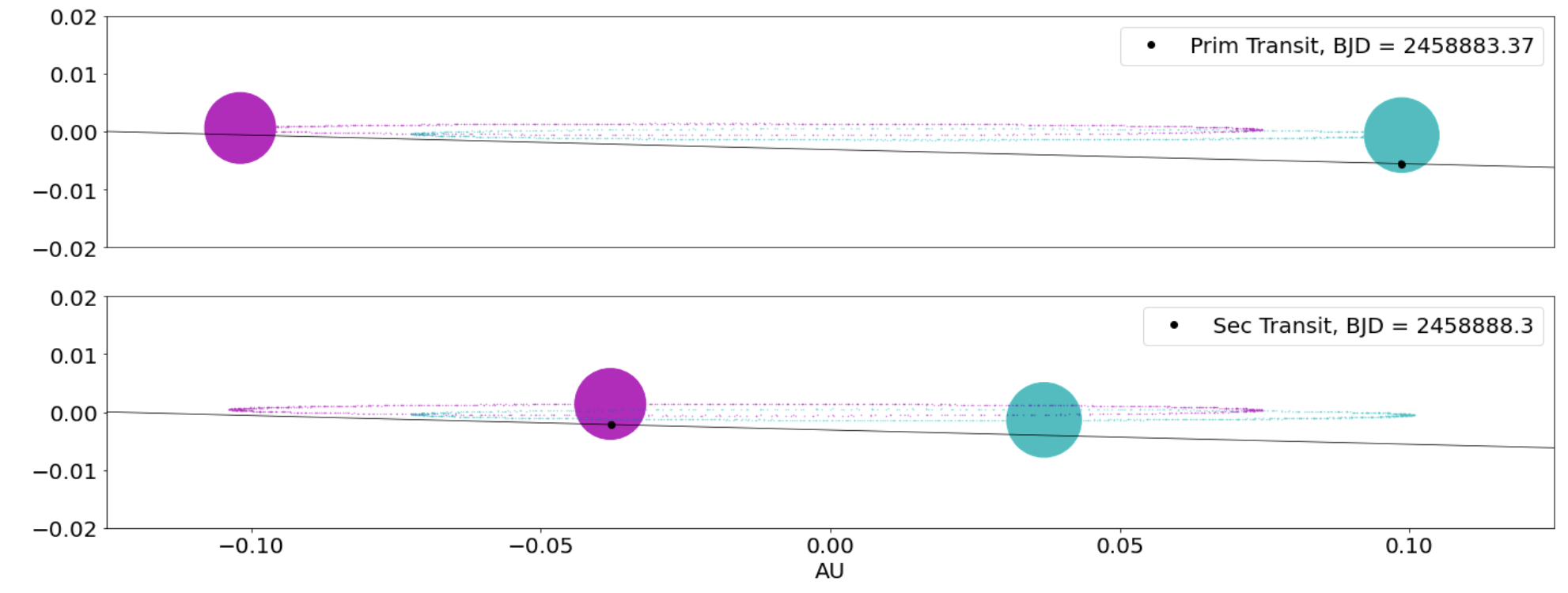}
    \caption{Upper panels, from left to right: the TESS data for the primary eclipse, the secondary eclipse, the planet transit of the primary, and the planet transit of the secondary. We note that the vertical scale on the transit panels is zoomed in by a factor of 30x. Also shown are the best-fitting model from Family 5 (described below), along with the residuals of the fit. Lower panels: To-scale configuration of the system at the times of the two transits. The primary star is in cyan color and the secondary is in magenta color.}\label{fig:showTESSall}
\end{figure*}

\begin{figure}
    \centering

    \includegraphics[width=0.95\textwidth]{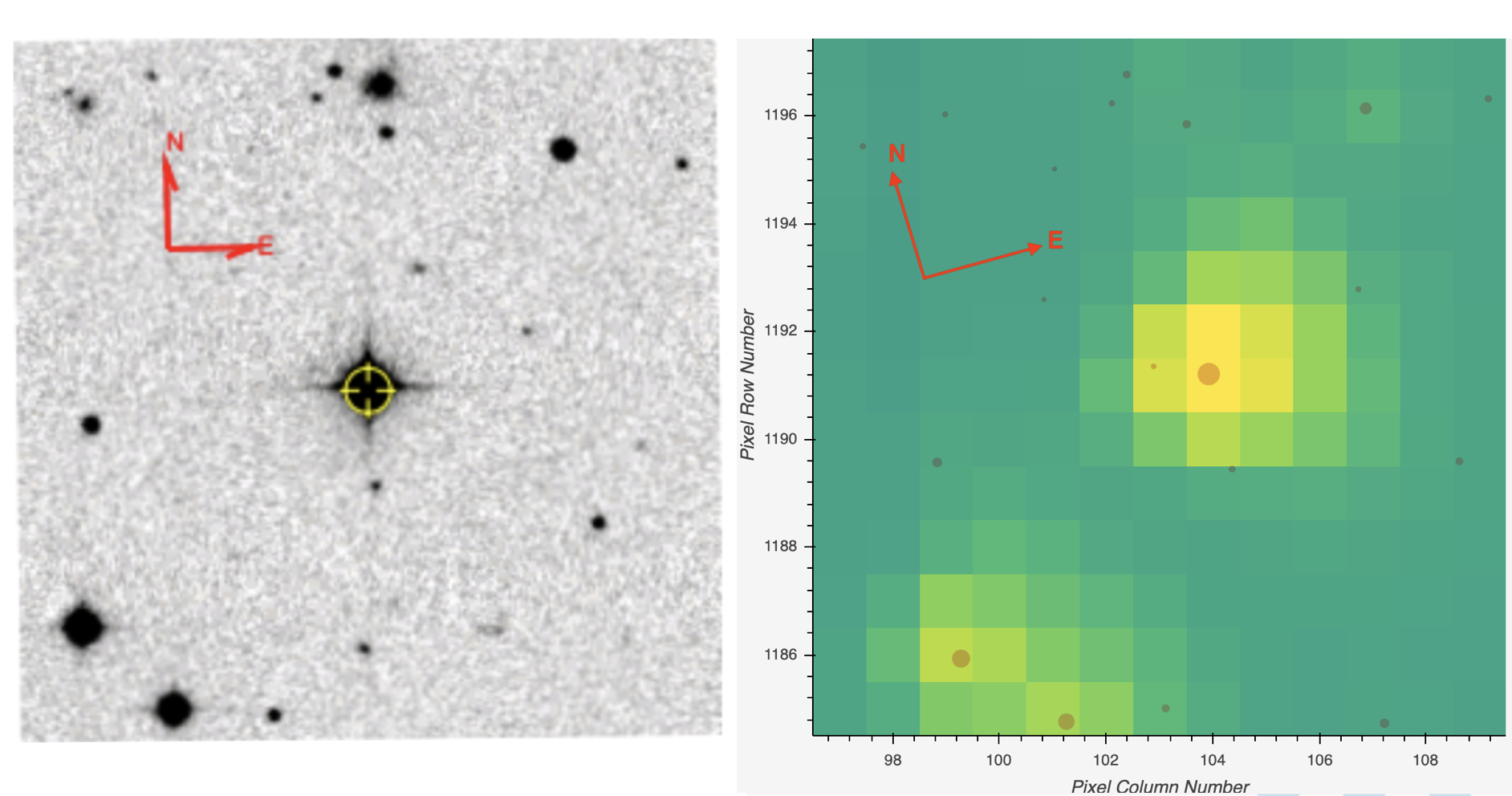}
    \caption{Left: $4.3\arcmin\times4.3\arcmin$ DSS red image of the target (highlighted with yellow cross-hair); Right: $4.3\arcmin\times4.3\arcmin$ Skyview image superimposed on the TESS pixels of the target for Sector 21, showing all nearby resolved {\it Gaia} sources down to G = 21 mag. For clarity in matching the images, we note that the bright source due N of the target and near the upper edge of the left image is the galaxy LEDA 1954290 and is not present in the Skyview image.}
    \label{fig:image_}
\end{figure}

The CBP transits, the stellar eclipses in TESS data, and the configuration of the system at the transit times are shown in more detail in Figure \ref{fig:showTESSall}. The two transits are present in all four light curve extractions from \texttt{eleanor} (``Raw'', ``Corrected'', ``PCA''\footnote{Based on principal component analysis.} and ``PSF''\footnote{Based on point-spread function analysis.} Flux), are independent of the aperture used for the extraction, and the contamination from nearby source is low (${\approx 2\times10^{-4}}$).  The closest and brightest known source inside the $13\times13$ target pixels, TIC 172900986, has an angular separation of $\approx 37\arcsec$ and is $\approx8$ magnitudes fainter (see Figure \ref{fig:image_}), resulting in a maximum flux contamination of $\approx 500$ ppm -- much smaller than the measured transit depths. The latter, combined with the catalog stellar radius ($2.02R_\odot$ from the TIC, $2.1R_\odot$ from {\it Gaia}\footnote{We note that these values do not take into account the binary-nature of the target.}), indicates that the size of the transiting body is $\approx 10R_\oplus$, further strengthening the CBP hypothesis. We note that there is a contact binary just at the edge of the default $13\times13$ target pixel array for \ticstar\ (TIC 172900982, near Column 101, Row 1185 on Figure \ref{fig:image_}) but it is not a source of contamination as it is separated from \ticstar\ by almost 10 pixels, does not overlap the PSF, and the timescale of the variability ($P = 0.23$ days) is unrelated to the CBP transits.

\subsection{Archival Photometry}

\begin{figure}
    \centering
    \includegraphics[width=0.95\linewidth]{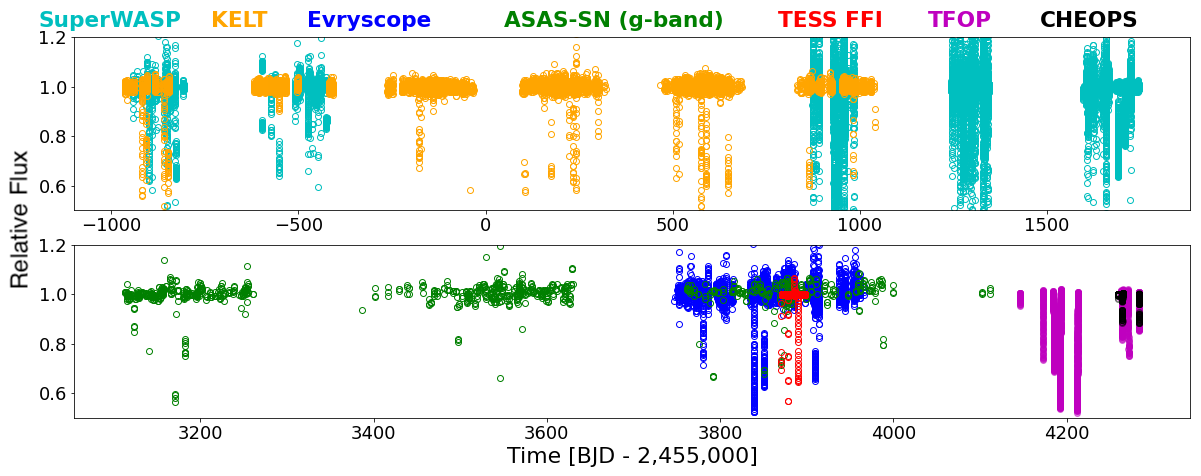}
    \caption{Archival and follow-up phototometric observations of \ticstar\ highlighting the baseline covered from the available data. The colors and symbols correspond to: SuperWASP (cyan circles), KELT (orange triangles), ASAS-SN (green upside-down triangles), Evryscope (blue crosses), TESS (red squares), and TFOP (magenta crosses). The binary eclipses are in detected in all datasets.}
    \label{fig:follow_up_}
\end{figure}

\begin{figure}
    \centering
    \includegraphics[width=1.\linewidth]{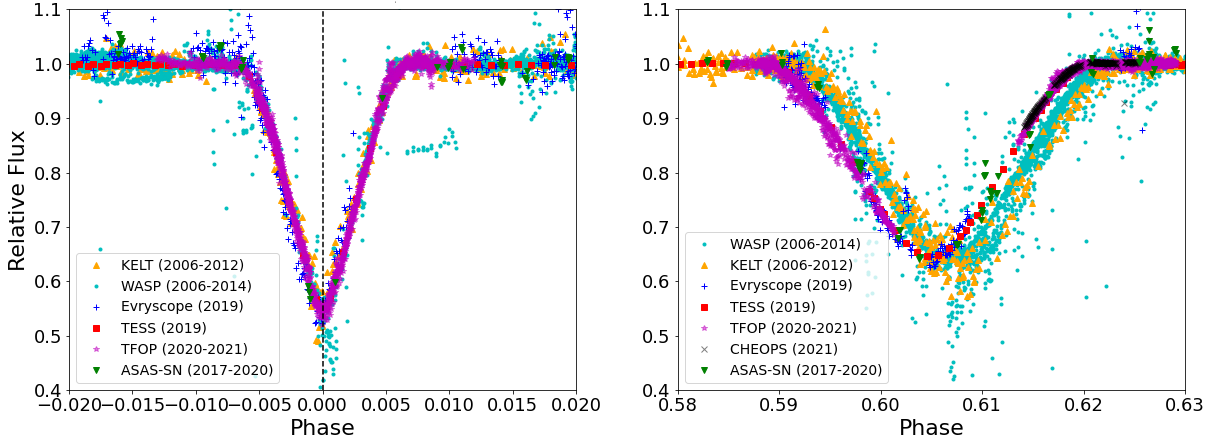}
    \includegraphics[width=1.\linewidth]{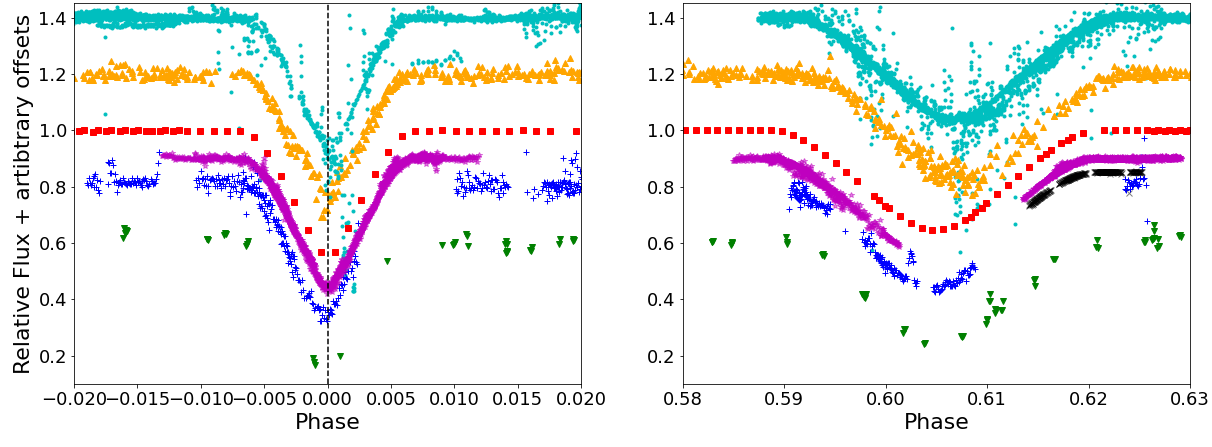}

    \caption{The photometric data shown in Figure \ref{fig:follow_up_} phase-folded on a linear period of $P = 19.65802$ days. The left panels shows the primary eclipse and the right panels show the secondary eclipse. The different data sets are vertically offset in the lower panels for clarity.  The phase change of the secondary eclipse relative to the primary---indicative of the apsidal motion of the binary---is clearly seen in the right panels.}
    \label{fig:archival_folded_post_DASCH}
\end{figure}

As a relatively bright target ($V = 10.141\pm 0.006$ mag), there is an extensive archive of photometric data on \ticstar\ from multiple sources and with long baselines. The target was observed by 
ASAS-SN (Shappee et al.\ 2014, Kochanek et al.\ 2017), 
Evryscope 
(Law et al.\ 2015; Ratzloff et al.\ 2019), 
KELT \citep{Pepper:2007}, and SuperWASP \citep{Pollacco2006}, 
and the stellar eclipses are detected in all datasets (see Figure \ref{fig:follow_up_}). Analysis of the archival data confirmed the orbital period of the EB estimated from TESS data and also showed a
clear change in the phase of the secondary eclipse relative to the primary eclipse
(see Figure \ref{fig:archival_folded_post_DASCH}). This phase change is indicative of apsidal precession where the argument of periaston
of the binary changes with time, and constrains the mass of the CBP.

SuperWASP-North was an array of 8 cameras located at the Roque de los Muchachos Observatory on La Palma. Each 200-mm/f1.8 lens was backed by 2k$\times$2k CCDs operating with a wide-band, 400--700-nm filter \citep{Pollacco2006}. TIC 172900988 was observed in 2006/07, accumulating 6000 photometric data points, then again in 2008, obtaining another 2000 points, and again from 2011/12 to 2014, accumulating 12\,000 more data points.  TIC 172900988 is the only bright star in the 48-arcsec extraction aperture.  

The Kilodegree Extremely Little Telescope \citep[KELT,][]{Pepper:2007, Pepper:2012} is a ground-based, small-aperture, wide-field photometric survey searching for transiting exoplanets, using a nonstandard passband equivalent to a broad $V+R+I$ filter.  It has a pixel scale of 23 arcsec/pixel, and the photometric reduction procedures are described in Siverd et al. (2012). The KELT-North telescope at Winer Observatory in Arizona observed \ticstar\ from 2006 October 27 to 2012 April 22, obtaining 8059 observations.

We used $g$'-band observations from Evryscope-North, located at Mount Laguna Observatory, approximately 44 miles (70 km) south-east of Palomar Observatory in southern California. The Evryscope pair of robotic all-sky telescopes, sited in Chile and California, are arrays of 6~cm telescopes designed to cover the entire accessible sky from each observing site in every exposure. Together, the Evryscopes survey an instantaneous field of view (FOV) of 16,512 square degrees at 2-minute cadence with 13\farcs2 pixel$^{-1}$ resolution. The Evryscope system design and survey strategy are described in Law et al.\ (2015) and Ratzloff et al.\ (2019). The data were filtered to exclude observations with signal-to-noise ratio less than 40. A total of 93 points were rejected (1.56\% of the observations). Eight additional points near HJD 2458909.84 were identified as outliers and omitted; unfortunately these were during a secondary eclipse. A total of 5656 points remained, and the HJD dates converted to BJD\_TBD using Jason Eastman's website tool\footnote{ \url{http://astroutils.astronomy.ohio-state.edu/time/utc2bjd.html}}
(Eastman et al. 2010).  The pipeline magnitudes were converted to normalized fluxes by dividing by the out-of-eclipse median, and the uncertainties in fluxes were obtained from the geometric mean of the upper and lower uncertainties in magnitudes.  The error bars were then scaled by a factor of 2.53 so that the uncertainties matched the rms scatter in the out-of-eclipse portions of the light curve.

\subsection{Follow-up Observations}

\subsubsection{Photometry}

With the archival data supporting our hypothesis that \ticstar\ was a CBP candidate based on the clear apsidal motion, we quickly turned our attention to obtaining follow-up observations. 
We contacted Dr.\ Dennis Conti of the American Association of 
Variable Star Observers (AAVSO)
who was able to rally the AAVSO observers on 2020 May 6 for observations later that night (May 7 UT; AAVSO Alert Notice 704). We obtained eight independent light curves from AAVSO members, plus an additional light curve from Mount Laguna Observatory. The observers were located across most of the North American continent and thus, 
despite the target setting early in the evening, we were able to record much of the eclipse. 
The observations were obtained in the BVRi bandpasses
and these data are discussed in Section 3.1. 
The details of the photometric observations are given in Table \ref{AAVSOtable}.

\begin{deluxetable}{llllll}
\tabletypesize{\scriptsize}
\tabletypesize{\tiny}
\tablecaption{Follow-Up Eclipse Observations\label{AAVSOtable}}
\tablewidth{0pt}
\tablehead{
\colhead{}  &
\colhead{} &
\colhead{}&
\colhead{} &
\colhead{} &
\colhead{}
}
\startdata
Observatory: & Astrosberge Private  & Boyce Astro Research & Lost Gold       & Terry Arnold Robotic  & RIT\\
             & Observatory          & Observatory (BARO)   & Observatory     & Observatory (TARO)    & Observatory\\
Observer:    & Serge Bergeron       & Pat Boyce            & Robert Buchheim & Scott Dixon           & Michael Richmond \\
Location:    & Casselman, Ontario, Canada & San Diego, CA  & Gold Canyon, AZ & Tierra Del Sol, CA    &	Rochester, NY \\
Aperture (m): & 0.305 & 0.431   & 0.400 & 0.400   & 0.300     \\   
Filter:       & V     & Sloan i & V     & Sloan i & Bessell R \\
Exposure time (s):& 45& 4.6     & 90    & 3       & 20        \\
Duration (hr):& 2.72  & 2.7     & 3.25  & 2.8     & 3.1       \\
CCD make/model: & SBIG ST10XME & FLI Proline 4710 Midband & ST-8XE & Apogee U16M 18803 & ATIK 11000 \\
Calibration     & AIJ +        & AIJ & MPO CANOPUS / & AIJ & XVista \\
 \ \ \ software: & VPhot 4.0.6 &     & PHOTORED      &     &        \\
\hline
Observatory: & Northbrook Meadow    & Paul and Jane Meyer  & Stellar Skies  & Mt.\ Laguna   &  \\
             & Observatory          &  Observatory (PJMO)  & Observatories  & Observatory   &  \\ 
Location:    & Northbrook, IL       & Waco, TX             & Pontotoc, TX   & Mt.\ Laguna, CA &  \\
Observer:    & Joe Ulowetz          & Bradley Walter       & Edward Wiley   & Mitchell Yenawine &\\
Aperture(m): & 0.235                & 0.610                & 0.280          & 1.00  & \\
Filter:      & V                    & R                    & B              & V   & \\
Exposure time (s): & 45             & 10                   & 60             & 3   & \\
Duration (hr): & 2.35               & 3.7                  & 3              & 4.2  &\\
CCD make/model:& QSI-583w          & Princeton Instruments & Moravian G1600 & e2v CCD42-40 & \\
               &                   & PIXIS 2048B eXcelon   &                &     &  \\
Calibration    & CCDStack2 +       & AIJ                   & AIJ            & AIJ  & \\
 \ \ \ software: & MaxIm DL5       &                       &                &     & \\
\enddata
\end{deluxetable}

We also acquired ground-based time-series follow-up photometry of TIC 172900988 as part of the TESS Follow-up Observing Program (TFOP)\footnote{\url{https://tess.mit.edu/followup}}. We used the {\tt TESS Transit Finder}, which is a customized version of the {\tt Tapir} software package \citep{Jensen2013}, to schedule our transit observations. The photometric data were extracted using {\tt AstroImageJ} \citep{Collins2017}. Observations were acquired using the Las Cumbres Observatory Global Telescope \citep[LCOGT;][]{Brown2013} network, Brigham Young Observatory, and Observatori Astron\`{o}mic Albany\`{a} as described in Table \ref{172900988sg1obs}.

\begin{table*}
\caption{TESS Follow-up Observing Program Observations\label{172900988sg1obs}}
\centering
\begin{tabular}{llcccccc}
\hline\hline
Observatory /  & Date & Filter & Exposure & Total &   Aperture & Pixel scale & FOV\\
Location  & (UTC) & & (sec) & (hrs) & (m) & (arcsec) & (arcmin)\\
\hline\\[-1.5mm]
{\it  Primary Observations }\\
 \cline{1-1}\\[-2mm]

LCOGT, Haleakala, Hawaii  & 2020 Nov. 19 & zs & 75 & 4.9 & 0.4 & 0.57 & $29.2\times19.5$ \\
LCOGT, Haleakala, Hawaii  & 2020 Dec. 09 & zs & 75 & 5.1 & 0.4 & 0.57 & $29.2\times19.5$ \\
Brigham Young Univ. Obs. & 2020 Dec. 09 & R & 30 & 6.8 & 0.2 & 1.42 & $25.7\times17.4$\\
LCOGT, Teide Obs., Canary Islands & 2020 Dec. 28 & zs & 75 & 7.5 & 0.4 & 0.57 & $29.2\times19.5$ \\
Observatori Astron\`{o}mic Albany\`{a}, Spain  & 2020 Dec. 28 & Ic & 120 & 5.6 & 0.4 & 1.44 & $36.0\times36.0$\\[2mm]

{\it  Secondary Observations }\\
 \cline{1-1}\\[-2mm]
 
LCOGT, Haleakala, Hawaii  & 2020 Oct. 23 & zs & 75 & 3.2 & 0.4 & 0.57 & $29.2\times19.5$ \\
LCOGT, CTIO, Chile  & 2020 Dec. 01 & zs & 18 & 2.5 & 1.0 & 0.39 & $26.5\times26.5$ \\
LCOGT, McDonald Obs., Texas  & 2020 Dec. 01 & zs & 18 & 6.4 & 1.0 & 0.39 & $26.5\times26.5$ \\

\hline
\end{tabular}
\end{table*}

As discussed below, our photodynamical model produced several solutions with nearly equal likelihood but a CBP orbital period that differs by a few days (other system parameters are mostly the same). Naturally, these solutions produced distinct predictions for the February 2021 conjunction of the CBP, separated by about a week and mutually exclusive. Catching these predicted transits from the ground is challenging due to their shallow depth, relatively wide window, and the usual diurnal and weather issues. Thus we pursued follow-up observations of the predicted transits with CHEOPS (Benz et al. 2020) through DDT program $\#004$. The combination of precision photometry, instrument stability, duty cycle and response time makes CHEOPS well-suited for these observations. The target was observed over the course of 
five visits, each about 5 hours long and partially covering an expected transit corresponding to four of the solutions mentioned above (see Figure \ref{fig:follow_up_}). The observations were processed with the CHEOPS data reduction pipeline (Hoyer et al. 2020), and their analysis is presented in Section 3. 

\subsubsection{High Angular Resolution Imaging}

As part of our standard process for validating transiting exoplanets to assess the effect of any contamination by bound or unbound companions on the derived planetary radii \citep{ciardi2015}, we observed TIC-1729 with infrared high-resolution adaptive optics (AO) imaging at Palomar Observatory.  The Palomar Observatory observations were made with the PHARO instrument \citep{hayward2001} behind the natural guide star AO system P3K \citep{dekany2013} on 2021 February 24 UT in a standard 5-point quincunx dither pattern with steps of 5\arcsec.  Each dither position was observed three times, offset in position from each other by 0.5\arcsec\ for a total of 15 frames. 

The camera was in the narrow-angle mode with a full field of view of $\sim25\arcsec$ and a pixel scale of approximately $0.025\arcsec$ per pixel.   Observations were made in the narrow-band $Br-\gamma$ filter $(\lambda_o = 2.1686; \Delta\lambda = 0.0326\mu$m) with an integration time of 5.6 seconds per frame (118 seconds total). 

\begin{figure}
    \centering
    \includegraphics[width=0.95\linewidth]{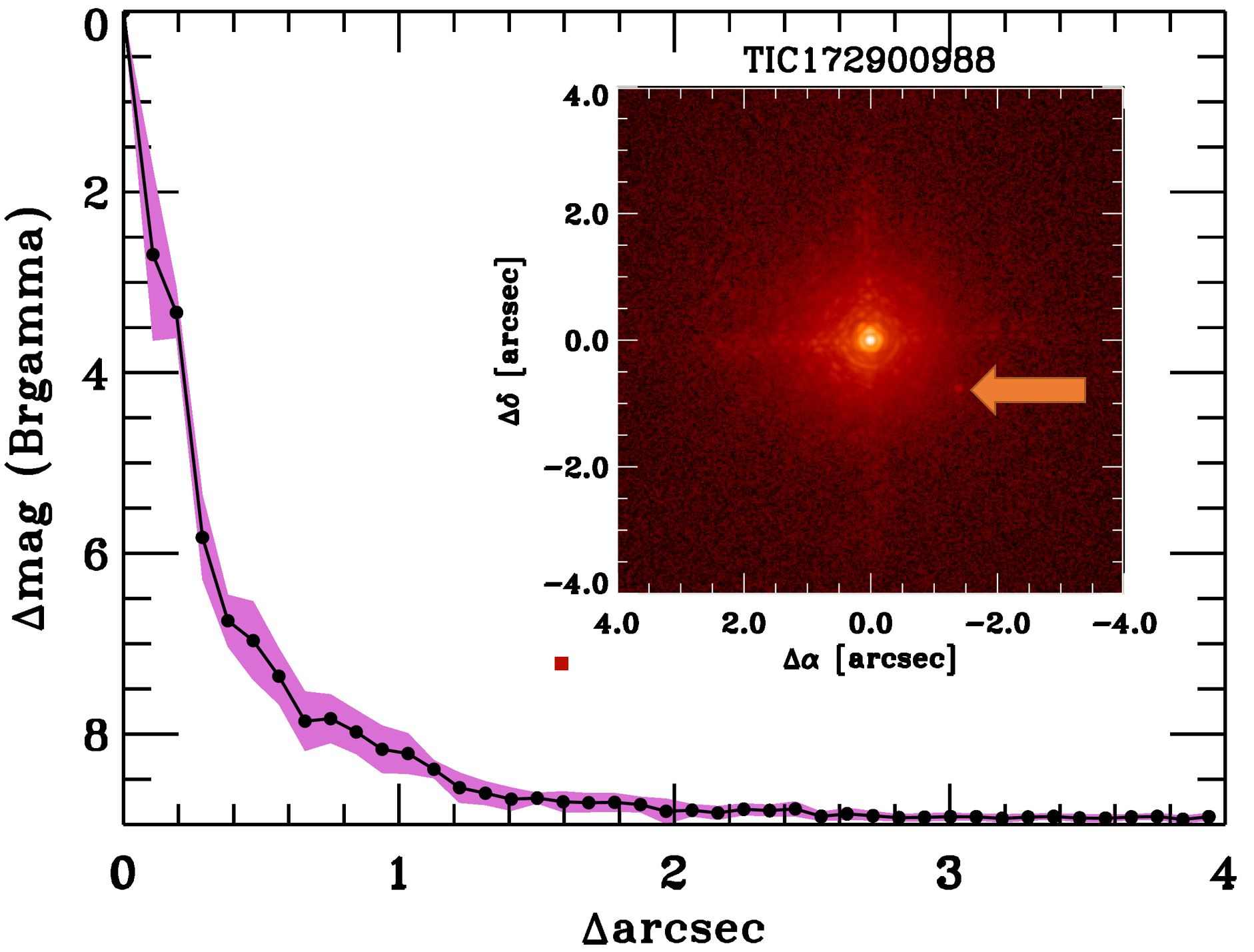}
    \caption{Companion sensitivity for the Palomar adaptive optics imaging.  The black points represent the 5$\sigma$ limits and are separated in steps of 1 FWHM ($\sim 0.097$\arcsec); the purple represents the azimuthal dispersion (1$\sigma$) of the contrast determinations (see text). The inset image is of the primary target. The detected companion is marked with an arrow and there are no additional companions to within 3\arcsec\ of the target. The location of the detected companion in the contrast plot ($\Delta$mag vs $\Delta$arcsec) is shown by the red square, such that the size of the square is approximately equal to uncertainty on the magnitude and separation.}
    \label{fig:ao_contrast}
\end{figure}

The AO data were processed and analyzed with a custom set of IDL tools.  The science frames were flat-fielded and sky-subtracted.  The flat fields were generated from a median average of dark subtracted flats taken on-sky.  The flats were normalized such that the median value of the flats is unity.  The sky frames were generated from the median average of the 15 dithered science frames; each science image was then sky-subtracted and flat-fielded.  The reduced science frames were combined into a single combined image using a intra-pixel interpolation that conserves flux, shifts the individual dithered frames by the appropriate fractional pixels, and median-coadds the frames.  The final resolution of the combined dither was determined from the full-width half-maximum of the point spread function to be $0.097\arcsec$ (Figure \ref{fig:ao_contrast}).

The sensitivities of the final combined AO image were determined by injecting simulated sources azimuthally around the primary target every $20^\circ $ at separations of integer multiples of the central source's FWHM \citep{furlan2017}. The brightness of each injected source was scaled until standard aperture photometry detected it with $5\sigma $ significance. The resulting brightness of the injected sources relative to the target set the contrast limits at that injection location. The final $5\sigma $ limit at each separation was determined from the average of all of the determined limits at that separation and the uncertainty on the limit was set by the rms dispersion of the azimuthal slices at a given radial distance (Figure \ref{fig:ao_contrast}).

A single faint companion was detected $1.59\pm0.01$\arcsec\ to the southwest at a position angle of $237^\circ\pm 1^\circ$.  The companion star is $\Delta Ks = 7.22 \pm 0.07$ mag fainter than the primary star, yielding an apparent magnitude of $K_2 = 16.0\pm0.1$ mag. 
Because the companion was only detected in the final reduced data, observations in a second filter were not obtained for a determination of the color temperature of the star. Further, the companion star is not detected in Gaia DR3. As such, we are unable to determine if the companion is bound to TIC 172900988 or if it is a background/foreground star. Based upon the analysis of the inner binary star, the two stars are a late-F and early-G main sequence star (see \S 4.1 below). Thus if the detected companion is bound, it has a relative K-band brightness consistent with the star being a late-M dwarf ($\sim$M8V). If it is indeed an M8V, the TESS magnitude to infrared color should be $Tmag-Kmag \approx 3.5$mag, making the real apparent TESS magnitude 10.5 magnitudes fainter than the combined brightness of the primary stars -- far too faint to be responsible for the observed planetary transits of $\approx 3000$ ppm. If the companion star is bound to the binary and the distance of the system is $\approx 240$pc ({\it Gaia} DR3 and see below), the M8V star ($M\approx 0.08 M_\odot$) has a projected separation of 380 au which would correspond to an orbital period of $\approx$ 5000 yr. If the companion star is not bound and has a typical color $Tmag-Kmag \approx 2$mag, it is still too faint to be responsible for the observed transits.

\subsubsection{Spectroscopy}

We obtained four spectra of \ticstar\ using the
Tull Coud\'e spectrograph (Tull et al.\ 1995)
on the Harlan J. Smith 2.7 m
telescope (HJST) at McDonald Observatory in west Texas.  Three spectra with net exposure times
of 1200 seconds each were obtained 2020 May 13 and one additional spectrum
with an exposure time of 1200 seconds was obtained 2020 May 14. The total
wavelength coverage was 3760 to 10,000 \AA\ with a resolving power of 
$R=60,000$.   An iodine ($I_2$) cell was used, and $I_2$ absorption lines are
superimposed on the stellar absorption lines between about 4992 \AA\ and
6200 \AA.  The spectra were reduced using custom scripts based on standard
IRAF\footnote{IRAF is distributed by the National Optical Astronomy Observatories, which are operated by the Association of Universities for Research in Astronomy, Inc., under cooperative agreement with the National Science Foundation.} 
routines.  The reduced spectra have signal-to-noise ratios of between
about 80 at 4400 \AA\ to 120 per pixel at 6500 \AA.   

Four spectra of \ticstar\ were 
obtained using the Tillinghast Reflector Echelle Spectrograph (TRES, 
Szentgyorgyi \& Fur{\'e}sz 2007; Buchave et al. 2010)
on the 1.5m telescope at the Fred L. Whipple Observatory in southern
Arizona, in October and December of 2020.
These spectra cover a wavelength range from about 3850
to 9100 \AA\ with a resolving power of $R\approx 44,000$, and have signal-to-noise
ratios of 30 to 35 per resolution element of 6.8~\kms\ near 5200 \AA.   The spectra were reduced
using the standard TRES data reduction pipeline described by Buchhave et al. (2010). 

We  obtained spectra of \ticstar\ on 2020 June 3 and
9 (one spectrum from each night) using the ARC Echelle 
Spectrograph \citep[ARCES;][]{Wang2003} on the Astrophysical Research 
Consortium 3.5-meter telescope at Apache 
Point Observatory (APO) in New Mexico. ARCES has a 
resolving power of $R\approx 32,500$ and a wide spectral coverage of 3200 \AA\
to 10,000 \AA. 
Each night we obtained three 5-minute exposures just after sunset at relatively high airmass ($\approx 2.5$--3). 
Quartz lamps were taken at the beginning of each half night, and thorium-argon lamps 
were taken at the beginning and end of each half night to perform wavelength calibration. 
The images were bias and dark subtracted using the Python package \texttt{ccdproc} (Craig et al.\ 2017). 
The spectral extraction, initial wavelength calibration, and normalization were 
performed using custom Python routines. The initial wavelength solution was performed by 
fitting a 4th-order polynomial individually for each order and 
then refined 
with a two dimensional 
dispersion solution using the \texttt{echelle} package of the IRAF software.  

Finally, eight spectra of the target were also obtained on the high-resolution spectrograph SOPHIE \citep{Purruchot2008}, which is mounted on the 193cm telescope at Observatoire de Haute-Provence (OHP), in France, and was designed and constructed to detect exoplanets \citep[e.g.][]{Bouchy2009}. The instrument has a long-term stability of $2\, {\rm m~s}^{-1}$. The spectra were all taken with an exposure time of 900s between 2020 November 15 and 2020 December 6 to cover several phases of the binary period. 

\begin{figure}
    \centering
    \includegraphics[width=0.95\linewidth]{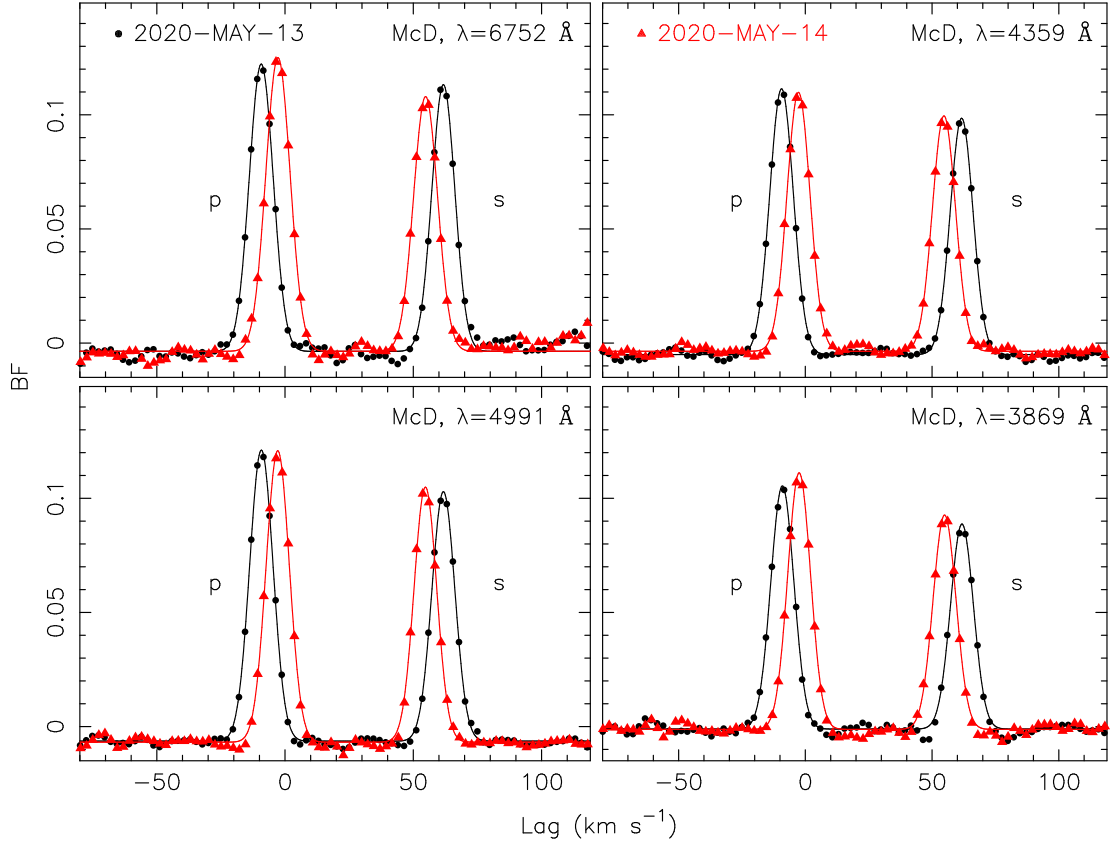}
    \caption{Broadening Functions (BFs) derived from the McDonald spectra
    are shown for four echelle orders (the central wavelengths are indicated 
    in each panel).  The BFs themselves are shown with the symbols, and the 
    solid lines show the best fitting double Gaussian functions. The peaks from the two components are clearly resolved and marked with a 'p' for the peak
    due to the primary star and an 's' for the peak due to the secondary star. 
    The Doppler 
    shift between the first
    spectrum from the night of 2020 May 13 (filled circles) and the spectrum from May 14 (filled triangles) is
    evident. No other peaks from additional stars are evident.}
    \label{fig:showBF_McD}
\end{figure}
\begin{figure}
    \centering
    \includegraphics[width=0.95\linewidth]{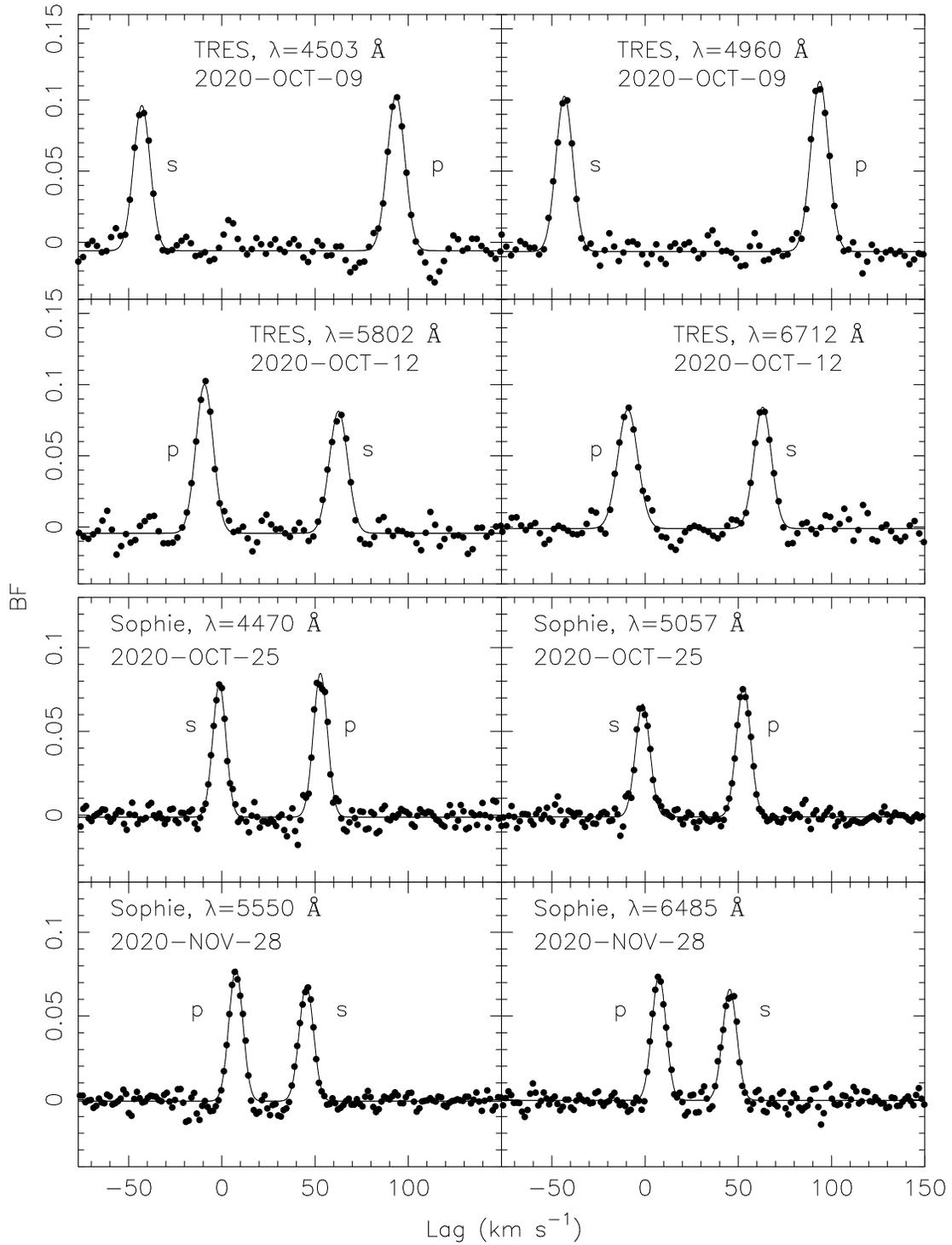}
    \caption{Similar to Figure \protect{\ref{fig:showBF_McD}}, but for the TRES spectra (top 4 panels) and the Sophie spectra (bottom 4 panels).}
    \label{fig:showBF_TRES}
\end{figure}
\begin{figure}
    \centering
    \includegraphics[width=0.95\linewidth]{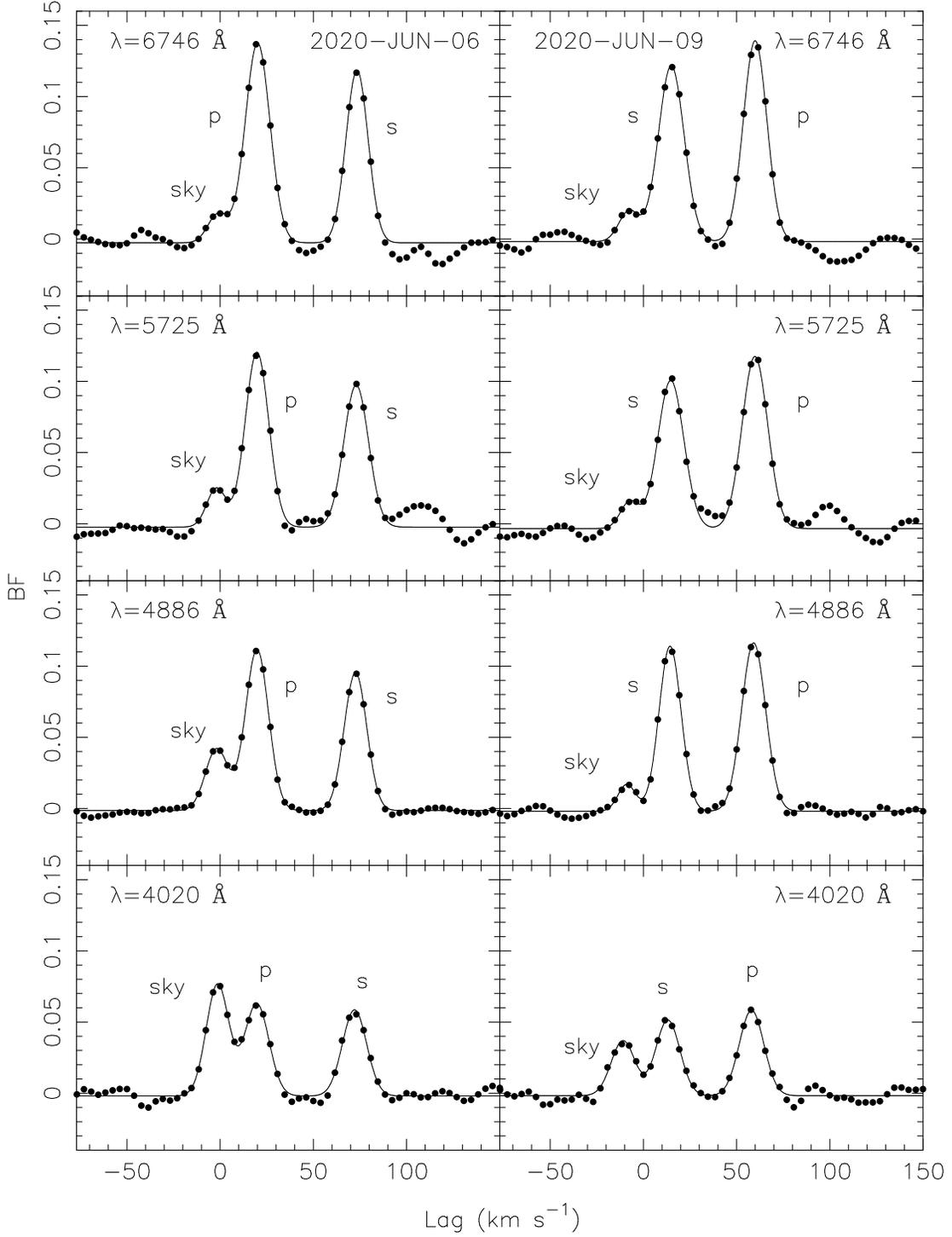}
    \caption{Similar to Figure \protect{\ref{fig:showBF_McD}}, but for the
    ARCES spectra.  In this case, an extra peak in the BFs due to sky contamination is seen near zero velocity.  The strength of that peak depends on
    the wavelength as it generally gets stronger for the bluer orders. 
    For these BFs, a triple Gaussian model was used to find the velocities of each peak.}
    \label{fig:showBF_ARCES}
\end{figure}

There are no indications of additional stars in any of the obtained spectra. To extract the radial velocities of each star from the spectra we used the ``broadening function'' (BF) technique (Rucinski 1992, 2002).  A high signal-to-noise spectrum of a slowly 
rotating star is needed as a template for the BF analysis.
For the McDonald spectra,
high signal-to-noise spectra
($\approx 1000$ per pixel) of diffused daytime sky near the zenith are routinely obtained, and
we used one such spectrum obtained on 2020 May 12 as the template. This spectrum was obtained with the ``Solar Port'', which is a ground quartz diffuser on the roof of the coude slit room.  This sends the light falling on it into the spectrograph.  That light is a combination of direct sunlight (the dominant component) and general daytime skylight. A spectrum of HD 182488 taken 2018 May 31 was used as the template for the TRES
spectrum. A spectrum of HD 185144 taken 2021 June 13 was used as the template
for the Sophie spectra.
Finally, a spectrum of the twilight sky was used as the template
for the ARCES spectra.    The BFs were extracted separately for each order 
(in preparation for this analysis, the spectra were normalized to their local
continua using spline fitting as implemented in IRAF).  
Representative BFs are shown 
for each instrument in Figures 
\ref{fig:showBF_McD},  \ref{fig:showBF_TRES}, and \ref{fig:showBF_ARCES}.  
With the exception of two Sophie observations
(those from 2020 October 31 and 2020 November 11),
the BF peaks from the primary and secondary stars are cleanly resolved, 
and no other peaks from additional stars (either bound to the system or along the 
line of sight) are evident.  The BFs from the ARCES spectra 
(Figure \ref{fig:showBF_ARCES}) show a third component near zero velocity from night sky background.

\begin{figure}
    \centering
    \includegraphics[width=0.95\linewidth]{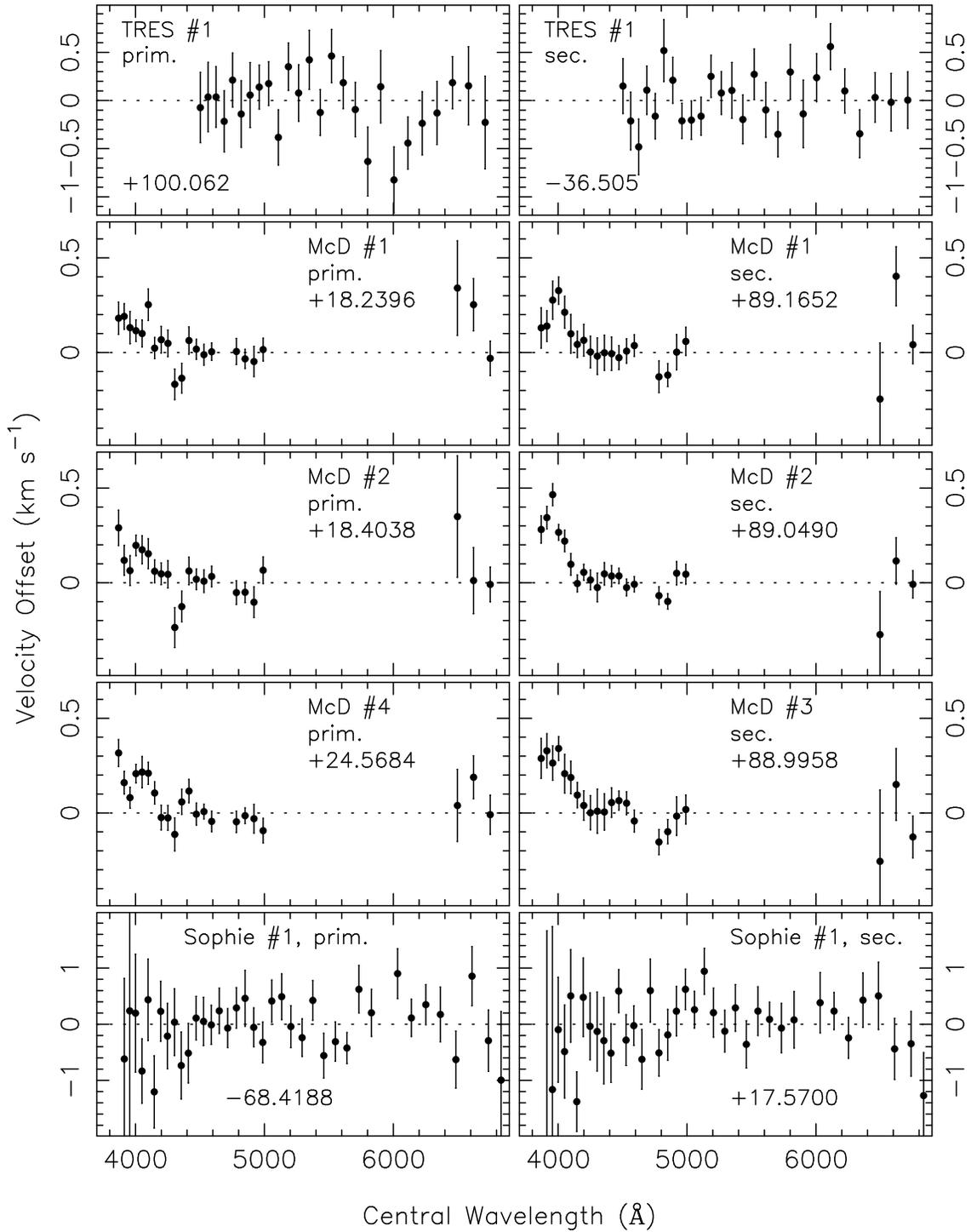}
    \caption{The radial velocities found from each echelle order as a function of the central wavelength,
    relative to the final adopted value for the first
    TRES spectrum
    (top panels), three of the McDonald spectra
    (middle panels), and the first Sophie spectrum
    (bottom panels). The velocities for
    the primary are on the left and the velocities for the secondary
    are on the right.}
    \label{fig:showwavetrendall}
\end{figure}

\begin{figure}
    \centering
    \includegraphics[width=0.95\linewidth]{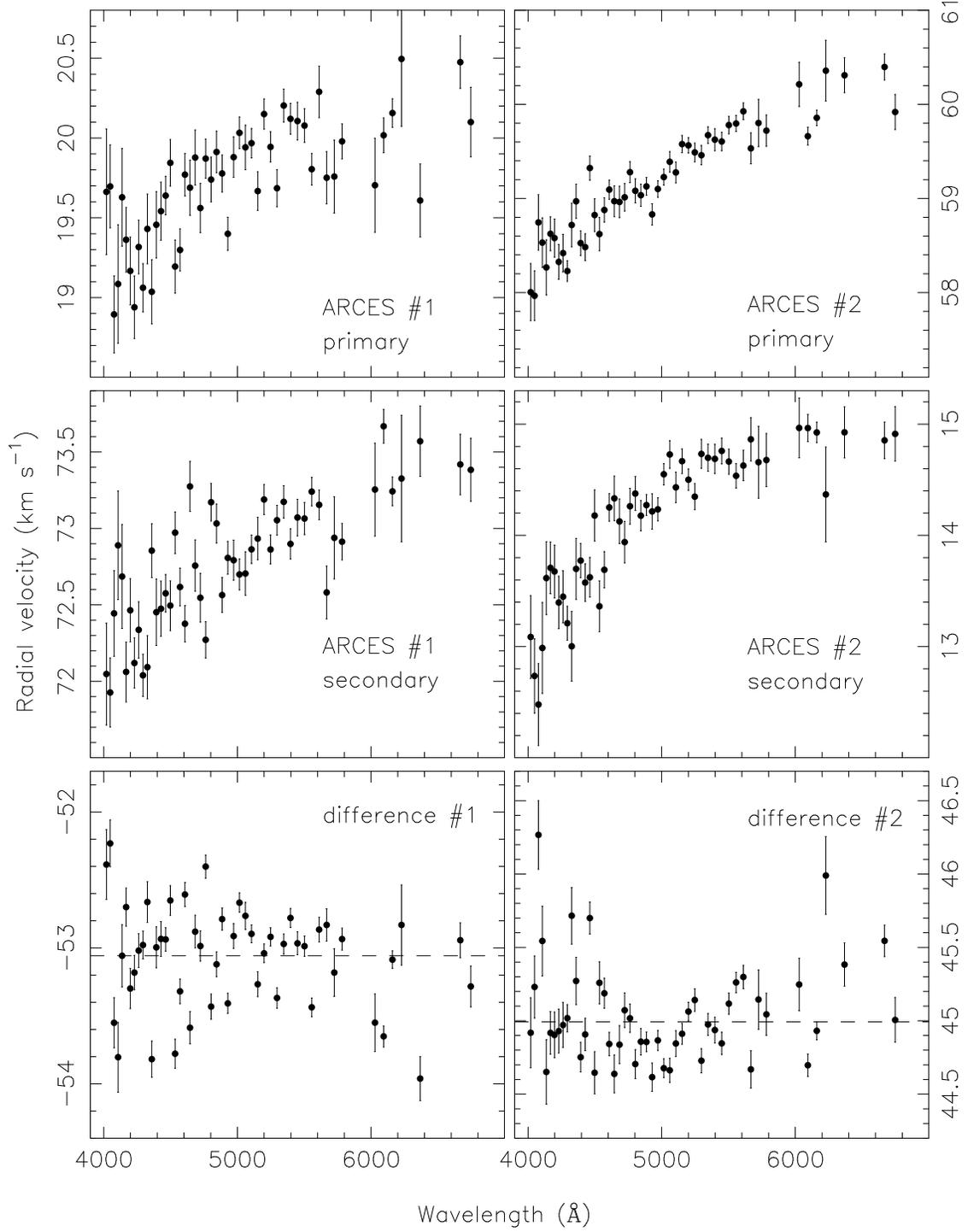}
    \caption{The radial velocities found from each echelle order as a function of the central wavelength for the 2020 June 3
ARCES spectrum (left panels) and the 2020 June 9 ARCES spectrum (right panels). There are clear trends in both the primary
and secondary velocities with wavelength (top two panels). In contrast, there is no apparent trend in the difference between the
primary and secondary velocity with wavelength (bottom panels). The dashed lines show the respective weighted averages of
the differences.}
    \label{fig:ARCES_RVs}
\end{figure}

\begin{figure}
    \centering
    \includegraphics[width=0.95\linewidth]{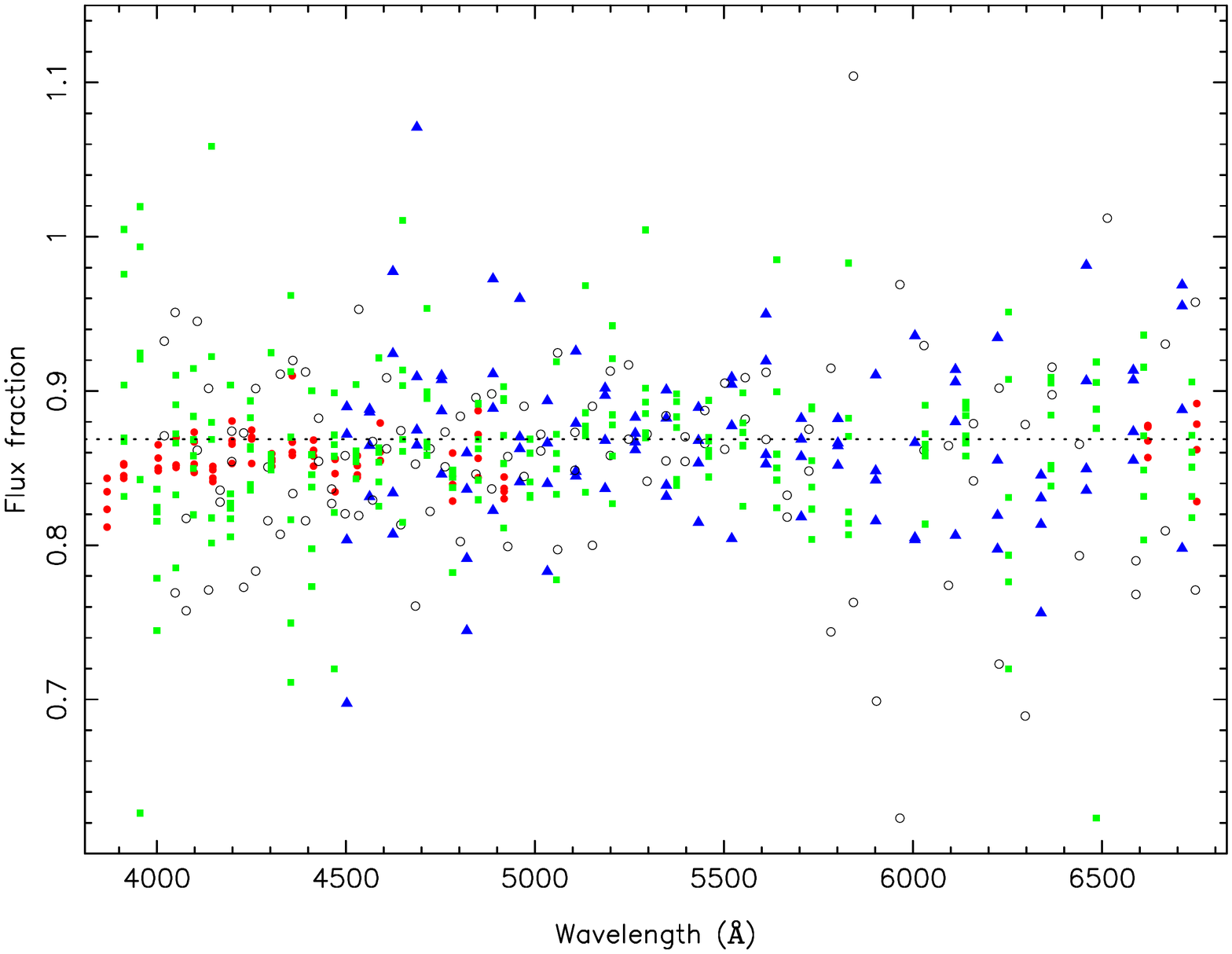}
    \caption{The flux fractions (the secondary flux divided 
    by the primary flux) as a function of wavelength for the
    McDonald spectra (red filled circles), the TRES spectra
    (blue filled triangles), the Sophie spectra
    (filled green squares) and the ARCES spectra (black open
    circles).  The two stars have similar spectra energy distributions as there are no trends with wavelength.  The dotted horizontal line is the adopted
    flux fraction of 0.8669 for all bandpasses used in the 
    photodynamical analysis.}
    \label{fig:plotfluxfrac}
\end{figure}

To obtain the final velocities for each spectrum, the following
procedures were used. For the TRES spectra,
we used 26 orders covering roughly 4500 to 6700 \AA.  The BFs were fitted with a
double Gaussian model to find the velocity shifts and widths. 
There were no apparent 
trends with central wavelength 
(Figure \ref{fig:showwavetrendall}).  A horizontal line was 
fitted to the velocities 
and the error bars on the individual measurements were scaled to give 
$\rm \chi^2_{min} = N_{dof}$ for this fit.  After the errors were scaled, a weighted average 
was computed to get the velocity for the primary and the velocity for the secondary
(Table \ref{RVtable}). A similar procedure was used
for the Sophie spectra, where we used 39 orders
covering roughly 3950 \AA\ to 6830 \AA.  The adopted radial
velocities for the six double-lined spectra are reported in Table \ref{RVtable}.

For the McDonald spectra, we
used 22 orders covering roughly 3900 to 6700 \AA.  Orders 42 and 43 were skipped owing to
a ``picket fence"
artifact caused by internal reflections, and orders 23 to 37 were skipped owing to the 
iodine lines (while the pipeline for extracting precise radial velocities for single-lined
spectra taken with the $I_2$ cell is well developed, a similar pipeline for double-lined
spectra is not quite perfected).  The BFs were fitted with double Gaussian models to find the 
velocity shifts and widths.  There was an odd trend with wavelength in that the six bluest 
orders had velocities that were shifted to slightly larger values 
(Figure \ref{fig:showwavetrendall}).  
This systematic 
shift seemed to be 
stronger in velocities for the secondary.  Given this 
systematic shift, we skipped the 
first six bluest orders when computing the final velocities.  
As before, a horizontal line was fitted 
to the remaining velocities, and the individual error bars were scaled to give $\rm \chi^2_{min} = N_{dof}$.  
After the errors were scaled, a weighted average was computed to get the velocity for the 
primary and the velocity for the secondary (Table \ref{RVtable}).

The ARCES spectra were the most challenging to extract radial velocities from.
There were 51 orders used, covering from about 
4000 \AA\ to about 6700 \AA\ (a few of the orders with strong
telluric absorption lines were skipped).
After the BFs were found, it was apparent that there was residual sky 
contamination,
and that the sky contamination was 
stronger in the bluer orders.  Given the extra peak, a 
triple Gaussian model was used to fit the BFs (Figure \ref{fig:showBF_ARCES}).  
There was a clear and strong trend seen in the velocities from individual orders 
(Figure \ref{fig:ARCES_RVs}).    
Although the wavelengths of the sky peaks seemed 
to be roughly constant with the echelle order, the velocities for both the primary 
and the secondary showed an almost linear trend with the central wavelength of the 
echelle order. This trend was most likely a result of atmospheric refraction at high air mass as the slit width (${\rm 1.6\arcsec}$) was larger than the seeing-limited image of the star (${\rm \approx1\arcsec}$) on the given nights, resulting in a chromatic shift in the stellar line spread function not seen in the residual sky spectrum which evenly filled the slit. The velocities at the reddest orders were nearly 2 km s$^{-1}$ larger than the velocities 
at the bluest orders.  However, the velocity difference between the two peaks stayed reasonably constant over wavelength. Thus, for each night,  
we measured the separation between the two peaks in a manner similar to what was done for the TRES spectra and used the velocity differences in the photodynamical fitting described below.

The flux fraction, which 
we define as the
secondary flux divided by the primary flux,
is a useful constraint on the photodynamical model.  The BFs
in each echelle order were modeled either with a double
Gaussian model (McDonald, Sophie, and TRES) or a triple Gaussian
model (ARCES), and the ratio of the areas under each peak was
taken to be the flux fraction for that echelle order.
Figure \ref{fig:plotfluxfrac} shows the individual
flux fractions as a function of the central wavelengths.
The two stars are pretty similar, and there are no
apparent trends in the flux fractions over these
wavelengths ($\approx 3900$~\AA\ to $\approx 6800$~\AA).  The average 
flux fractions are $ 0.8687 \pm 0.0052$ for TRES, 
$0.8555 \pm  0.0018$ for the McDonald spectra, $0.8632\pm 0.0043$ for Sophie,
and $0.8544\pm 0.0079$
for ARCES. To adopt a flux fraction for the
photodynamical model, we
combined the flux fractions from the TRES and Sophie instruments
and computed a weighted average of $0.8669\pm 0.0037$.
The McDonald flux fractions were excluded since
the iodine-free echelle orders have very little overlap
with the filters used for the photometry 
(the $B$-band with a peak near $\approx  4650$ \AA\
and filters redward of this).  The ARCES flux fractions
were excluded owing the contamination from night sky lines.
Since the stars are very similar, we adopted
a flux fraction of $0.8669\pm 0.0037$ for all bandpasses
in the photodynamical analysis
described below.


\section{Photodynamical Analysis of the Light and Velocity Curves}
\label{sec:ELC}

We used the Eclipsing Light Curve
(ELC) code of Orosz \& Hauschildt (2000) to model the light
and radial velocity curves of \ticstar\ (see Orosz et al.\ 2019 for a recent
discussion of the modifications relevant for the analysis of circumbinary 
systems).  We performed a full  photodynamical analysis in which the sky 
positions of
the various bodies as a function of time are found by using a numerical integrator to solve
the Newtonian equations of motion.  
The equations of motion have
extra terms to account for precession due to General Relativity, for precession due to tidal bulges on each
star (Eggleton, Kiseleva, \& Hut 1998; Mardling \& Lin 2002), and light-travel time across the orbits.  The numerical integrator that ELC uses is a symplectic,
12th-order Gaussian Runge–Kutta (GRK12) routine based on
methods and codes devised by Hairer et al. (2006).  For \ticstar,
the stepsize for the GRK12 routine was 0.05 days.  After the integrator is
run, we have the sky positions of each body as a function of time.
Then, given the sky positions and
information about the radiative properties of the components, the observed
light curves can be computed using the algorithm discussed
in Short et al.\ (2018), assuming spherical bodies with quadratic
limb darkening laws.

\subsection{Observational Data Through 2020}

We have light curves in 8 distinct bandpasses:  the TESS bandpass, $B$, $V$, $R$, Sloan $i$ (see Table \ref{AAVSOtable}), Sloan $g$ (ASAS-SN and Evryscope),
KELT, and WASP.  A secondary eclipse was partially covered 
at the start of the TESS observations, and these data were not used in the fitting owing to difficulties with normalizing the light curve. We only want to fit photometric data near an
eclipse or a transit event, and the selection of observations
near these events from the archival data requires a good
ephemeris.  Fortunately it
was not difficult to find a suitable ephemeris
($P=19.658025$, $T_0=3878.33860$, where the 
time convention is BJD $-$ 2,455,000)
using the TESS data, the RV data, and some of the early
WASP data where partially covered primary eclipses are 
seen.  Using this ephemeris, 
we selected data with phases within 0.02 of 
phase zero (e.g.\ the primary eclipse) and also with phases between 0.575 and 0.645
(e.g.\ the secondary eclipse) from
the KELT, WASP, and Evryscope data sets. All of the photometric data were converted to
flux units as necessary, and then
normalized to have an out-of-eclipse flux of 1.0.  In an effort to facilitate
the fitting, we also fit for the measured times of the primary eclipse, the secondary
eclipse, and the two transits in the TESS data.  Experience has shown us that
it is important to include the transit times in the fitting
because the $\chi^2$ penalty of the light curve fit does not change
once the predicted model transit does not overlap with the observed transit. When including the times, the likelihood of a model increases as the predicted
transits get closer to the observed ones.
We also have
5 ``sets'' of radial velocities:  McDonald, TRES, Sophie,
ARCES \#1, 
and ARCES \#2 (in the case of ARCES, we are fitting for the velocity
difference between the two components).   
The systemic velocities for each set were fit separately as nuisance parameters, with primary and secondary velocities initially forced to have the same systemic velocity offset within each set.  

For the model setup for \ticstar, we have a total of 54 free parameters.
There are parameters related to the orbit of the planet:  the orbital period $P_2$, the time of barycentric (inferior) conjunction $T_{\rm conj,2}$, the eccentricity parameters $\sqrt{e_2}\cos\omega_2$ and
$\sqrt{e_2}\sin\omega_2$, the inclination angle $i_2$, and the  nodal
angle $\Omega_2$.  There are similar parameters related
to the binary orbit: the orbital period $P$, the time of barycentric
conjunction $T_{\rm conj}$, the eccentricity parameters
$\sqrt{e}\cos\omega$ and $\sqrt{e}\sin\omega$, and the inclination angle
$i$ (the nodal angle of the binary is set to $\Omega=0$).  The
mass and radius of the planet are specified by the mass ratio $Q_3\equiv(M_1+M_2)/M_3$
(i.e., the ratio of the binary mass to the planet mass) and the radius ratio $R_1/R_3$ (primary radius to planet radius).  The component
masses and radii for the binary are specified by the primary mass
$M_1$, the binary mass ratio $Q\equiv M_2/M_1$, the primary
radius $R_1$, and the radius ratio $R_1/R_2$.  The effective temperatures of the two
stars are specified by the primary temperature $T_1$ and the temperature
ratio $T_2/T_1$.  We have light curves in eight bandpasses, so consequently
we have 32 sets of limb darkening coefficients with two per star
per bandpass.  The
standard quadratic limb darkening law given by
$$
I/I_0 = 1 -  u_1(1-\mu) + u_2 (1-\mu)^2
$$
was used (where $\mu = \cos\theta$ is the projected distance from the
center of the stellar disk), but with the ``triangular'' sampling
technique of Kipping (2013) with coefficients given by
$q_1=(u_1+u_2)^2$ and $q_2=0.5u_1(u_1+u_2)^{-1}$.
The remaining free parameters are the apsidal constants
for the primary ($k_{2,1}$) and secondary ($k_{2,2}$), and a 
contamination parameter to account for the other sources of light in
the TESS aperture.  

ELC has several fitting algorithms available.  For \ticstar, 
the algorithms we used were a simple MCMC algorithm
outlined in Tegmark et al.\ (2004), a ``differential
evolution'' MCMC algorithm (DE-MCMC, Ter Braak 2006), and the
nested sampling algorithm outlined by Skilling (2006).  In the case of
\ticstar, the size of the parameter space to search is relatively large
given that most of the orbital parameters of the planet are not immediately
known. The planet's orbit is presumably close to edge-on given the two observed
transits, but other than that the range of possibilities for the planet's orbital period,
eccentricity, and argument of periastron are fairly large.  Regarding the orbital
period of the planet, we can compute
the minimum orbital period based on dynamical stability, and we find 
$P_{\rm 3,min}\approx 129.7$
days \citep{Quarles2018}.  As discussed in Kostov et al.\
(2020b), the orbital period of the planet can be estimated from the timings and
durations of the two transits and the detailed knowledge of the binary's orbit. Using this method, we estimated that the planet's orbital period is likely in
the range between 150 and 340 days for eccentricity smaller than 0.2 (details presented below).

Our basic strategy was to use the nested sampling algorithm to find possible solutions, and then to use the MCMC and the DE-MCMC algorithms to refine them.  In
the nested sampling algorithm, one defines the parameter space by giving
minimum and maximum values of the free parameters.  One then randomly 
selects ``live points'',
where the number of them ($N_{\rm live}$) should be at least twice the number of
free parameters ($N_{\rm parm}$).  The likelihood values for each live point are computed.  The
``sampling'' part of the nested sampling algorithm is relatively straightforward.
Select a point in parameter space at random, and evaluate the likelihood.  If that
likelihood is better than the likelihood of the worst live point, then replace that live point with
the new point.  The discarded live point goes into the list of ``dead points''.  
If the new point has a worse likelihood than that of the worst live point, then
that new point is
discarded and not considered again.  To make the sampling practical
in a reasonably short period of time, the 
$N_{\rm parm}$-dimensional hyper-ellipsoid that bounds the live points is
computed, and new possible points are randomly selected from the 
hyper-ellipsoid (for
large dimensions, the volume of the hyper-ellipsoid can be orders of magnitude smaller than the volume of the hyper-cube defined by the ranges
of each fitting parameter).  As better live points are found, the volume
of the hyper-ellipsoid shrinks and potential replacement points are sampled from increasingly relevant regions of parameter space.  
Skilling (2006) showed that if the volume of the hyper-ellipsoid is successively scaled by a factor proportional to $1/N_{\rm live}$,
then the live points and dead points can be used to compute the Bayesian evidence
and also to construct a posterior sample.  
Unfortunately, when the number of parameters $N_{\rm parm}$
is larger than about 20, the statistically scaled volume of the hyper-ellipsoid
is usually orders of magnitude larger than the actual volume of the hyper-ellipsoid.  
As a result, in can take a very long time (a few weeks of wall-clock time or longer
for cases like \ticstar)
to run the nested sampling algorithm
to completion, even when using a parallelized version that runs on multiple 
computer cores   since
the random samples usually come from regions of parameter space that are far from
the region that has the largest likelihood.  
In our implementation of the nested sampling algorithm, we have the option of
omitting the statistical scaling of the volume of the hyper-ellipsoid.  The advantage
of this is that it is much faster to find replacements for the live points since
the volume from which the random samples are drawn is generally
much smaller.  Although
we forfeit the ability to compute a posterior sample or to compute the evidence,
the increased speed at which the algorithm can find good solutions as starting
points for other algorithms makes up for this shortcoming.  

\begin{figure*}
    \centering
    \includegraphics[width=0.71\linewidth,angle=-90]{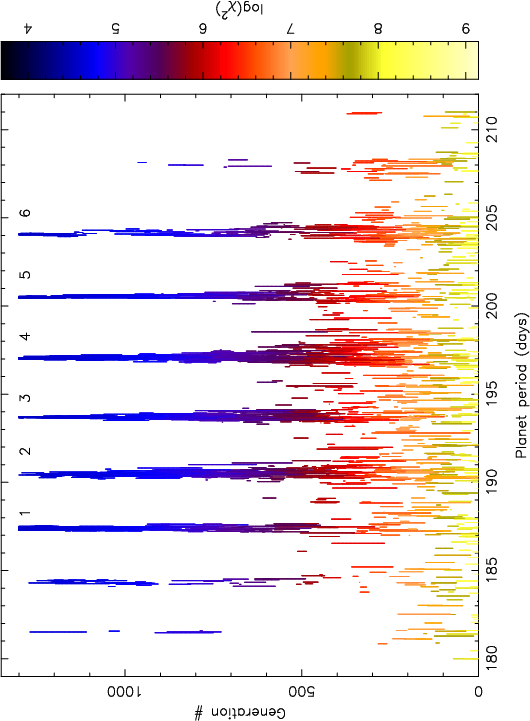}
    \caption{Nested sampling photodynamical solutions showing the evolution of the live points as a function of the CBP period. 
    The colors of the points give the $\chi^2$ values of the points, as indicated in the color bar.  Points with lighter colors have relatively large values of $\chi^2$, and points with the darker colors have relatively small values of $\chi^2$. As better solutions are found by the algorithm, the live points evolve to regions of higher likelihood. For the CBP period, there are six families (``fingers'') that remain populated and have comparable likelihood.  The number next to the right of each finger indicates the family number shown in Tables \ref{fitted} and \ref{derived}. The fingers with periods near $\approx 181$, 184, 208, and 211 days have many fewer
    live points in them, and the live points that remain there
    have
    lower likelihoods than the live points in the fingers
    labeled 1 through 6.  For these reasons these outlying
    fingers were not considered further.
    Note that this is a representative figure based on the 2020 data; the ``fingers'' persist after the inclusion of the 2021 data but recreating this figure with all available data proved to be computationally prohibitive. The best-fit planet periods shift by about a day between the numbers in the figure and those listed in Tables \ref{fitted} and \ref{derived}. See text for details.   
    }
    \label{fig:fingers_}
\end{figure*}

Our quest to find good model solutions proceeded by iteration.  When running the nested sampling algorithm, we kept the limb darkening
coefficients fixed, 
which left us with 22 free parameters and 44 live points.
It is usually the case that when modeling circumbinary systems, reasonably good
parameters for the binary can be found quickly, whereas good parameters for
the planet (its orbital parameters and its mass) take longer to find.  
Once a good solution for the binary's parameters was
found in the first run of the nested sampling, we used the MCMC and the DE-MCMC algorithms to refine the solution, which
included the limb darkening coefficients.    With better values for the limb
darkening parameters, the nested sampling was run again with the limb darkening
coefficients fixed at better values.  For the second iteration of
the nested sampling, we used 66 live points.  As the progress of the
second run was monitored, it was found that not all values of the orbital
period of the planet near the best value were possible.  Instead, there
are distinct families of solutions as shown in Figure \ref{fig:fingers_}.  The figure shows the evolution of the live points as they are replaced with better models. We note that this is a representative figure based on the 2020 data. The ``fingers'' persist after the inclusion of the 2021 data (discussed below) but recreating this figure with all available data proved to be computationally prohibitive. However, we confirmed the pattern by using brute-force steppers in the planet period and found that regions of high ${\rm \chi^2}$ quickly emerge between the fingers, and the best-fit planet periods shift by about a day (compare location of fingers here to the periods listed in Table 6). At first (small $y$-values in the plot), the planet's period among the live points fills the range between about 180 and 210 days nearly uniformly.  As better live points are found (moving up the $y$-axis), distinct ``fingers'' develop, and at the end, six families of solutions remain (top of the $y$-axis). This represents a problem for the nested sampling algorithm. The range of the planet periods is still fairly large, so the hyper-ellipsoid cannot become very small. Moreover, many, if not most, of the randomly selected points will fall in between the
fingers where the likelihood is not very high.  As a result, the convergence
of the algorithm is slowed.  In cases like this, the ``multinest'' routine
discussed by Feroz, Hobson, \& Bridges (2009) might offer relief, but
we do not have an implementation of multinest within ELC.

As an aside, we do not have a good explanation of why we see these families of solutions
with distinct values of the planet's period, as opposed to a smooth distribution
of points with a range of $\approx 10$ to 20\% around the best value
that one might expect based on the discussion in Kostov et al.\ (2020b). It is tempting to think that these discrete solutions are somehow analogous to some kind of aliasing caused by cycle count ambiguities over the relatively long time baseline we have (about 5600 days or about 15 years). However, the fingers still appear when we only include the TESS data and the follow-up photometric data that were taken after the transits were identified. We note that the CBP Kepler-34 produced a ``1-2 punch'' where a pair of transits about 5 days apart were observed with {\em Kepler} (Welsh et al.\ 2012), but when we fit only those two transits (along with the stellar eclipses and radial velocity data) using the nested sampling in a similar manner that was used on \ticstar, we found only a broad range of possible planet periods with no distinct and separate families of solutions. Thus, it appears that the appearance of these distinct families of solutions is related to the peculiarities of \ticstar\ itself rather than being a generic feature only having a pair of observed transits with a few days between them.

These distinct families of solutions with regions of low likelihood
between them also present problems for the MCMC and the DE-MCMC algorithms.
We therefore broke up the parameter space into six regions where the
range of the planet's period was restricted in such a way to isolate
each family, and leaving the ranges of the other 53 parameters to be the same (the identifying numbers for each family are
shown in Figure \ref{fig:fingers_}).
Using the best nested sampling solution from each family as an initial
seed (see Orosz et al.\ 2019 for a discussion on how the chains are
initialized), the DE-MCMC algorithm was run on each family.  
After these preliminary runs, we noticed two minor issues.
First, the residuals for
the secondary star radial velocities were systematically larger
than the residuals for the primary star radial velocities by
$\approx 75$ m s$^{-1}$.  Second, a comparison of the stellar
masses and radii with stellar evolution models
(see Section \ref{sec:stellarproperties} showed a poor fit.
We therefore made two modifications.  First, with a physical interpretation in mind such as convective blueshift and gravitational redshift, we allowed the
systemic radial velocities of the primary and secondary
star to vary independently for each radial velocity set.  
With this modification we can no longer use the ARCES velocities
since we were fitting for the difference between the primary and
secondary velocities.
Second,
we used the flux fraction as an observed constraint.  As discussed
previously, there is no apparent trend in the flux fractions
with wavelength, so we adopted a flux fraction of
$0.8669 \pm 0.0037$ for each bandpass. 

After these modifications, the DE-MCMC code was 
run seven times for each 
family, with each
run lasting 15,000 generations.  
After a generous burn-in period of 7500 generations, posterior samples were taken
from every 1500th generation for each run and combined.
For each posterior distribution (each of which had 
8064 samples), the numbers are sorted and we computed the median value
$X_{\rm med}$, the value $X_{15.85}$ that marks the point where 15.85\%
of the distribution is smaller, and the value of $X_{84.15}$ that marks
the point where 15.85\% of the distribution is larger
(i.e., 68.3\% of the distribution is between
$X_{15.85}$ and $X_{84.15}$).  The adopted parameter value
is taken to be the median $X_{\rm med}$, and the
$1\sigma$ uncertainty is taken to be the larger of $(X_{\rm med}-X_{15.85})$ or
$(X_{84.15}-X_{\rm med})$. 
The results for the fitted parameters are shown in Table
\ref{fitted}, and for some derived parameters of interest in Table
\ref{derived}.  From looking at Table \ref{fitted}, the solution for
Family 5 (where $P_3=200.452$ days) has the smallest overall 
$\chi^2$, with $\chi^2_{\rm min}=11,348.96$.  The solutions for the
other families are nearly as good with $\chi^2_{\rm min}$ values 
within $\approx 280$.
At this time it is difficult to choose between these these
families with high
confidence.  

\begin{figure*}
    \centering
    \includegraphics[width=0.91\linewidth,angle=0]{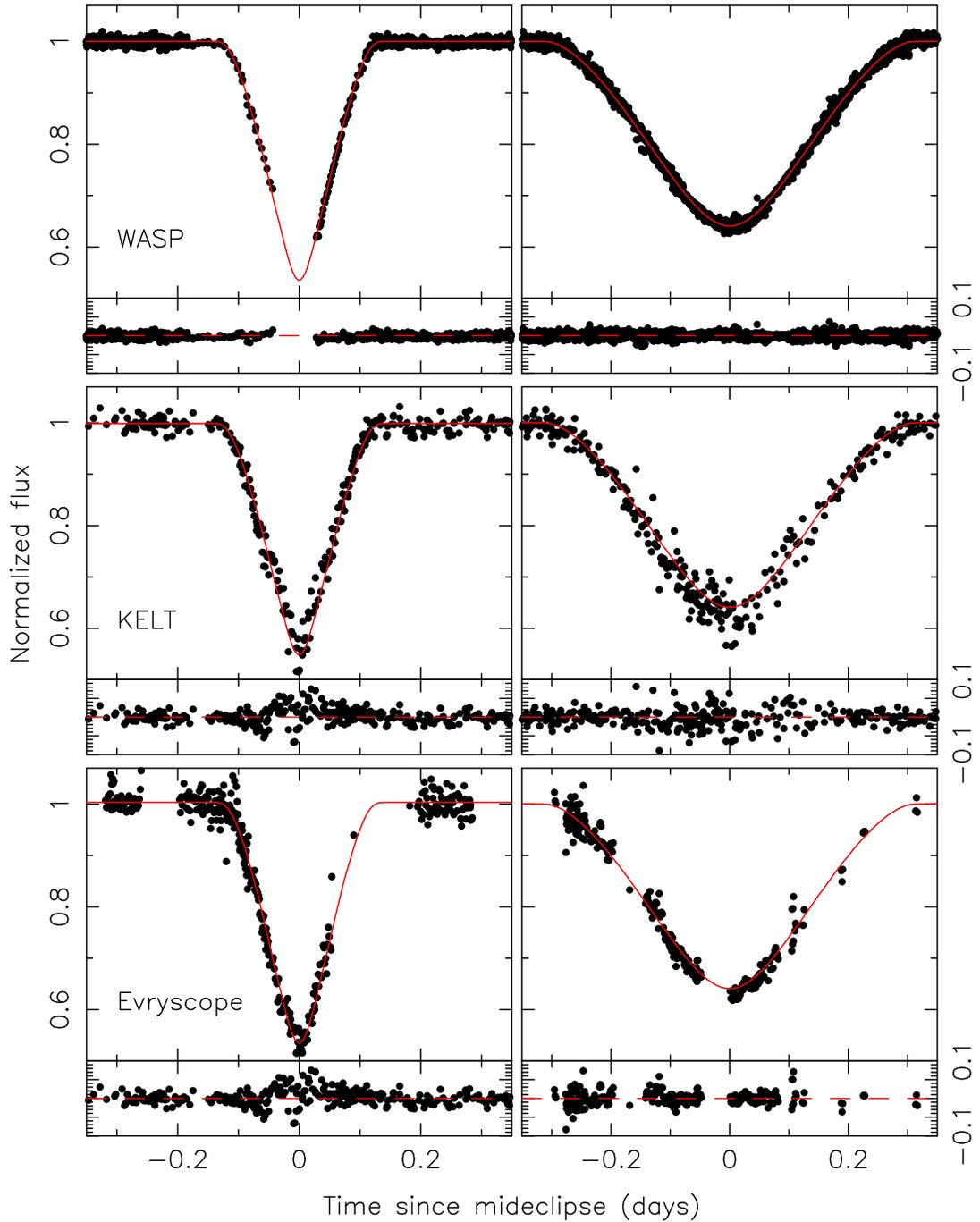}
    \caption{The archival photometric data during primary eclipse (left column) and
    secondary eclipse (right column).  From top to bottom, the WASP data, the KELT
    data, and Evryscope data are shown, where the time of each observation has been converted to the difference in time from the nearest corresponding model
    (Family 5)
    eclipse.  
    Comparing this figure to Figure \protect\ref{fig:archival_folded_post_DASCH}
    we see that the effects of the apsidal motion on the phase of the secondary
    have been accounted for in the model.}
    \label{fig:showarchive}
\end{figure*}
\begin{figure*}
    \centering
    \includegraphics[width=0.91\linewidth,angle=0]{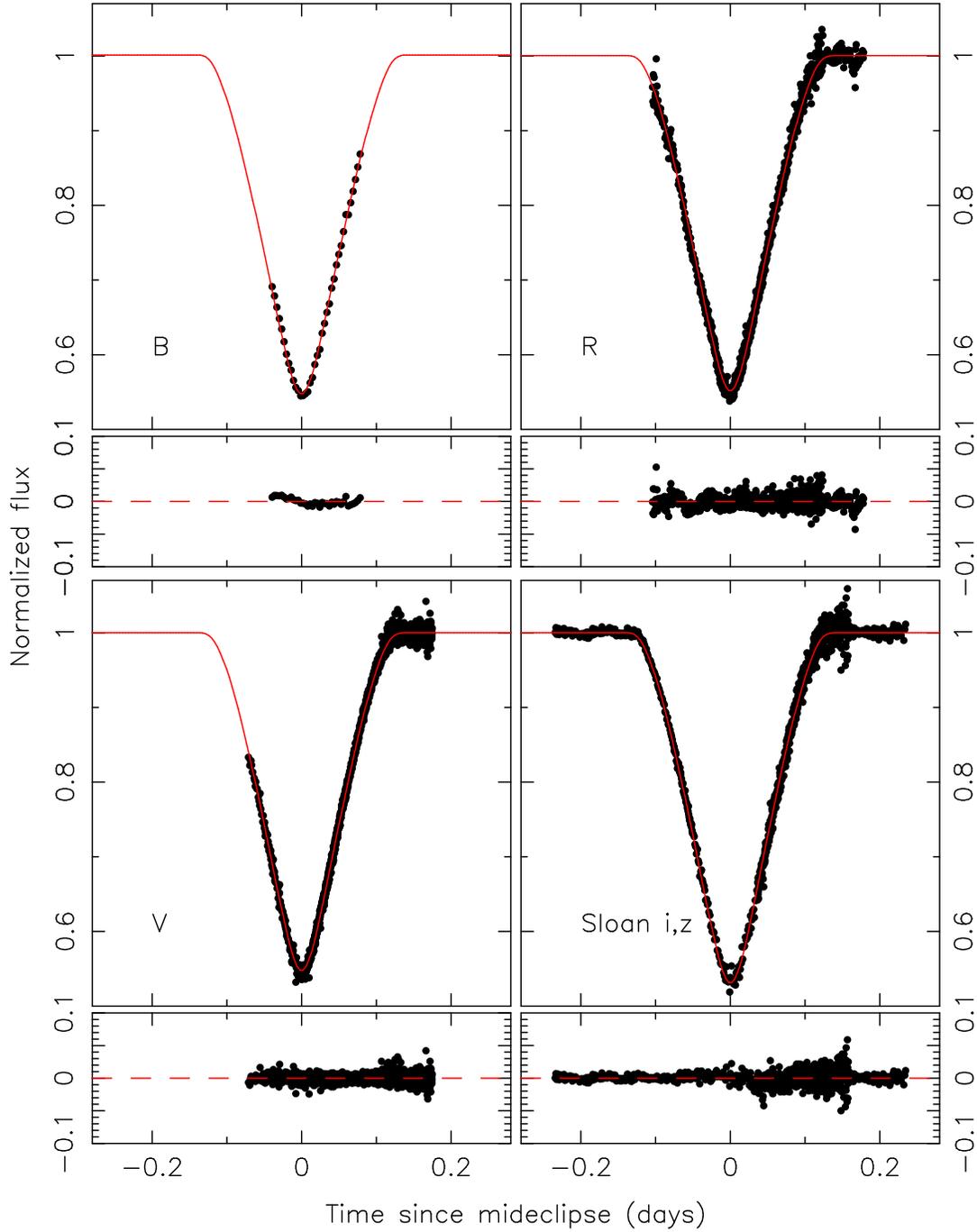}
    \caption{The ground-based follow-up photometry obtained after the transits
    were identified are shown for $B$ (top left), $V$ (bottom left),
    $R$ (top right), and Sloan $i$ and $z$ (bottom right) (see Tables 1 and 2 for details). The model curves
    are the best-fitting solution from Family 5.}
    \label{fig:showground}
\end{figure*}
\begin{figure*}
    \centering
    \includegraphics[width=0.91\linewidth,angle=0]{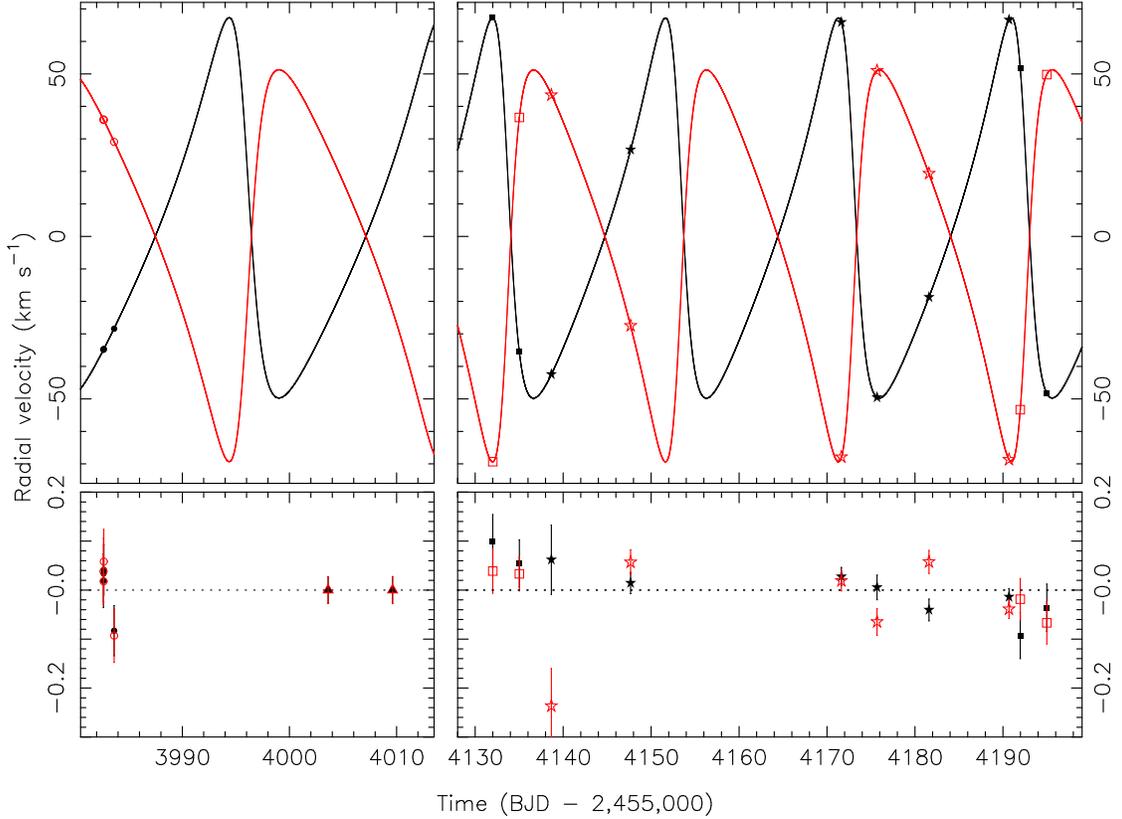}
    \caption{The radial velocities for the primary (black symbols) and the secondary
    (red symbols) and the best-fitting models from Family 5.
    The filled circles and open circles are the primary and 
    secondary velocities from the McDonald spectra, respectively;
    the filled and open triangles are the primary and secondary
    velocities from the ARCES spectra, respectively (these were not
    used in the final fitting);
    the filled and open squares are the primary and secondary velocities
    from the TRES spectra, respectively; and the filled and open stars
    are the primary and secondary velocities from the SOPHIE spectra,
    respectively.}
    \label{fig:showRV}
\end{figure*}

Figures \ref{fig:showTESSall}, \ref{fig:showarchive}, and \ref{fig:showground}
show the various photometric data and the best-fitting models (for Family 5).  
There do not appear to be any large systematic problems with the model fits.  Comparing Figure
\ref{fig:archival_folded_post_DASCH}, where the data are phased with a linear ephemeris, and Figure \ref{fig:showarchive}, where the times are relative to 
the nearest corresponding eclipse, we see that the apsidal motion is
accounted for in the model.
Finally, Figure \ref{fig:showRV} shows the radial velocities  and the best-fitting model (for Family 5).  The velocity residuals are all less than 200 meters
per second.  The residuals for the McDonald spectra are even tighter still, with a scatter of around 20 meters per second.   

In each of the three spectroscopic data sets the fitted center-of-mass velocities
for the primary and secondary components are systematically different, and are
always larger for the secondary. The differences (secondary minus primary) are
$+0.030 \pm 0.023$~\kms\ for McDonald, $+0.082 \pm 0.005$~\kms\ for TRES, and
$+0.076 \pm 0.008$~\kms\ for SOPHIE.
While in principle this could result from instrumental effects, the fact that the
sign is the same for all three instruments, and the magnitudes quite similar,
suggests an alternate explanation of an astrophysical nature.
The component properties (mass, radius, temperature) differ by less than 5\%, yet
this is enough to cause perceptible differences in the gravitational redshift,
which we expect to be larger for the secondary by 0.010~\kms. Although convective
blueshifts are not as easy to quantify, based on
the prescription by Dravins et al.\ (1999) we expect the effect to be
roughly 0.060~\kms\ smaller for the secondary, resulting in a more positive velocity
for that component. The two effects combine to give a larger expected velocity for
the secondary by about 0.070~\kms, which appears to explain the bulk of the shifts
mentioned above.


As noted earlier, the solution for Family 5 is formally the best.  
With one exception, there does
not seem to be any pattern in how the various parameters change from family
to family.  That one exception is the mass of the planet:  it is the smallest
%
($M_3 = 823.9\pm 4.8\,M_{\oplus}$ for Family 1 (where the planet's period
$P_3$ is the shortest) and largest 
%
%
($M_3 = 981.3\pm 5.7\,M_{\oplus}$)
for Family 6 (where the planet's period $P_3$ is the largest).  This trend is
easy to understand.  The planet is responsible for most of the apsidal motion
seen in the binary.  As the planet's orbital period gets longer, the planet's mass
needs to go up to have the same effect on the binary.

\begin{figure*}
    \centering
    \includegraphics[width=0.85\linewidth]{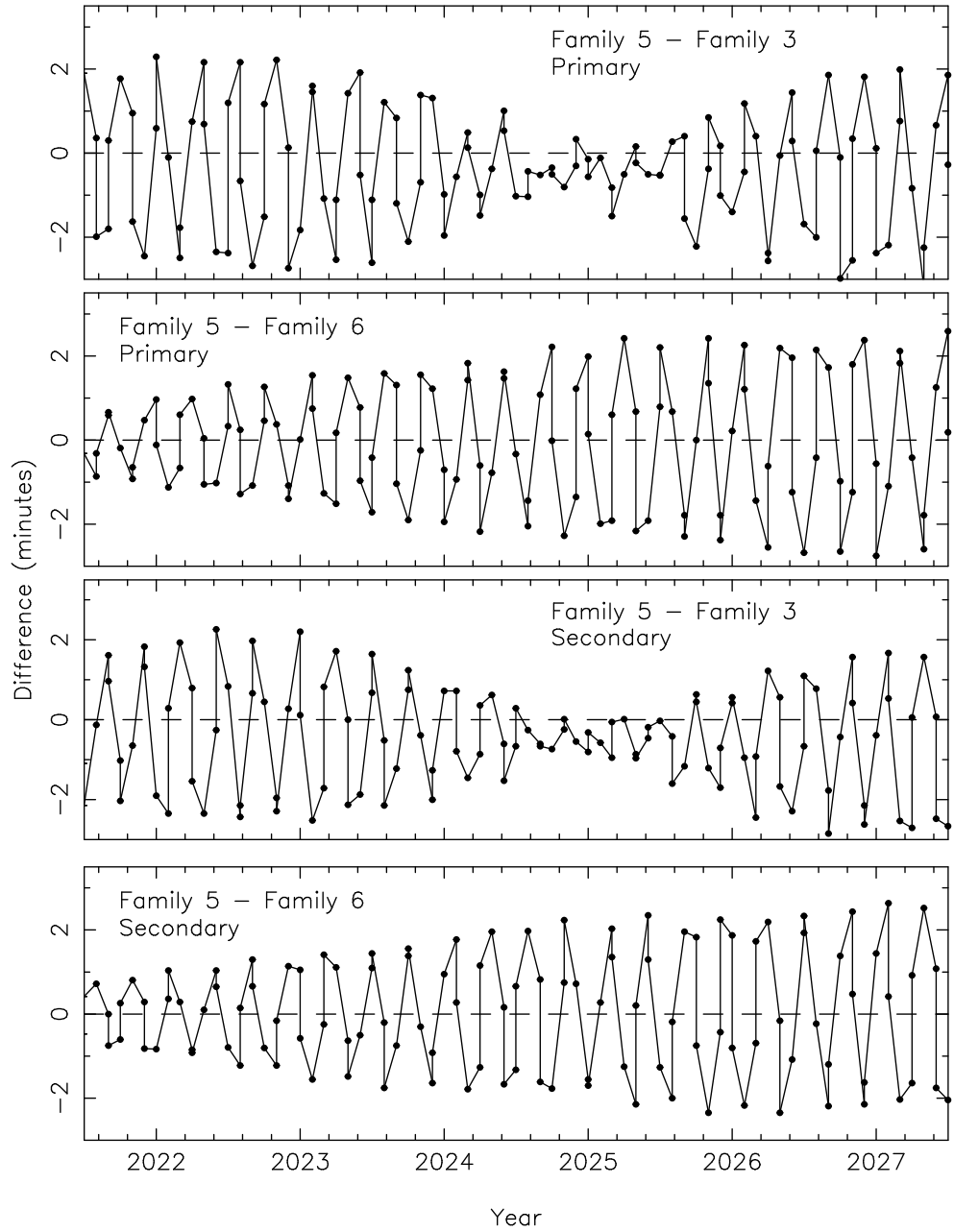}
    \caption{Differences in the times of predicted eclipses between the Family 3, Family 5, and Family 6 solutions out to the middle of the year 2027. 
    The differences can be up to $\approx 2$ minutes (the timing depends on the
    combination of which families are in the difference) and should be  measurable.}
    \label{fig:ploteclipsediff}
\end{figure*}

Future transit and eclipse observations will allow us to determine which
family represents the correct solution.  We used the models from the posterior
samples for each family to compute the times (and their uncertainties)
of future transits (see Table \ref{futuretransit}) and eclipses (see Tables
\ref{family1primary} through
\ref{family6secondary}
in the Appendix).  
As one might expect, the 
predicted times of the next pair of observable transits depends on the orbital
period of the planet.  
A clear detection of one or more transits in the future will unequivocally
tell us which family represents the true solution.    If future transit observations
are not available, then observations of future eclipses will eventually allow
us to discriminate between the various families.  Figure 
\ref{fig:ploteclipsediff} shows differences in the predicted eclipse times
between Family 5 and Family 3, and between Family 5 and Family 6.  The differences between the 
predicted times can  be on the order of one to
two minutes, which should be easy to detect.

Finally, we note that there is a distinction between the initial conditions
of the best fitting models for each family, and the parameters given in
Table \ref{fitted}, which come from the medians of the posterior distributions.  
The parameters in Table \ref{fitted} represent our best estimates of the parameters
and their statistical uncertainties.  However, those parameters collectively will
not necessarily produce an optimal fit to the data owing to the complexity of the model and correlations among 
the various parameters.  For the purposes of computing an optimal model, the
initial conditions for the best-fitting models should be used.  The initial
conditions for the best-fitting models
for all of the families
are given in Tables
\ref{Family1init} through
\ref{Family6init}
in the Appendix.  

\subsection{Observational Data for Predicted Transits in 2021 Feb-Mar}

\begin{figure*}
    \centering
    \includegraphics[width=0.85\linewidth]{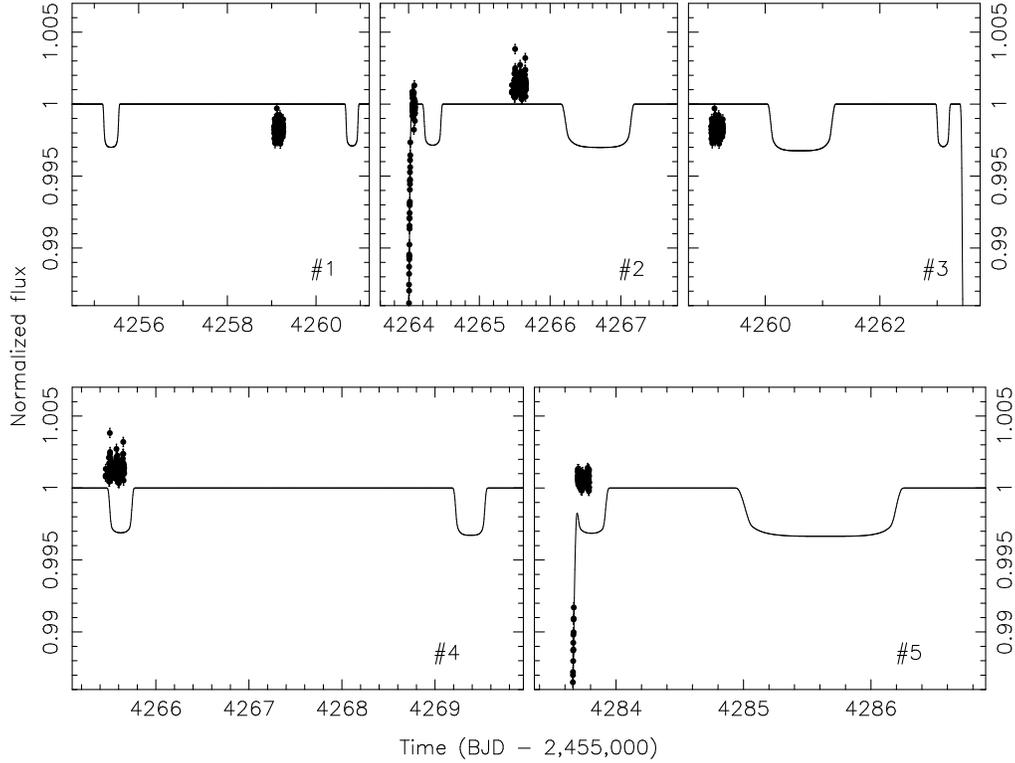}
    \caption{The five CHEOPS visits from February and March, 2021 along
    with the predicted eclipses from Families 1 through 5 (note the data near
    day 4259 are shown twice).  The egress part of an eclipse and some
    of the out-of-eclipse light curve were observed near day 4264, and
    the data from all five visits were normalized to that out-of-eclipse part
    of the light curve.  Small but significant zero-point offsets
    are seen.}
    \label{fig:cheopsfig1}
\end{figure*}

The photodynamical model predicted transits of the planet occurring during a conjunction in early 2021 (Feb-Mar, depending on the family). To detect the transits, we obtained four observations from CHEOPS via a generous allocation of Director's Discretionary Time. These data are shown in Figure \ref{fig:cheopsfig1}. We note that these observations were scheduled to be centered on the predicted transit ingress corresponding to the best-fit model for the respective family based on the radial velocity data available at the time, i.e. less than half of the RV measurements. As we received more measurements, the model fit to the binary star in particular improved and, as a result, some of the times of the predicted transits changed. This is the reason the current best-fit transit models for Family 1, 2, and 3 do not overlap with the CHEOPS observations. 

The egress part of an eclipse was observed near day 4264, and this observation
includes a small part of the out-of-eclipse light curve.  A fit to this
partial eclipse was used to normalize all of the CHEOPS data. CHEOPS data were obtained near the expected transits for the Family 4 and Family 5 solutions. Unfortunately, there are two problems that make the interpretation of these data difficult.  First, there are small but significant flux 
offsets between each visit as seen in Figure \ref{fig:cheopsfig1} (e.g.\ the observations covering out-of-eclipse phases in
panels 1, 2, and 3 have fluxes that are different by 
a few tenths of a percent). These offsets might be produced by either a genuine stellar variability (which would be undetectable in TESS data), or could represent a systematic instrumental effect. Second, the duration of each visit was relatively short. As a result, the data near day 4266 could be consistent with being at the bottom of the Family 4 transit, or could be consistent with the out-of-transit light curve for all other families. Likewise,
the predicted transit from the Family 5 solution partially overlaps with
the egress of an eclipse.  The light curve at the end of the
corresponding CHEOPS visit is flat, but that could be consistent with
being the out-of-eclipse light curve for Families 1 through 4 and Family 6,
or it could be consistent with the bottom of the transit for the Family 5
solution.

We attempted to separately fit the CHEOPS data near the Family 4 and Family
5 transits.  ELC is currently limited to 8 distinct filter band-passes, so we removed
the data for the $B$ filter and replaced the specific intensities
in the model atmosphere table with intensities for CHEOPS using
the appropriate filter response.  

\begin{figure*}
    \centering
    \includegraphics[width=0.85\linewidth]{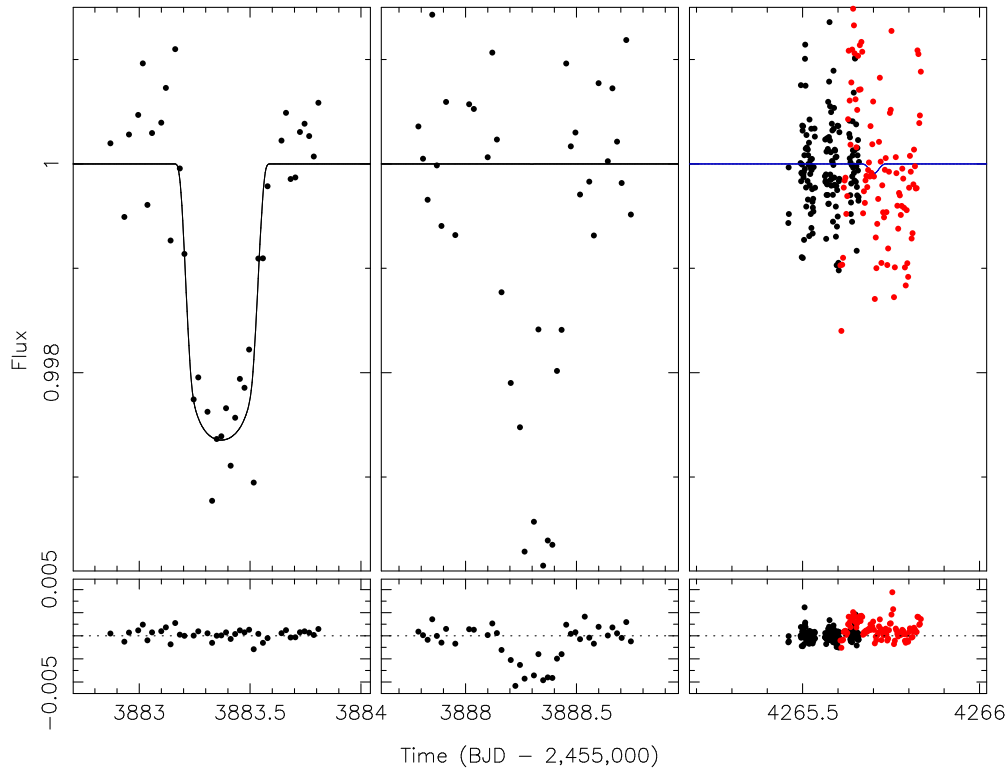}
    \caption{Left: The transit of the primary star observed by TESS according to the best-fitting Family 4 model and using only the CHEOPS data. Middle: same as the left panel but for the transit of the secondary star. Right: CHEOPS data (black points) and KeplerCam data (red points,
    binned to $\approx 2$ minutes) during a potential transit of the secondary for the Family 4 solution.  The best-fitting model is the solid line.  The transit of the secondary observed by TESS is not matched and thus Family 4 can be likely ruled out (see text for details).}
    \label{fig:showCHEOPS1}
\end{figure*}

Thanks to a fantastic response from the community, we also obtained several ground-based observations of the predicted transits. In particular, for the potential Family 4 transit we obtained observations from KeplerCam in the Sloan i filter. 
These data are consistent with a flat line, as seen in Figure \ref{fig:showCHEOPS1}, where we also show the results of the fitting of the potential Family 4 solution using both the CHEOPS (black points) and KeplerCam (red points). We note that it is possible to orient the planet's orbit in such a way that no transit is observable near day 4265.5, so the model near the CHEOPS data and KeplerCam data is nearly flat. However, this occurs at the expense of the fit to the transit of the secondary star observed by TESS (middle panel in the figure). As a result, we can likely rule out the Family 4 solution, although the confidence level is not as high as one would normally like.

\begin{figure*}
    \centering
    \includegraphics[width=0.85\linewidth]{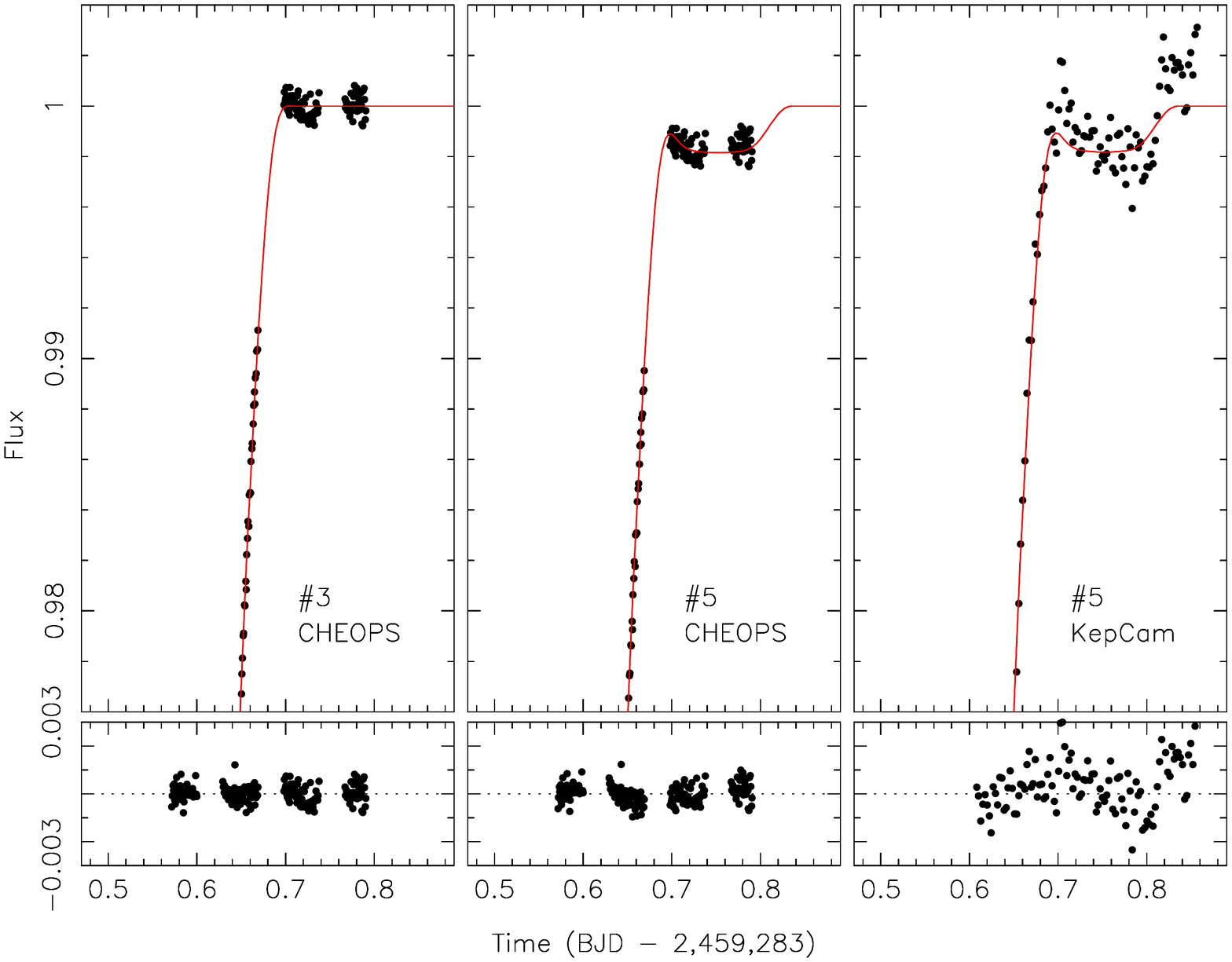}
    \caption{Left:  The CHEOPS data from day 4283 with the best-fitting
    Family 3 model.
    Middle:  The same CHEOPS data with the best-fitting Family 5 model.
    Right:  The nearly simultaneous KeplerCam data (binned to $\approx 2$
    minutes), with the best-fitting Family 5 model. Overall, Family 5 remains a viable solution. }
    \label{fig:showCHEOPS4}
\end{figure*}

For the potential Family 5 transit, we also have KeplerCam observations that are nearly simultaneous with the CHEOPS observations, so we fit these data in a similar manner as above. The results are shown in Figure \ref{fig:showCHEOPS4}. Here, the families where there is no transit blended with the egress of the eclipse fit the CHEOPS data better than does the Family 5 model ($\chi^2\approx 249$ vs.\ $\chi^2=290$ 
for Family 5). However, the Family 5 model fits the KeplerCam data better than the other families ($\chi^2 = 4089$ for Family 5 vs.\ $\chi^2\approx4381$ for the others). Given the relatively short time span of the follow-up observations and the noise present in the KeplerCam data (especially near the end when morning twilight was approaching), Family 5 remains a viable solution.

We also attempted to catch some of the predicted transits from TRAPPIST-North but the observations did not reach the necessary precision.

\section{Discussion}
\label{sec:discussion}

\subsection{Stellar properties and comparison with stellar evolution models}
\label{sec:stellarproperties}

We discuss here the determination of the effective temperatures of the binary components as well as the system metallicity, based on the TRES spectra of \ticstar\ described earlier. We followed a cross-correlation procedure similar to the one described by \cite{Torres:2002} to optimize the match of the observed spectrum against synthetic spectra, except that in this case the spectrum is double-lined, so instead of 1-D cross-correlations we used TODCOR \citep{Zucker:1994}, which is a two-dimensional correlation algorithm. This allows us to account for the presence of the secondary star in determining the stellar properties. Calculated templates were taken from a library of pre-computed spectra based on model atmospheres by R.\ L.\ Kurucz \citep[see][]{Nordstrom:1994, Latham:2002} covering the region centered on the \ion{Mg}{1}\,b triplet at 5187~\AA. We ran a grid of correlations solving for a mean effective temperature (strictly, a luminosity-weighted mean) and the rotational broadening of components, $V_{\rm rot} \sin i$, assumed to be the same for the two stars. Initial tests indicated the rotational broadening is below our spectral resolution, so we adopted $V_{\rm rot} \sin i = 0$~\kms. As discussed by \cite{Torres:2012}, the temperatures determined in this way tend to be strongly correlated with the surface gravity and with metallicity, with higher values of these parameters leading to hotter temperatures. Given that in this case the surface gravities are well determined from the photodynamical analysis, we fixed them to a common value of $\log g = 4.25$, near the average for the two components. On the assumption that the metallicity is solar, the resulting mean temperature we obtained is $\langle T_{\rm eff}\rangle = 5770$~K. Increasing the metallicity to ${\rm [Fe/H]} = +0.5$ gave a hotter mean temperature of $\langle T_{\rm eff}\rangle = 6140$~K, as expected. Estimated uncertainties for these values are 100~K. Individual temperatures for the components were then computed by using the temperature ratio as measured from the photodynamical analysis.

\begin{figure}
  \epsscale{0.75}
  \plotone{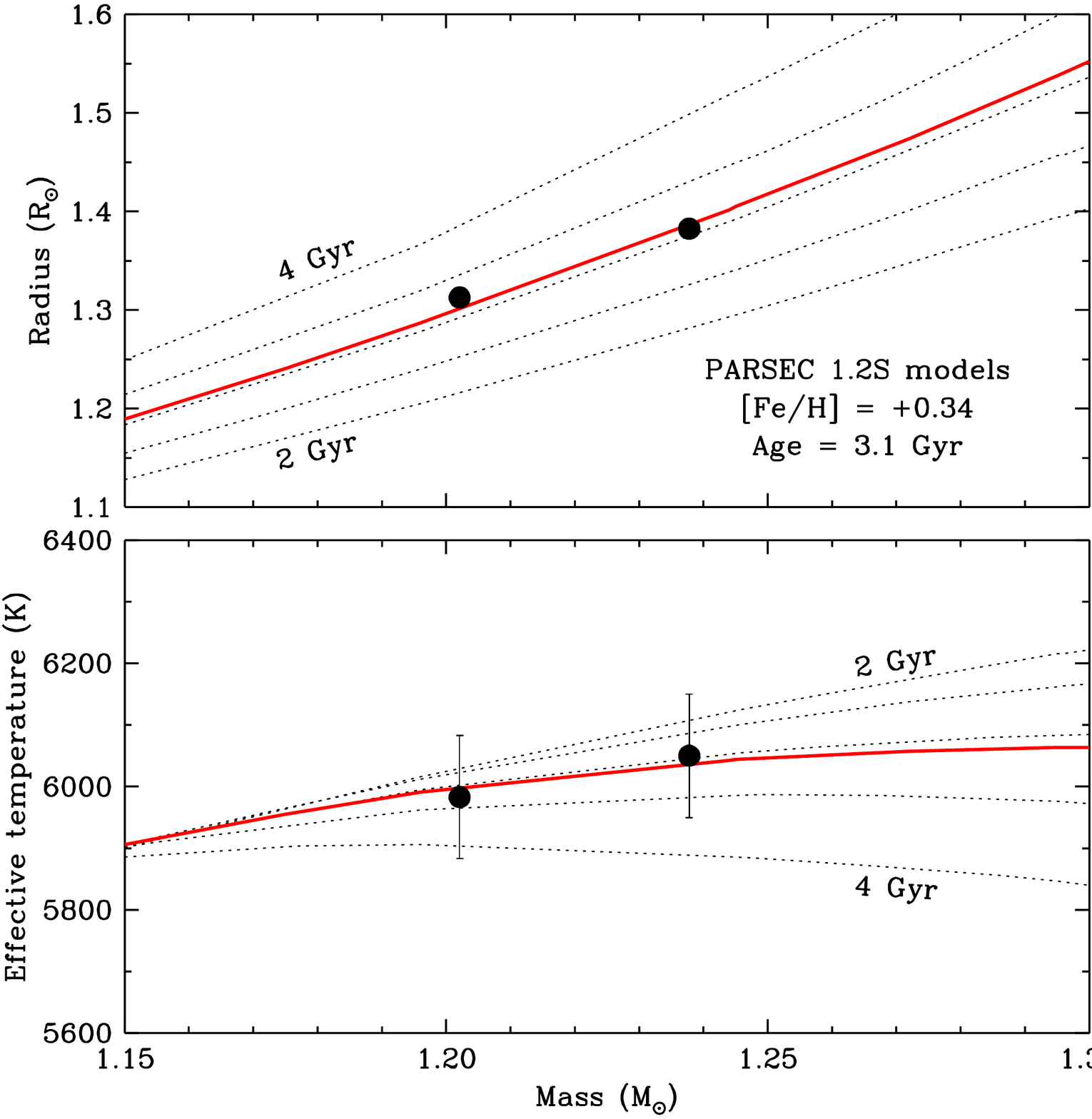}
  \figcaption{Comparison of the mass, radius, and temperature determinations for the components of \ticstar\ against stellar evolution models from the PARSEC series \citep{Chen:2014} for the best-fitting composition of ${\rm [Fe/H]} = +0.34$, and solar-scaled abundances (${\rm[\alpha/Fe]} = 0.0$).  \emph{Top:} Mass-radius diagram with isochrones from 2 to 4~Gyr every 0.5~Gyr, indicated with dotted lines. The best fit for an age of 3.1~Gyr is shown by the red line. Errorbars are smaller than the symbol size. \emph{Bottom:} Mass-temperature diagram showing the same isochrones as above.\label{fig:parsec}}
\end{figure}

A preliminary comparison against stellar evolution models from the PARSEC 1.2S series \citep{Chen:2014} indicated a satisfactory fit in the mass-radius plane for either metallicity (at different ages), but revealed that the predicted temperatures were hotter than we obtained when assuming solar metallicity, and cooler than we obtained when adopting the higher composition of ${\rm [Fe/H]} = +0.5$. We then explored intermediate metallicity values, and found good agreement in both the mass-radius and mass-temperature diagrams for a composition of ${\rm [Fe/H]} = +0.34 \pm 0.10$. The mean system temperature at this composition is 6030~K, and the individual values are 6050 and 5983~K for the primary and secondary, respectively, corresponding to spectral types of approximately F9 and G0. This fit is shown in Figure~\ref{fig:parsec}, in which model isochrones are plotted for ages between 2 and 4~Gyr, in steps of 0.5~Gyr. The best simultaneous agreement with all observations is achieved for an age of about 3.1~Gyr, according to these models.

As an independent test on the absolute temperatures, we gathered photometry for \ticstar\ from the literature and constructed 14 different but non-independent color indices. We de-reddened them using an estimate of the interstellar reddening of $E(B-V) = 0.03$ from \cite{Schlafly:2011}, and then applied 19 different empirical color-temperature calibrations from \cite{Casagrande:2010} and \cite{Huang:2015}. Adopting ${\rm [Fe/H]} = +0.34$ from the spectroscopic analysis above, we obtained a mean photometric temperature of $\langle T_{\rm eff}\rangle = 5940 \pm 100$~K, marginally lower than the spectroscopic value, but within the uncertainties.

\begin{figure}
  \includegraphics[width=\linewidth, trim=100 75 80 90,clip]{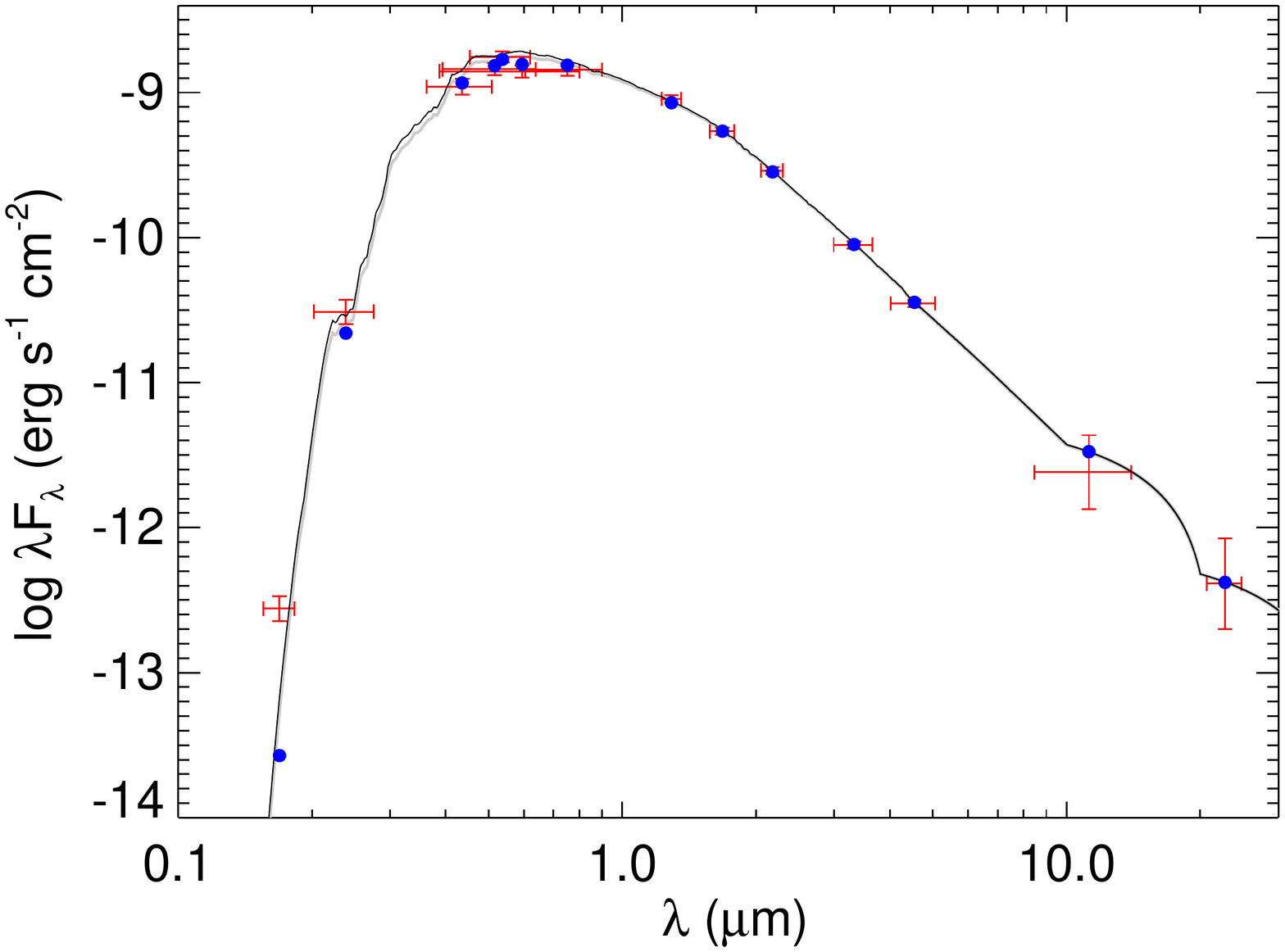}
  \figcaption{Spectral energy distribution of \ticstar\ showing the fit obtained by combining two stellar components with effective temperatures and radii as described in the text. 
  The solid line is the best-fit Kurucz atmosphere model (reddened in gray, unreddened in black), blue dots are the predicted fluxes in each observed bandpass (see the text), and the measurements are indicated with red error bars (the horizontal ones representing the wavelength range). 
  \label{fig:sed}}
\end{figure}

We ran an additional test on the spectroscopic parameters by performing a fit of the spectral energy distribution (SED) of \ticstar\ using the methods of \citet{Stassun:2016}. Brightness measurements in the 
{\it GALEX\/} NUV,
Tycho-2 ($B_{\rm T}$, $V_{\rm T}$), {\it Gaia\/} ($G, G_{\rm BP}, G_{\rm RP}$),
2MASS ($J$, $H$, $K_{\rm S}$), and WISE ($W1$--$W4$) passbands were gathered from the VizieR database, and cover the range from 0.2 to 22~$\mu$m (Figure~\ref{fig:sed}). 
In addition, we pulled the {\it GALEX\/} FUV flux to provide an estimate of the chromospheric activity. 
With the $\log g$ and effective temperature values as above, we obtained an excellent fit with $\chi_\nu^2 = 0.85$ (excluding the FUV flux which appears in excess) and best-fit extinction $A_V = 0.08 \pm 0.03$. The best fit metallicity is [M/H] = $+0.25 \pm 0.05$, strongly constrained by the NUV flux in particular; this is somewhat lower but comparable to the estimate of +0.34 above. The distance inferred from comparing the integrated bolometric flux to the bolometric luminosity given by the radii and temperatures is $241.0 \pm 3.6$~pc, consistent with the distance of $245.0 \pm 2.7$~pc from {\it Gaia\/} EDR3, adjusted by the parallax offset recommended by \citet{Lindegren:2020}. 

The above results clearly point to a supersolar composition for \ticstar, which is consistent with expectations for a massive planet and late F- to early G-type stars of the masses we determine \citep[see, e.g.,][]{Santos:2017}.

\begin{figure}
  \epsscale{0.75}
  \plotone{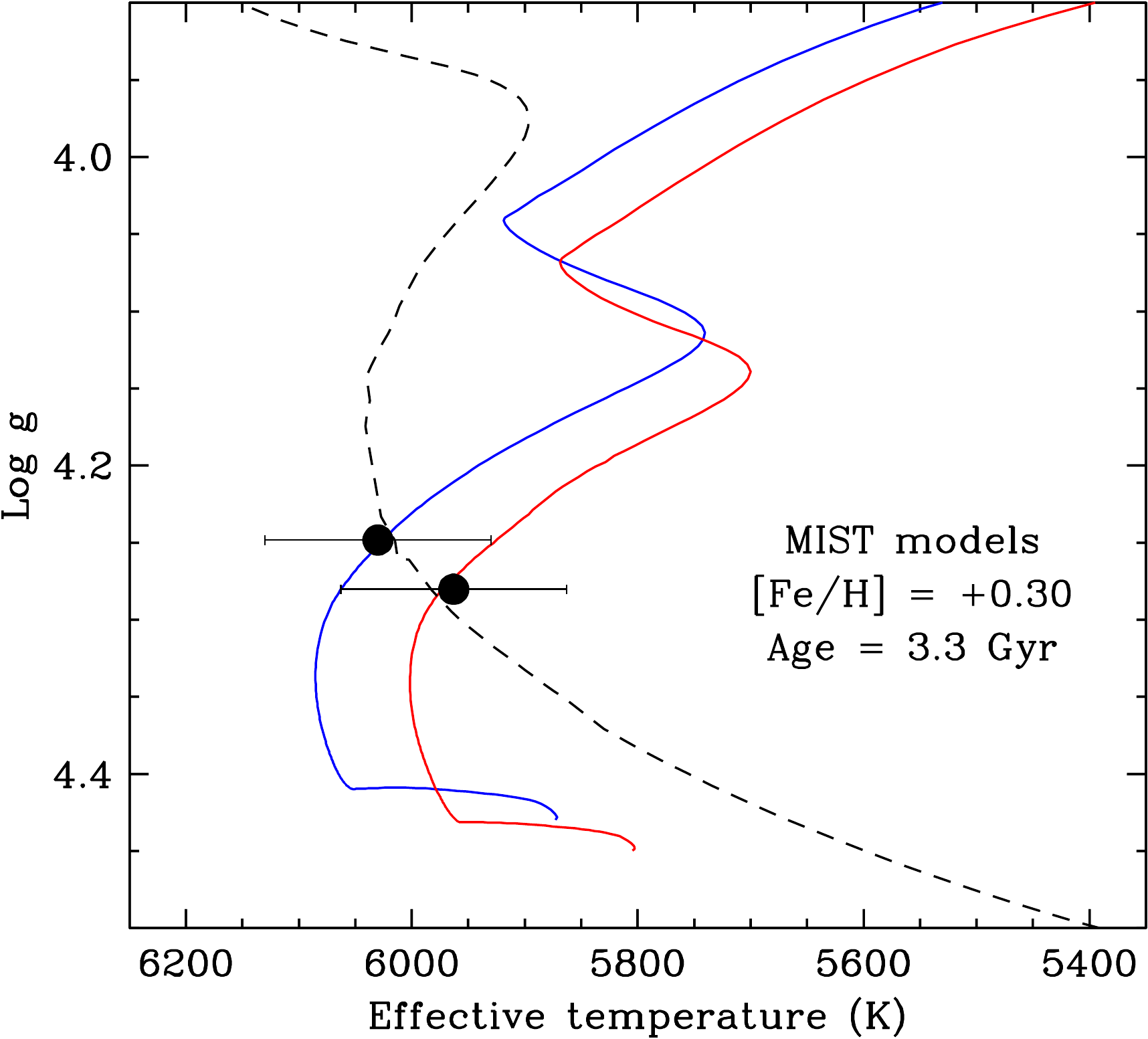}
  \figcaption{Observations for \ticstar\ compared against evolutionary tracks from the MIST series of models \citep{Choi:2016}, for the exact masses we measured. The models match the observations for a metallicity of ${\rm [Fe/H]} = +0.30$ (solar-scaled abundances), and an age of 3.3~Gyr indicated by the dashed-line isochrone. The blue/red lines correspond to the primary/secondary respectively.\label{fig:mist}}
\end{figure}

To illustrate the evolutionary state of the stars, Figure~\ref{fig:mist} compares the surface gravity and temperature determinations against a different set of models from the MIST series \citep{Choi:2016}. Evolutionary tracks are shown for the measured masses, and a metallicity of ${\rm [Fe/H]} = +0.30 \pm 0.10$ that best fits the observations, at which the individual spectroscopic temperatures are 6030 and 5963~K for the primary and secondary (6000~K for the luminosity-weighted mean). The best-fit age is 3.3~Gyr. These results are consistent with those obtained from the PARSEC models. Both stars are seen to be evolved, and are near the midpoint of their main-sequence lifetimes. For the final stellar parameters we adopt the average of the determinations based on the PARSEC and MIST models ($T_1 = 6040$~K, $T_2 = 5970$~K, ${\rm [Fe/H]} = +0.32$), and report them in Table \ref{tab:EBparameters}.

We end this section with a note about the expected rotation of the stars.
Given the 19.7~d orbital period, one might expect the stars to have been tidally spun-down and thus rotating slowly. The timescale for reaching spin-orbit synchronicity is approximately 1.5 Gyrs, significantly shorter than the age of the system (though the timescale for orbital circularization is much longer). However, the binary is quite eccentric ($e=0.45$) and this means the equilibrium pseudosynchronous spin period \citep{Hut81} is only about 8.3 d. The corresponding
$V_{\rm rot}\sin{i}$ is 8.4 \kms\ for the primary star. This somewhat relatively rapid spin may induce stellar activity that could be detected (e.g.\ starspot-induced photometric modulations, or the 
Mount Wilson \ion{Ca}{2} H and K $S$-index).
Such enhanced activity, persistent on a long timescale, is reminiscent of the ``forever young'' effect \citet{Mason2013}, where the stars remain more rapidly rotating, and hence active, than their ages would imply.

The observed FUV excess in the SED (Figure~\ref{fig:sed}) can be used as a check on these ideas. For simplicity we assume that both of the (nearly identical) stars in the system contribute equally to the observed excess. We then infer a chromospheric activity of $\log R'_{\rm HK} = -4.8 \pm 0.2$ via the empirical relations of \citet{Findeisen:2011}, which in turn predicts a rotation period of 10.8$\pm$2.0~d via the empirical rotation-activity relations of \citet{Mamajek:2008}. The latter relations also imply an age of $3.5 \pm 1.4$~Gyr, consistent with that determined above. 


\subsection{Apsidal Motion, Transit Durations, {\it Gaia} EDR3 Astrometry}
Based on the analysis of all available data, the observed phase change amounts to a change in the binary's argument of periastron of $2.88\times 10^{-3}$ degrees per cycle. Precession due to the effects of General Relativity accounts for a change of $1.70\times 10^{-4} $ degrees per cycle, and precession due to tidal bulges (assuming zero obliquity) accounts for $2.32\times 10^{-5}$ degrees per cycle, assuming apsidal constants of $k_2=0.01$ for both stars (Torres, Andersen, \& Gim\'enez 2010).  Thus we see that most of the apsidal precession must be accounted for by another source, which is the circumbinary body of planetary mass.

\begin{figure}
    \centering
    \includegraphics[width=0.95\textwidth]{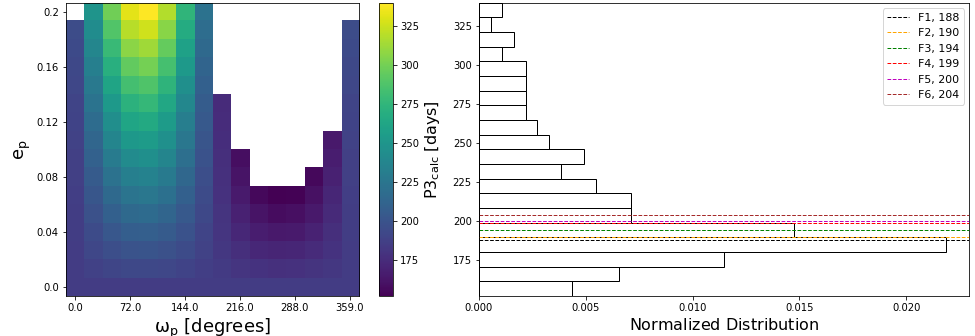}
    \caption{Left panel: Calculated CBP orbital period ($P_{\rm 3,calc}$) based on the duration of the detected transits and using the RV-derived EB parameters (Kostov et al. 2020b), as a function of ${(e_3, \omega_3)}$, and for $e_3$ = [0.0, 0.2]. The white area represents $P_{\rm 3,calc}$ that is ruled out based on Quarles et al. (2018) critical eccentricity, i.e., 
    $0.8(1-(P_{3,\rm stab}/P_{\rm 3,calc})^{(2/3)})$, 
    where $P_{\rm 3,stab} = 129.7$ days. Right panel: Normalized distribution of $P_{\rm 3,calc}$. The horizontal dashed lines represent the true value $P_3$ from the photodynamical modeling for the six families of solutions presented above.}
    \label{fig:12_punch_calc}
\end{figure}

As mentioned above, the measured times of the two CBP transits, combined with the RV-derived stellar masses and EB orbital parameters allow an analytical estimate of the planet's orbital period (e.g. Kostov et al. 2020b) independent of  photodynamical analysis. Assuming $e_3$ = [0.0, 0.2] for the eccentricity of the planet -- the range for the known transiting CBPs (Welsh \& Orosz 2018) -- and taking into account orbital stability, the calculated period of the CBP, $P_{3,calc}$, ranges from $\approx$152 days to $\approx340$ days, with an average of 209.8 days and a median of 194.1 days (see Figure \ref{fig:12_punch_calc}). This is in line with the CBP orbital periods corresponding to the families of solutions listed in Table \ref{derived}, and further strengthens their validity. 

{\it Gaia} measured an astrometric excess noise above a single star model for \ticstar\ of 0.069 mas, with a significance of $\sigma = 5.707$ ({\it Gaia} EDR3, Gaia Collaboration 2020). According to the catalog, there are 291 good observations ($N = astrometric\_n\_good\_obs\_al$) so the target falls in the large 
$N$ limit and the excess noise is significant. At the measured parallax of 4.0608 mas, the astrometric excess noise for \ticstar\ corresponds to a measured physical displacement of $\Delta a_{\rm measured} = 0.017$ au (Belokurov et al.\ 2020). We note that the binary has high eccentricity ($e = 0.45$) and the two stars spend most of the time near apoastron ($r_{\rm max} = a(1+e) = 0.27$ au). Accounting for the argument of periastron of the binary (69.6 degrees), the projected apoastron is 0.25 au and thus the expected photocenter-induced physical displacement is $\Delta a_{\rm expected} = 0.01$ AU\footnote{Taking into account the binary mass ratio of 0.97 and luminosity ratio of 0.87.} -- comparable to the measured displacement. Additionally, \citet{Stassun:2021} have shown using a set of benchmark eclipsing binaries that the renormalized unit weight error (RUWE) statistic can be quantitatively used as an estimator of photocenter motion. Their empirical relation for the RUWE value for TIC 1729 (0.882) suggests a photocenter semi-major axis of $\sim$0.064~mas, or 0.016~au at the distance of the system. Thus there is evidence for binarity based on {\it Gaia}'s astrometric excess noise as well as on RUWE.

\subsection{TIC 172900988 in the Circumbinary Planet Context}

\begin{figure}
    \centering
    \includegraphics[width=0.75\textwidth]{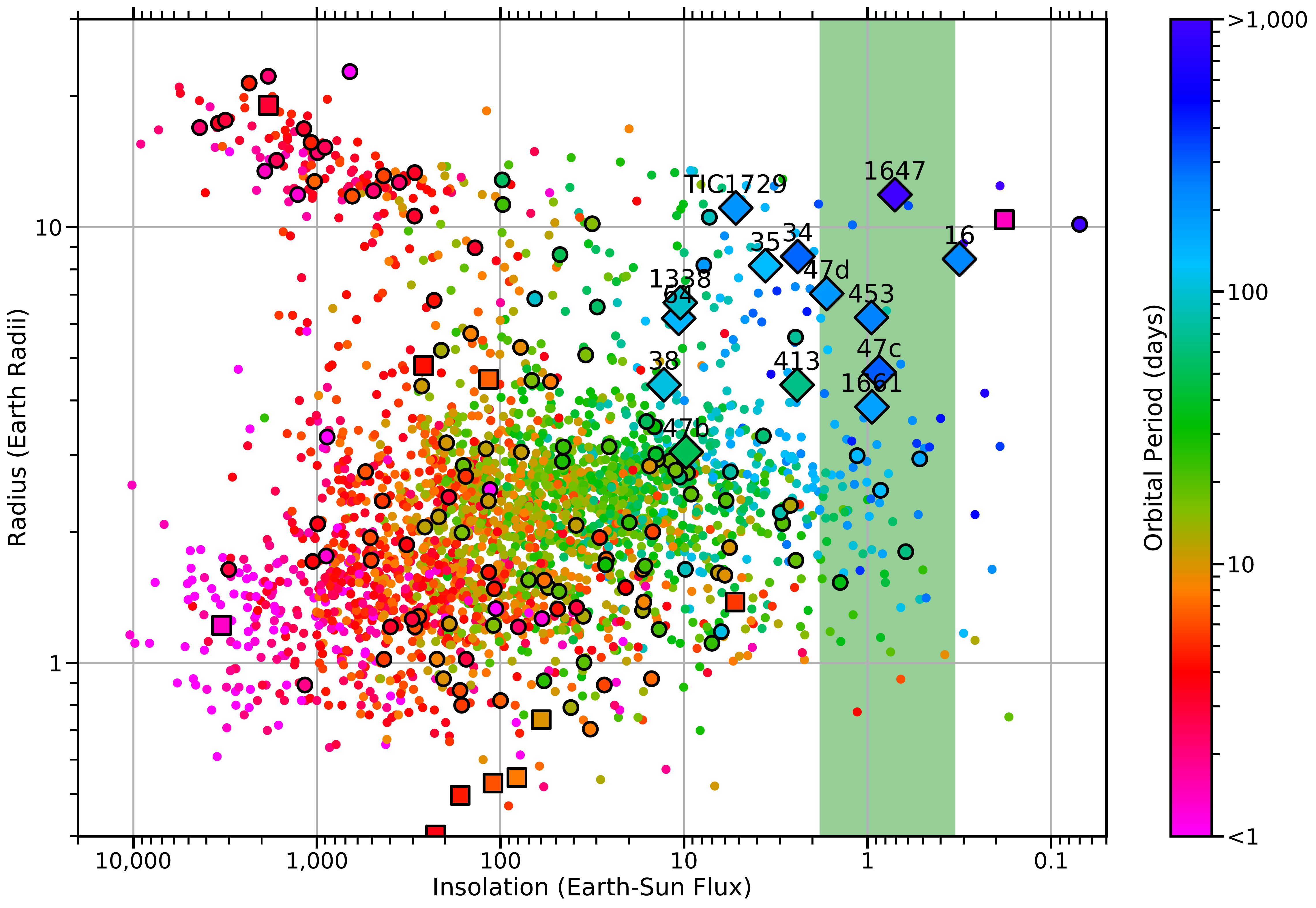}
    \caption{Planet radius versus insolation, using the NExSci Exoplanet Archive confirmed transiting planet sample (as of 2021 June 27). The transiting circumbinary planets are shown as diamond symbols, and tend to lie in the large-radius and low-insolation portion of the diagram.  Shown in green is the approximate position of the habitable zone, based on the ``late Venus'' and ``early Mars'' insolations \citep{Kopparapu2014}. Note that the insolation increases towards the left. For comparison, non-CBP planets that reside in binary star systems are outlined with circles, and planets in triple or higher systems are outlined in squares.
   }
    \label{fig:radius-vs-insolation}
\end{figure}

\begin{figure}
    \centering
    \includegraphics[width=0.5\textwidth]{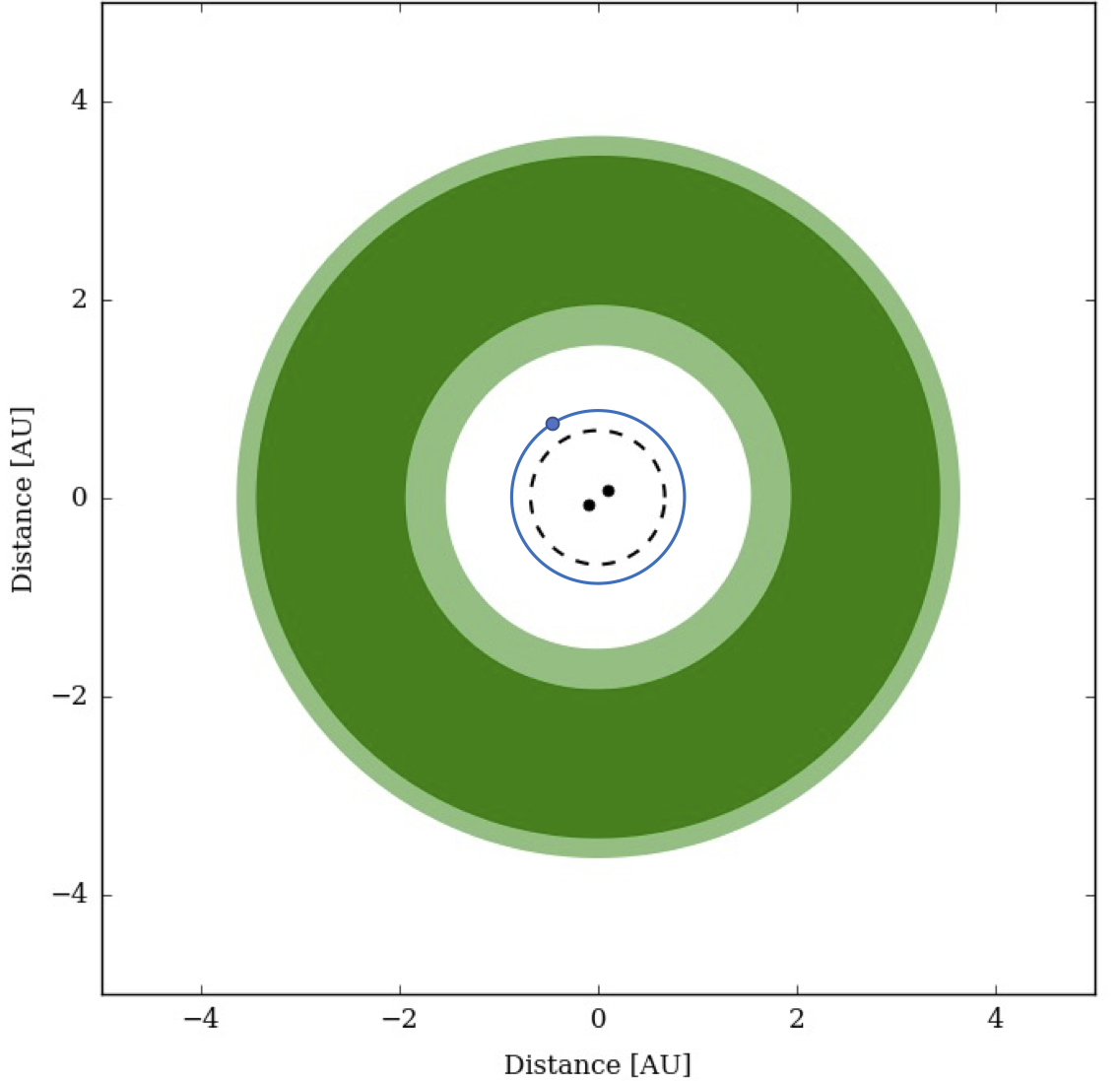}
    \caption{Conservative (dark green) and optimistic (light green) habitable zones of the \ticstar \ system. The orbit of the CBP is shown in blue. The dashed black circle represents the planet's critical instability orbit. (Figure obtained using Binary-HZ calculator (${http://astro.twam.info/hz/}$, Mueller \& Haghighipour 2014))}
    \label{fig:hz}
\end{figure}

Compared with single-star transiting exoplanets, the CBPs are required to have longer orbital periods and larger planetary radii. The longer period is due to the stability requirement around the binary, coupled with the lack of CBPs around the shortest-period binaries (\cite{Welsh2014}, Armstrong et al. 2014, Martin \& Triaud 2014). The larger radius is possibly a detection bias against finding smaller planets due to the CBPs being intrinsically more difficult to detect than planets orbiting a single star (e.g.\ see Windemuth et al. 2019, Martin \& Fabrycky 2021). Yet the CBPs have smaller radii, on average, than the hot Jupiter subset. The observed radius distribution of the known CBPs may be an interesting combination of orbital migration history and observational bias against detecting planets that have not migrated or have been scattered to longer periods 
(e.g., see Pierens \& Nelson (2008, 2013), Kley \& Haghighipour (2014, 2015), \citet{Kley2019}, \citet{Penzlin2020}). 
In general, the currently known CBPs experience low insolation, and a sizeable fraction of them are in the habitable zone of their host binaries,
likely a fortuitous consequence of the bias of the Kepler sample towards G- and K-type stars.
A comparison of the radii and insolations of the CBPs versus other transiting exoplanets is shown in Figure \ref{fig:radius-vs-insolation} (for CBP habitability see also Kane \& Hinkel 2013). As seen in the figure, CBPs tend to reside in the larger-radius and lower-insolation portion of the diagram. However, it is difficult to know if the CBPs have a different parent population because of the different detection biases and small sample size.
For comparison, the figure also marks planets that have two or more host stars but are not circumbinary (i.e., S-type planets). S-type planets in binary star systems are outlined with circles and those in triple or higher-order stellar systems are outlined with squares. Of the 2761 planets with radius and insolation estimates shown in the figure, 137 (5.0\%) are in multiple star systems. Currently there are 125 (4.5\%) non-circumbinary planets in multiple star systems, roughly a factor of 10 more than the circumbinary planets.
Unlike the circumbinary planets, these S-type planets do not appear to favor any particular region in the diagram; they seem to follow the single-star distribution. Certainly this population does not show a tendency for experiencing low insolation or residing in the habitable zone.
\ticstar~b itself is too hot to be in the habitable zone, given its mean insolation of $\sim$5.2 
$S_{\earth}$. Figure \ref{fig:hz} shows a schematic birds-eye view of the \ticstar \ system, with the habitable zone region shown in green.

\begin{figure}
    \centering
    \includegraphics[width=0.75\textwidth]{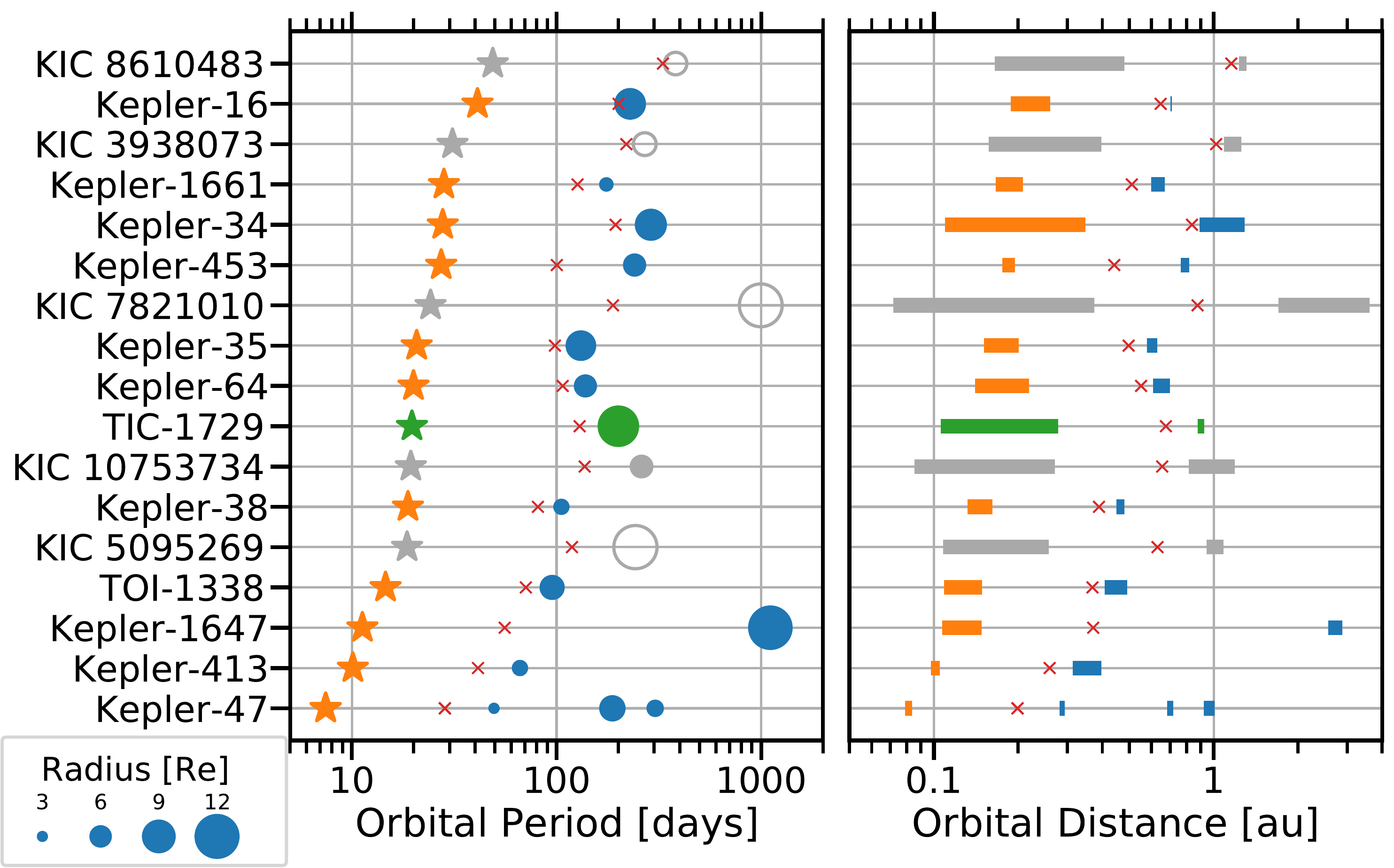}
    \caption{Periods and semi-major axes for all the currently known Kepler and TESS circumbinary planets and host stars. The orange and blue colors represent the binary stars and planets, respectively, with \ticstar\ 
    shown in green, and the gray color representing the CBP candidates. The red ``x'' denotes the critical stability orbit. In the left-hand panel, the size of the blue dots represent the radius of the planet; the non-transiting planets are shown as gray circles with a radius estimated via a mass-radius relation. In the right-hand panel, the horizontal bars denote the range in orbital distance spanned, due to the eccentricity of the orbits.
    }
    \label{fig:CBP_ladder}
\end{figure}

With solar-like stars in a moderately-high eccentricity ($e=0.45$) 19.7-day orbit, the binary hosting \ticstar~b falls neatly into the distribution of known CBP systems, albeit with a higher than typical metallicity. The planet's orbit has a small mutual inclination and the $\sim$200-day period is typical of the known CBP population. The planet is close to the critical (in)stability radius: 
$a/a_{\rm crit}= 1.34$ ($P/P_{\rm crit}=1.55$) using either the more sophisticated
interpolation scheme in \citet{Quarles2018}\footnote{We note that the stability limit is not sharp boundary and has dependencies on the planet eccentricity, mutual inclination and locations of mean motion resonances, see Figure \ref{fig:param_stab}.} or the approximation given in
\citet{Holman1999}. 
In Figure \ref{fig:CBP_ladder} we show \ticstar~b in comparison with the other known CBPs. With the notable exception of Kepler-1647 and KIC 7821010, the figure shows the accumulation of the orbits of the currently known CBPs near their boundaries of stability. This figure includes five candidate CBPs presented at various conferences going back as far as the Extreme Solar Systems II meeting\footnote{2011 September 11-17, Grand Teton National Park, WY}. One of the candidates, KIC 10753734, is transiting \citep{Orosz2016} and four others do not transit: 
KIC 7821010, KIC 8610483, KIC 3938073 
(presented as a group for the first time at the ``Towards Other Earths II'' 
conference\footnote{Slides of the talk presented at ``Towards Other Earths II, The Star-Planet Connection'', 2014 September 15-19, Porto, Portugal are available at \url{https://www.astro.up.pt/investigacao/conferencias/toe2014/files/wwelsh.pdf}}), 
and KIC 5095269 \citep[discovered by][]{Getley2017}. For consistency with the other 16 systems shown in Figure \ref{fig:CBP_ladder}, we use our own preliminary photodynamical values for KIC 5095269 (Orosz et al. in prep.).
The non-transiting planets are detected by their gravitational perturbation of the binary star's orbit, as manifested by eclipse timing variations (ETVs). While such a planet's radius is unknown, its mass is measurable by the pattern it produces in the ETVs (e.g.~Borkovits et al. 2015). 
For the non-transiting cases, we used an approximate radius obtained by employing the mass-radius relation given in \citet{Bashi2017}. 
While \ticstar~b has the second-largest radius of the published circumbinary planets, exceeded only by Kepler-1647 b, its radius is in no way atypical.
Where \ticstar~b does differ from the published CBP is its mass: with a mass $\approx 2.9 \, M_{\rm Jup}$, it is the most massive transiting CBP known, a factor of 2 times larger than the next most massive planet, Kepler-1647b. However, both candidate CBPs KIC 7821010 and KIC 5095269 likely have comparable or larger masses.

\subsection{Near Term ``Transitability'' and Long-term Dynamical Stability}

Models from the posterior samples for the 6 families were integrated using
the GRK12 routine in ELC
starting at day $-1700$ and going  
out to day 200,000 (i.e., from BJD 2,453,300 to 
BJD 2,655,000, which is about 530 years into the future) 
in order to
investigate near-term variations in the orbital elements and transit probabilities.
For each model, the conjunction times and impact parameters
for various combinations of the bodies over that time span
are computed (e.g.\ primary eclipse, secondary eclipse, transits of the primary, etc.) and saved.  In addition, the instantaneous values of the orbital elements and the
Cartesian coordinates of all of the bodies are recorded every 10 days.  
A Lomb-Scargle periodogram was computed for the planet's orbital inclination time
series to find the precession period of the planet's orbital plane.  The
results are given in Table \ref{derived}.  The precessional periods are between
about 44 years for Family 1 to about 54 years for Family 6.  Both the binary's inclination and the planet's
orbital inclination vary with the same $\approx 50$ year period (depending
on the family), but are out of phase.  The binary's inclination varies by only
about $0.5^{\circ}$, and as such the binary will always be eclipsing.

\begin{figure*}
    \centering
    \includegraphics[width=0.95\linewidth]{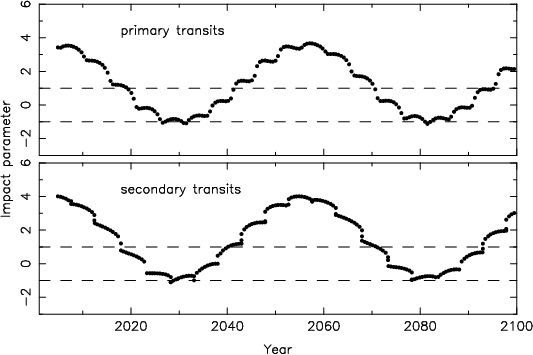}
    \caption{Evolution of the CBP impact parameter $b$ over the course of 100 years for the primary (upper panel) and secondary transits (lower) for Family 5. Transits occur  when $b$ is in between the two dashed lines, e.g.\ $|b| < 1$.
    The planet produces transits for $\approx 40\%$ of its precession cycle
    for this solution. 
    }\label{fig:plot_impact}
\end{figure*}

On the other hand, the inclination of the planet's orbit varies by $\approx 5^{\circ}$, and this is a large enough change that transits will not always
be possible (see also Schneider 1994, Kostov et al. 2014, Martin 2017).  Figure \ref{fig:plot_impact} shows the impact parameter $b$ of
the (inferior) conjunctions of the planet  with the primary and with the
secondary over the span of about 100 years
(using the Family
5 solution).  Only $\approx 40\%$ of
the conjunctions have $|b| < 1$, which is a necessary condition to have
an observable transit.  As seen from Table \ref{derived}, the transit fractions
$\eta_{\rm prim}$ and $\eta_{\rm sec}$
are about 15\% and 40\%.  This transit fraction is consistent with
the rough average of $\approx 36\%$ for the
first 10 known CBP systems (Martin
2017). Given that
TESS observed \ticstar\ for only one sector out of thirteen in the Northern Hemisphere
sky during its second year of operation, 
we were fortunate that TESS was able to observe the transits. Overall, the expected yield of TESS CBPs producing two (or more) transits during a single conjunction from Cycles 1 and 2 is on the order of a hundred (Kostov et al. 2020b).

\begin{figure*}
    \centering
    \includegraphics[width=\textwidth]{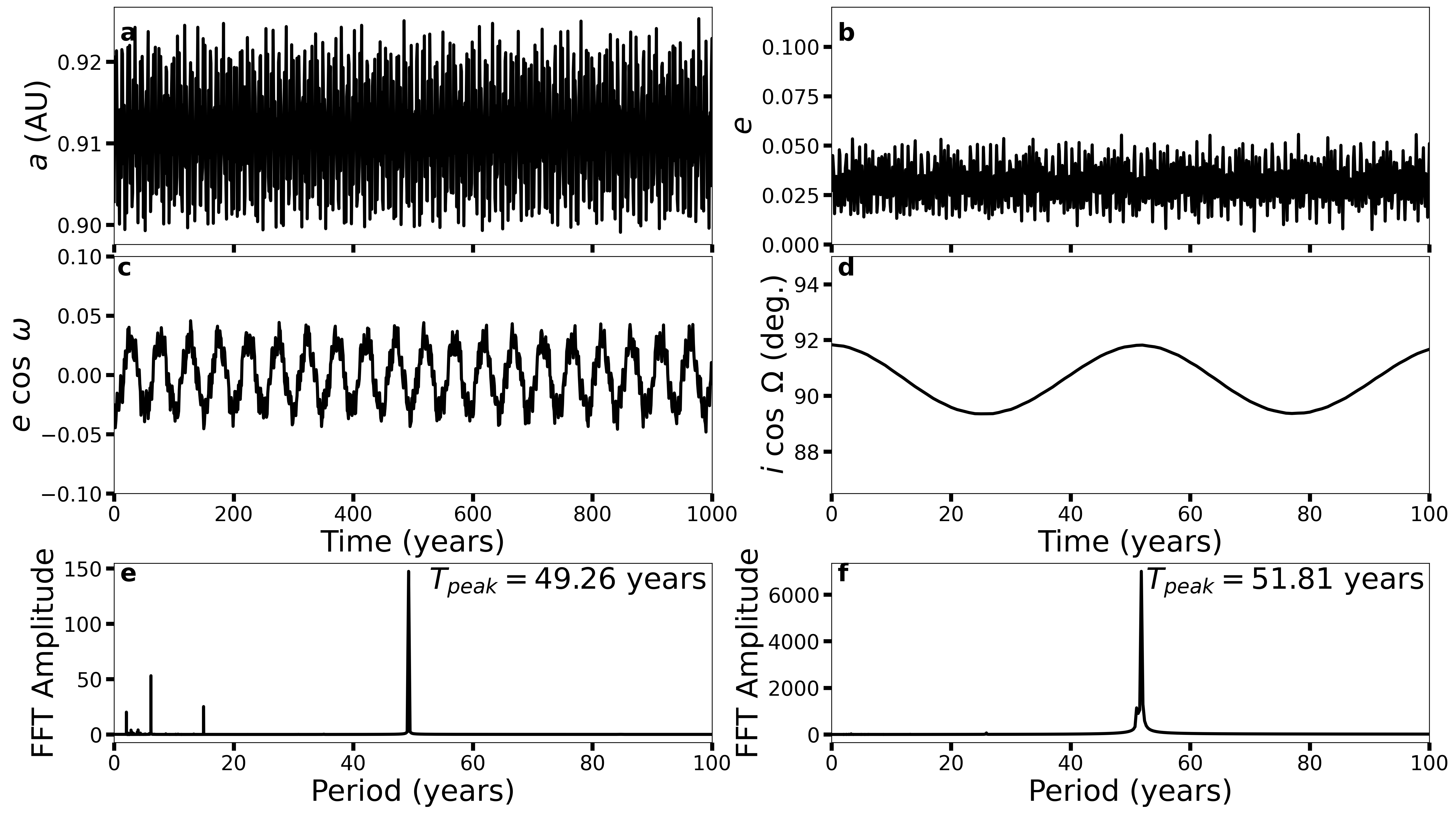}
    \caption{Short-term (1000 years) orbital evolution of the planet using Family 5 for the: (a) semimajor axis, (b) eccentricity, (c) x-component of eccentricity vector, and (d) x-component of (sky) inclination vector.  The simulations are extended to 10,000 years to calculate the periodogram of the (e) apsidal and (f) nodal precession.}
    \label{fig:short_stab}
\end{figure*}

We performed 3-body integrations to understand both the short- and long-term evolution of Family 5 better.  We used the ias15 integrator in REBOUND \citep{Rein2012} for the short-term (1000 years) integrations, which shows the planetary semimajor axis and eccentricity vary in time (Figure \ref{fig:short_stab}a \& Figure \ref{fig:short_stab}b).  These variations are bounded with a small amplitude, which is indicative of long-term stability (i.e., no secular growth).  As a result the $x$-component of the eccentricity (Figure~\ref{fig:short_stab}c) and inclination (\ref{fig:short_stab}d) vectors also show a well-behaved evolution, where the small scale variations in the eccentricity (Figure~\ref{fig:short_stab}c) are due to the 3-body interaction with the inner binary.  As mentioned in the discussion of ``transitability'', the orbital plane of the planet precesses over several decades. Thus, we extended the initial short-term integrations to 10,000 years and identified the dominant secular frequencies (transformed to periods) using the \verb|fft| module from \texttt{scipy}.  Figures \ref{fig:short_stab}e \& \ref{fig:short_stab}f illustrate the periodograms associated with the evolution of the planetary eccentricity and inclination vectors, respectively.  The secular cycle for the planetary eccentricity using Family 5 is $\approx $49.3 years, while the inclination is a little longer ($\approx $51.8 years).  There is a longer variation in the inclination over $\approx $4,000 years (not shown in Figure~\ref{fig:short_stab}d), where this envelope expands slightly in the range of inclination variation to $\approx 88.8^{\circ}-92.3^{\circ}$ without changing the mean value.

\begin{figure*}
    \centering
    \includegraphics[width=\textwidth]{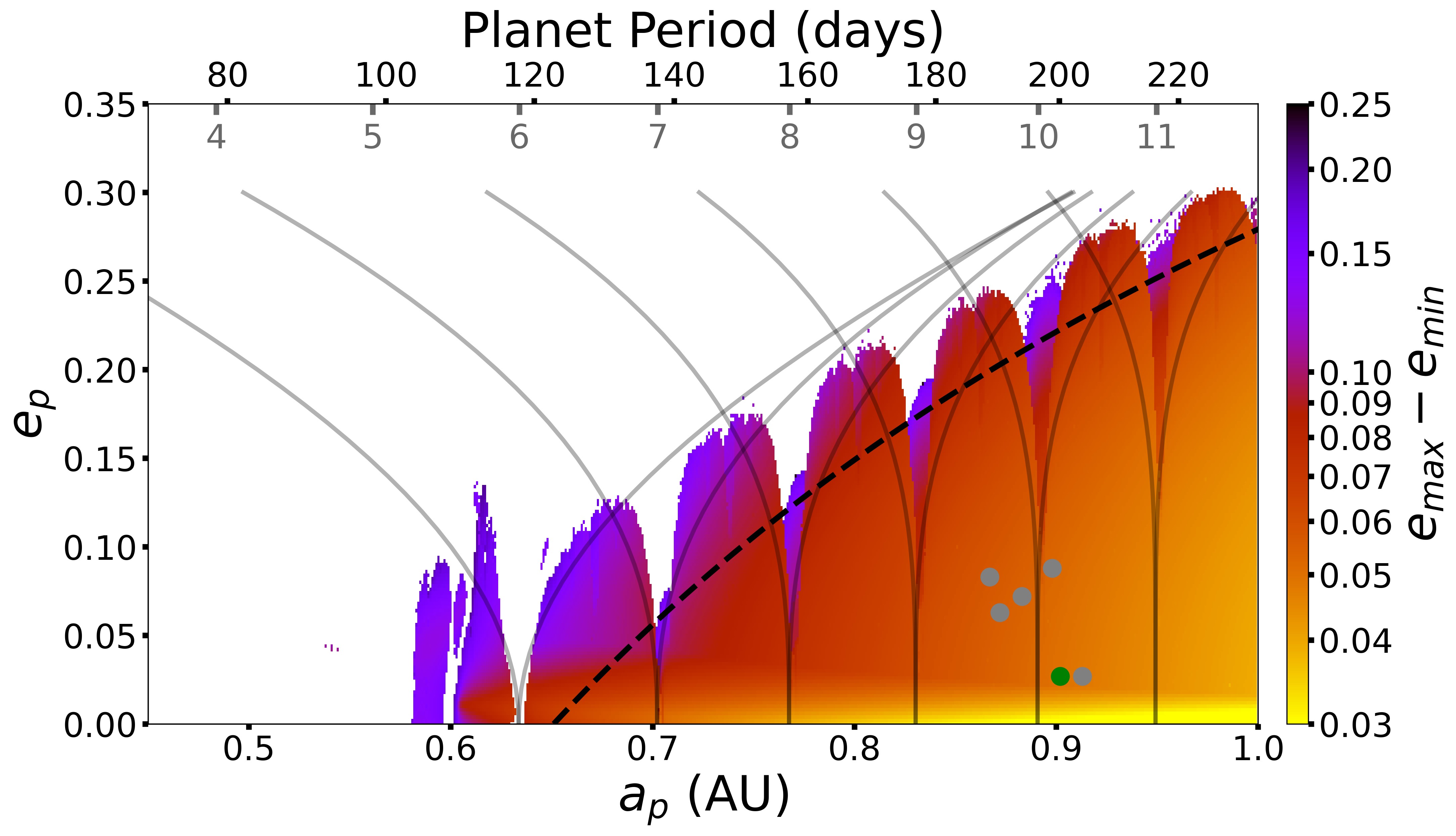}
    \caption{Amplitude of eccentricity variation (colored cells) for a planet using the Family 5 solution, where the white cells denote unstable initial conditions.  The black dashed curve marks the stability limit using the fitting function from \cite{Quarles2018}.  The upper axis shows corresponding planet period for a given semimajor axis. The nominal location of the $N:1$ mean motion resonances are marked (gray ticks) and the potential libration regions for each resonance are indicated (gray curves). }
    \label{fig:param_stab}
\end{figure*}

The long-term evolution of Family 5 used a different approach, where we investigated the stability of many initial conditions using $10^5$ yr integrations with the well-established \texttt{mercury6} code that is modified to efficiently evolve planets in binary systems \citep{Chambers2002}.  These simulations keep the binary solution for Family 5 fixed while varying the planetary semimajor axis and eccentricity in the planetary solution from 0.45--1.0 AU (0.001 AU steps) and 0.0--0.36 (0.002 steps), respectively.  A simulation is terminated if the planet collides with the inner binary (i.e., pericenter crosses the inner binary orbits), escapes the system (i.e., distance to the center of mass is greater than 10 au), or the full simulation time ($10^5$ yr) elapses.  We use these simulations to measure the amplitude of eccentricity variation ($e_{max}-e_{min}$) of a planet for a given initial condition in semimajor axis $a_p$ and eccentricity $e_p$.  This technique is similar to using the chaos indicator MEGNO \citep{Cincotta2000}, but is less computationally expensive.  Figure \ref{fig:param_stab} illustrates the nominal location of the binary $N:1$ mean motion resonances (MMRs) with gray curves, which typically shape the parameter space for which stable CBPs can inhabit \citep{Mardling2013}.  The white cells denote initial conditions that undergo an instability, where the colored cells show the amplitude of eccentricity oscillations.  The dashed curve marks the estimate for the stability boundary accounting for planetary eccentricity using \cite{Quarles2018}.  The green dot marks the nominal parameters for Family 5 (just outside the 10:1 MMR), where the parameters for other Families (gray dots) also lie in the stable regions around the 10:1 MMR.  Note that solutions for interior to Family 5 have elevated eccentricity. All 6 Families represent stable 3-body solutions as long as the planetary eccentricity remains low ($e_p\lesssim 0.15$) despite the forced eccentricity ($\sim$0.02, yellow strip in Figure \ref{fig:param_stab}) from the inner binary.

\section{Conclusions}
\label{sec:conclusions}

We presented the discovery of a new transiting CBP from TESS, TIC 17290098 b. The target was observed in a single sector and the planet produced just two transits -- one across each star -- during the same conjunction. The host system is a double-lined spectroscopic binary exhibiting primary and secondary eclipses. Extensive archival data from multiple surveys (ASAS-SN, KELT, SupeWASP) shows clear apsidal motion of the EB caused by the gravitational perturbation from the CBP. The host binary star has an orbital period of 19.7 days, eccentricity of 0.45, stellar masses of 1.24 and $1.2 M_{\odot}$ respectively, and stellar radii of 1.38 and $1.31 R_{\odot}$ for the primary and secondary stars respectively. We estimated an age for the system of about 3.1 Gyr and a metallicity of [Fe/H] = +0.34. The CBP is slightly larger than Jupiter ($R_3 = 11.07 R_{\oplus}$). The orbital period and mass of the CBP cannot be uniquely-constrained from the available data, and are between about 190 to 205 days and about 820 to $980 M_{\oplus}$, respectively. 

We note that TESS will observe the target again in Sectors 44 through 47 (2021 October to 2022 January). Unfortunately, it will miss the predicted transits for the corresponding conjunctions by several weeks. Thus follow-up observations from other instruments are key for strongly constraining the orbit and mass of the CBP. In particular, observing the predicted 2022 February-March conjunction of the CBP is critical for solving the currently-ambiguous orbit of the planet. As a relatively bright target ($V=10.141$ mag), the system is accessible for high resolution spectroscopy, e.g.~Rossiter-McLaughlin effect, transit spectroscopy. TIC 172900988 demonstrates the discovery potential of TESS for circumbinary planets with orbital periods greatly exceeding the duration of the observing window.

\acknowledgments
This paper includes data collected by the \emph{TESS} mission, which are publicly available from the Mikulski Archive for Space Telescopes (MAST). Funding for the \emph{TESS} mission is provided by NASA's Science Mission directorate. 

VBK is thankful for support from NASA grants 80NSSC20K0054 and 80NSSC20K0850. This material is based upon work supported by the National Science Foundation under Grant NSF AST-1617004 to WFW and JAO. We are also deeply grateful to John  Hood, Jr.\ for his generous support of exoplanet research at San Diego State University. DJS. is supported as an Eberly Research Fellow by the Eberly College of Science at the Pennsylvania State University. The Center for Exoplanets and Habitable Worlds is supported by the Pennsylvania State University, the Eberly College of Science, and the Pennsylvania Space Grant Consortium. NH acknowledges support from NASA XRP through grant number 80NSSC18K051. Support for this work was provided by NASA through the NASA Hubble Fellowship grant HF2-51464  awarded by the Space Telescope Science Institute, which is operated by the Association of Universities for Research in Astronomy, Inc., for NASA, under contract NAS5-26555.

Resources supporting this work were provided by the NASA High-End Computing (HEC) Program through the NASA Center for Climate Simulation (NCCS) at Goddard Space Flight Center, and through the NASA Advanced Supercomputing (NAS) Division at Ames Research Center for the production of the SPOC data products.  Personnel directly supporting this effort were Mark L. Carroll, Laura E. Carriere, Ellen M. Salmon, Nicko D. Acks, Matthew J. Stroud, Bruce E. Pfaff, Lyn E. Gerner, Timothy M. Burch, and Savannah L. Strong. This research has made use of the Exoplanet Follow-up Observation Program website, which is operated by the California Institute of Technology, under contract with the National Aeronautics and Space Administration under the Exoplanet Exploration Program. Resources supporting this work were provided by the NASA High-End Computing (HEC) Program through the NASA Advanced Supercomputing (NAS) Division at Ames Research Center for the production of the SPOC data products.  This research was supported in part through research cyberinfrastructure resources and services provided by the Partnership for an Advanced Computing Environment (PACE) at the Georgia Institute of Technology.

This work has made use of data from the European Space Agency (ESA) mission {\it Gaia} (\url{https://www.cosmos.esa.int/gaia}), processed by the {\it Gaia} Data Processing and Analysis Consortium (DPAC, \url{https://www.cosmos.esa.int/web/gaia/dpac/consortium}). Funding for the DPAC has been provided by national institutions, in particular the institutions participating in the {\it Gaia} Multilateral Agreement. This work makes use of observations from the LCOGT network. LCOGT telescope time was granted by NOIRLab through the Mid-Scale Innovations Program (MSIP). MSIP is funded by NSF. Based in part on observations obtained with the Apache Point Observatory 3.5-meter telescope, which is owned and operated by the Astrophysical Research Consortium. Based in part on observations made at Observatoire de Haute Provence (CNRS), France. The SOPHIE observations were obtained under an OHP DDT programme (PI Triaud) and this work received funding from the European Research Council (ERC) under the European Union's Horizon 2020 research and innovation programme (grant agreement n$^\circ$ 803193/BEBOP) as well as from the Leverhulme Trust (research project grant n$^\circ$ RPG-2018-418). This work makes use of observations from CHEOPS which is an ESA mission in partnership with Switzerland with important contributions to the payload and the ground segment from Austria, Belgium, France, Germany, Hungary, Italy, Portugal, Spain, Sweden, and the United Kingdom.  We thank Kate Isaak, the ESA CHEOPS Project Science Scientist, and the CHEOPS Science Operations Centre for their help and support with the CHEOPS observations.

The ongoing AAVSO Photometric All-Sky Survey (APASS) is funded by the Robert Martin Ayers Sciences Fund.

\facilities{
\emph{Gaia},
CHEOPS,
MAST,
TESS,
ASAS-SN,
Evryscope,
KELT,
WASP,
NCCS,
LCOGT}

\software{
AstroImageJ \citep{Collins2017}, 
ELC (Orosz et al.\ 2019),
eleanor (Feinstein et al.\ 2019),
EXOFASTv2 \citep{Eastman:2019},
TAPIR (Jensen et al.\ 2013), 
mercury6 \citep{Chambers2002},
Rebound \citep{Rein2012,Rein2015}
%
\url{http://astroutils.astronomy.ohio-state.edu/time/}
%
}

\newpage
\begin{deluxetable}{l r r r }[!ht]
\tabletypesize{\normalsize}
\tablecaption{Stellar parameters for TIC 172900988\label{tab:EBparameters}}
\tablewidth{0pt}
\tablehead{
\colhead{Parameter} & \colhead{Value} & \colhead{Error} &\colhead{Source}
}
\startdata
\multicolumn{4}{l}{\bf Identifying Information} \\
\hline
TIC ID & 172900988 & & TIC \\
{\it Gaia} ID & 709709093301608448 & & {\it Gaia} EDR3 \\
Tycho Reference ID & 2483-160-1 & & TRC \\
$\alpha$  (hh:mm:ss) & 08:34:38.799192 & 0.000013  & {\it Gaia} EDR3 \\
$\delta$  (dd:mm:ss) & +31:33:14.314250 & 0.00010  & {\it Gaia} EDR3 \\
$\mu_{\alpha}$ (mas~yr$^{-1}$) & $-6.5329$ & 0.0145 & {\it Gaia} EDR3 \\
$\mu_{\delta}$ (mas~yr$^{-1}$) & $-22.8868$ & 0.0115 & {\it Gaia} EDR3 \\
$\pi$ (mas) & 4.0607 & 0.0149 & Gaia EDR3 \\
Distance (pc) & 246.263 & 0.904 & Gaia EDR3\\
\\
\multicolumn{4}{l}{\bf Photometric Properties} \\
\hline
$V$ (mag) & 10.141 & 0.006 & TIC \\
$Gaia$ (mag) & 10.0477 &  0.0006 & {\it Gaia} EDR3 \\
$T$ (mag) & 9.6319 & 0.0078 & TIC \\
Contamination & 0.000453 & & TIC \\
\\
\multicolumn{4}{l}{\bf Stellar Properties} \\
\hline
Teff (K) & 6030 & 100 & TIC\\
Teff, 1 (K) & 6050 & 100 & This work\\
Teff, 2 (K) & 5983 & 100 & This work\\
$\lbrack $Fe/H$ \rbrack$  & +0.34 & 0.10 & This work \\
Age (Gyr) & 3.1 & 0.1 & This work\\
\enddata
\end{deluxetable}

\begin{deluxetable}{lrrrr}
\tabletypesize{\normalsize}
\tablecaption{Radial Velocities for TIC 172900988\label{RVtable}}
\tablewidth{0pt}
\tablehead{
\colhead{Time}    &
\colhead{Date} & 
\colhead{Primary}  &
\colhead{Secondary}  &
\colhead{Instrument} \\
\colhead{(BJD - 2,455,000)}    &
\colhead{} & 
\colhead{velocity (km s$^{-1}$)}  &
\colhead{velocity (km s$^{-1}$)}  &
\colhead{} 
}
\startdata
   3982.63491 & 2020-May-13T03:14:15.8 & $ -9.248\pm 0.060$ & $ 61.677\pm 0.067$  &  Tull \\ 
  3982.65341 & 2020-May-13T03:40:54.9 & $ -9.114\pm 0.064$ & $ 61.534\pm 0.043$  &  Tull \\ 
  3982.67077 & 2020-May-13T04:05:54.7 & $ -8.995\pm 0.057$ & $ 61.460\pm 0.064$  &  Tull \\ 
  3983.63222 & 2020-May-14T03:10:24.2 & $ -2.758\pm 0.056$ & $ 54.759\pm 0.054$  &  Tull \\ 
  4003.61687 & 2020-Jun-03T02:48:17.6 & $  0.000\pm 0.027$ & $ 53.057\pm 0.027$\tablenotemark{a}  &  ARCES \\ 
  4009.62264 & 2020-Jun-09T02:56:35.8 & $  0.000\pm 0.027$ & $-44.992\pm 0.028$\tablenotemark{a}  &  ARCES \\ 
  4131.99887 & 2020-Oct-09T11:58:22.1 & $ 93.471\pm 0.111$ & $-43.072\pm 0.066$  &  TRES \\ 
  4135.02023 & 2020-Oct-12T12:29:08.2 & $ -9.340\pm 0.098$ & $ 62.841\pm 0.051$  &  TRES \\
\enddata
\tablenotetext{a}{Velocity difference between primary and secondary.}
\end{deluxetable}

\newpage

\begin{deluxetable}{lrrrrrr}
\rotate
\tabletypesize{\scriptsize}
\tablecaption{Fitted parameters for TIC 172900988\tablenotemark{a}\label{fitted}}
\tablewidth{0pt}
\tablehead{
\colhead{Parameter}   & 
\colhead{Family 1}    &
\colhead{Family 2} & 
\colhead{Family 3}  &
\colhead{Family 4}  &
\colhead{Family 5} &
\colhead{Family 6}   
}
\startdata
$\chi^2_{\rm min}$ & 11410.26 &11192.97 &11174.33 &11164.23 &11141.53 &11187.00\\
\hline
\multicolumn{7}{c}{Planet orbital parameters} \\
\hline
$P_2$ (days) & $188.835\pm0.092$ & $190.375\pm0.053$ & $193.965\pm0.078$ & $198.856\pm0.099$ & $200.452\pm0.011$ & $204.046\pm0.057$ \\
$T_{\rm conj,2}$\tablenotemark{b} & $3976.3\pm2.2$ & $3932.5\pm1.1$ & $3877.5\pm2.8$ & $3954.2\pm3.1$ & $3937.5\pm0.5$ & $3932.0\pm0.9$ \\
$\sqrt{e_2}\cos\omega_2$ & $-0.2779\pm0.0048$ & $-0.2510\pm0.0070$ & $-0.2677\pm0.0068$ & $-0.2810\pm0.0045$ & $-0.1536\pm0.0063$ & $-0.1150\pm0.0126$ \\
$\sqrt{e_2}\sin\omega_2$ & $-0.0734\pm0.0091$ & $-0.0058\pm0.0176$ & $ 0.0329\pm0.0144$ & $-0.0951\pm0.0083$ & $ 0.0578\pm0.0149$ & $ 0.1168\pm0.0213$ \\
$i_2$ (deg) & $88.20\pm0.21$ & $88.18\pm0.19$ & $89.67\pm0.25$ & $91.84\pm0.20$ & $91.79\pm0.19$ & $90.50\pm0.32$ \\
$\Omega_2$ (deg) & $-0.88\pm0.22$ & $-0.74\pm0.17$ & $-2.18\pm0.24$ & $ 0.72\pm0.17$ & $ 0.32\pm0.11$ & $ 1.65\pm0.15$ \\
\enddata
\tablenotetext{a}{Instantaneous parameters valid at BJD 2,453,250.0}
\tablenotetext{b}{BJD - 2,455,000}
\tablenotetext{c}{fixed}
\end{deluxetable}

\newpage

\begin{deluxetable}{lrrrrrr}
\tabletypesize{\scriptsize}
\tablecaption{Derived parameters for TIC 172900988\tablenotemark{a}\label{derived}}
\tablewidth{0pt}
\tablehead{
\colhead{Parameter}   & 
\colhead{Family 1}    &
\colhead{Family 2} & 
\colhead{Family 3}  &
\colhead{Family 4}  &
\colhead{Family 5} &
\colhead{Family 6}   
}
\startdata
\multicolumn{7}{c}{Planet orbital parameters} \\
\hline
\multicolumn{7}{c}{Planet orbital parameters} \\
\hline
$P_2$ (days) & $188.835\pm0.092$ & $190.375\pm0.053$ & $193.965\pm0.078$ & $198.856\pm0.099$ & $200.452\pm0.011$ & $204.046\pm0.057$ \\
$T_{\rm conj,2}$\tablenotemark{b} & $3976.3\pm2.2$ & $3932.5\pm1.1$ & $3877.5\pm2.8$ & $3954.2\pm3.1$ & $3937.5\pm0.5$ & $3932.0\pm0.9$ \\
$a_2$ (au) & $0.86762\pm0.00033$ & $0.87235\pm0.00023$ & $0.88328\pm0.00027$ & $0.89809\pm0.00034$ & $0.90281\pm0.00016$ & $0.91359\pm0.00023$ \\
$e_2$ & $0.0826\pm0.0038$ & $0.0631\pm0.0037$ & $0.0728\pm0.0028$ & $0.0880\pm0.0037$ & $0.0271\pm0.0011$ & $0.0269\pm0.0021$ \\
$\omega_2$ (deg) & $194.8\pm  1.6$ & $181.3\pm  4.0$ & $173.0\pm  3.2$ & $198.7\pm  1.4$ & $159.4\pm  5.5$ & $134.6\pm  8.4$ \\
$i_2$ (deg) & $88.20\pm0.21$ & $88.18\pm0.19$ & $89.67\pm0.25$ & $91.84\pm0.20$ & $91.79\pm0.19$ & $90.50\pm0.32$ \\
\enddata
\tablenotetext{a}{Instantaneous parameters valid at BJD 2,453,250.0}
\tablenotetext{b}{BJD - 2,455,000}
\tablenotetext{c}{True anomaly, mean anomaly, mean longitude}
\tablenotetext{d}{Mutual inclination between orbital planes}
\end{deluxetable}

\newpage

\begin{deluxetable}{lrrrrr}
\tabletypesize{\scriptsize}
\tablecaption{Future transit times for TIC 172900988\label{futuretransit}}
\tablewidth{0pt}
\tablehead{
\colhead{Type}   & 
\colhead{Time}    &
\colhead{Date} & 
\colhead{Impact}  &
\colhead{Duration}  &
\colhead{Fraction\tablenotemark{a}} \\
\colhead{}   & 
\colhead{(BJD - 2,455,000)}    &
\colhead{} & 
\colhead{Parameter}  &
\colhead{(hr)}  &
\colhead{} 
}
\startdata
\hline
\multicolumn{6}{c}{Family 1 } \\
\hline
secondary &  $  4443.242\pm  0.055$ & 2021-Aug-16T17:47:51.6 & $-0.202\pm 0.181$ & $ 6.90\pm  0.22$   & 0.999  \\ 
primary   &  $  4446.986\pm  0.047$ & 2021-Aug-20T11:40:07.1 & $ 0.364\pm 0.258$ & $ 5.84\pm  0.93$   & 0.980  \\ 
primary   &  $  4627.972\pm  0.075$ & 2022-Feb-17T11:20:20.0 & $-0.032\pm 0.174$ & $ 7.50\pm  0.17$   & 0.999  \\ 
secondary &  $  4633.473\pm  0.076$ & 2022-Feb-22T23:20:52.3 & $ 0.423\pm 0.284$ & $ 6.55\pm  1.43$   & 0.935  \\ 
secondary &  $  4815.934\pm  0.090$ & 2022-Aug-24T10:24:29.3 & $ 0.172\pm 0.268$ & $ 6.60\pm  0.59$   & 0.993  \\ 
primary   &  $  4819.611\pm  0.121$ & 2022-Aug-28T02:39:53.9 & $ 0.688\pm 0.192$ & $ 6.79\pm  2.36$   & 0.748  \\ 
primary   &  $  5000.753\pm  0.096$ & 2023-Feb-25T06:03:38.0 & $ 0.300\pm 0.249$ & $ 5.89\pm  0.73$   & 0.987  \\
\enddata
\tablenotetext{a}{Fraction of posterior sample with observable transits.}
\end{deluxetable}

\clearpage

\appendix

\begin{deluxetable}{lrrr}[h]
\tablecaption{Future primary eclipses for TIC 172900988, Family 1\label{family1primary}}
\tablewidth{0pt}
\tablehead{
\colhead{Cycle}   & 
\colhead{Time}    &
\colhead{Date} & 
\colhead{Duration}  \\
\colhead{}   & 
\colhead{(BJD - 2,455,000)}    &
\colhead{(UT)} & 
\colhead{(hr)}  
}
\startdata
313 &  $  4369.78998\pm  0.00004$ &    2021-Jun-04T06:57:34.2 & $ 6.1608\pm 0.0052$ \\
314 &  $  4389.44763\pm  0.00007$ &    2021-Jun-23T22:44:35.2 & $ 6.1621\pm 0.0052$ \\
315 &  $  4409.10436\pm  0.00004$ &    2021-Jul-13T14:30:16.3 & $ 6.1616\pm 0.0052$ \\
316 &  $  4428.76285\pm  0.00004$ &    2021-Aug-02T06:18:30.2 & $ 6.1604\pm 0.0052$ \\
317 &  $  4448.42189\pm  0.00005$ &    2021-Aug-21T22:07:31.5 & $ 6.1604\pm 0.0052$ \\
318 &  $  4468.08001\pm  0.00005$ &    2021-Sep-10T13:55:13.2 & $ 6.1612\pm 0.0052$ \\
319 &  $  4487.73732\pm  0.00005$ &    2021-Sep-30T05:41:44.6 & $ 6.1617\pm 0.0052$ \\
320 &  $  4507.39453\pm  0.00005$ &    2021-Oct-19T21:28:07.3 & $ 6.1613\pm 0.0052$ \\
321 &  $  4527.05248\pm  0.00004$ &    2021-Nov-08T13:15:34.2 & $ 6.1604\pm 0.0052$ \\
322 &  $  4546.71157\pm  0.00005$ &    2021-Nov-28T05:04:39.3 & $ 6.1600\pm 0.0052$ \\
\enddata
\end{deluxetable}

\begin{deluxetable}{lrrr}
\tablecaption{Future secondary eclipses for TIC 172900988, Family 1\label{family1secondary}}
\tablewidth{0pt}
\tablehead{
\colhead{Cycle}   & 
\colhead{Time}    &
\colhead{Date} & 
\colhead{Duration}  \\
\colhead{}   & 
\colhead{(BJD - 2,455,000)}    &
\colhead{(UT)} & 
\colhead{(hr)}  
}
\startdata
313 &  $  4362.01453\pm  0.00011$ &    2021-May-27T12:20:55.3 & $14.8352\pm 0.0106$ \\
314 &  $  4381.67141\pm  0.00011$ &    2021-Jun-16T04:06:49.5 & $14.8338\pm 0.0106$ \\
315 &  $  4401.33021\pm  0.00011$ &    2021-Jul-05T19:55:30.2 & $14.8329\pm 0.0107$ \\
316 &  $  4420.98881\pm  0.00010$ &    2021-Jul-25T11:43:52.8 & $14.8345\pm 0.0107$ \\
317 &  $  4440.64583\pm  0.00011$ &    2021-Aug-14T03:29:59.7 & $14.8355\pm 0.0107$ \\
318 &  $  4460.30277\pm  0.00011$ &    2021-Sep-02T19:15:58.9 & $14.8349\pm 0.0107$ \\
319 &  $  4479.96048\pm  0.00011$ &    2021-Sep-22T11:03:05.7 & $14.8338\pm 0.0107$ \\
320 &  $  4499.61886\pm  0.00011$ &    2021-Oct-12T02:51:09.1 & $14.8335\pm 0.0107$ \\
321 &  $  4519.27721\pm  0.00011$ &    2021-Oct-31T18:39:11.2 & $14.8346\pm 0.0107$ \\
322 &  $  4538.93466\pm  0.00011$ &    2021-Nov-20T10:25:54.3 & $14.8359\pm 0.0107$ \\
\enddata
\end{deluxetable}

\begin{deluxetable}{lrrr}
\tablecaption{Future primary eclipses for TIC 172900988, Family 2\label{family2primary}}
\tablewidth{0pt}
\tablehead{
\colhead{Cycle}   & 
\colhead{Time}    &
\colhead{Date} & 
\colhead{Duration}  \\
\colhead{}   & 
\colhead{(BJD - 2,455,000)}    &
\colhead{(UT)} & 
\colhead{(hr)}  
}
\startdata
313 &  $  4369.78969\pm  0.00005$ &    2021-Jun-04T06:57:09.5 & $ 6.1560\pm 0.0058$ \\
314 &  $  4389.44783\pm  0.00008$ &    2021-Jun-23T22:44:52.9 & $ 6.1572\pm 0.0058$ \\
315 &  $  4409.10459\pm  0.00006$ &    2021-Jul-13T14:30:36.6 & $ 6.1574\pm 0.0058$ \\
316 &  $  4428.76238\pm  0.00004$ &    2021-Aug-02T06:17:49.7 & $ 6.1562\pm 0.0058$ \\
317 &  $  4448.42138\pm  0.00005$ &    2021-Aug-21T22:06:47.1 & $ 6.1557\pm 0.0058$ \\
318 &  $  4468.07998\pm  0.00006$ &    2021-Sep-10T13:55:10.0 & $ 6.1563\pm 0.0058$ \\
319 &  $  4487.73758\pm  0.00006$ &    2021-Sep-30T05:42:06.9 & $ 6.1570\pm 0.0058$ \\
320 &  $  4507.39469\pm  0.00007$ &    2021-Oct-19T21:28:20.9 & $ 6.1570\pm 0.0058$ \\
321 &  $  4527.05226\pm  0.00005$ &    2021-Nov-08T13:15:15.0 & $ 6.1562\pm 0.0058$ \\
322 &  $  4546.71104\pm  0.00005$ &    2021-Nov-28T05:03:53.7 & $ 6.1554\pm 0.0058$ \\
\enddata
\end{deluxetable}

\newpage

\begin{deluxetable}{lrrr}
\tablecaption{Future secondary eclipses for TIC 172900988, Family 2\label{family2secondary}}
\tablewidth{0pt}
\tablehead{
\colhead{Cycle}   & 
\colhead{Time}    &
\colhead{Date} & 
\colhead{Duration}  \\
\colhead{}   & 
\colhead{(BJD - 2,455,000)}    &
\colhead{(UT)} & 
\colhead{(hr)}  
}
\startdata
313 &  $  4362.01482\pm  0.00010$ &    2021-May-27T12:21:20.1 & $14.8254\pm 0.0117$ \\
314 &  $  4381.67151\pm  0.00010$ &    2021-Jun-16T04:06:58.7 & $14.8244\pm 0.0117$ \\
315 &  $  4401.32977\pm  0.00011$ &    2021-Jul-05T19:54:52.4 & $14.8231\pm 0.0117$ \\
316 &  $  4420.98866\pm  0.00009$ &    2021-Jul-25T11:43:40.4 & $14.8240\pm 0.0117$ \\
317 &  $  4440.64622\pm  0.00009$ &    2021-Aug-14T03:30:33.7 & $14.8255\pm 0.0117$ \\
318 &  $  4460.30304\pm  0.00010$ &    2021-Sep-02T19:16:22.8 & $14.8255\pm 0.0117$ \\
319 &  $  4479.96037\pm  0.00010$ &    2021-Sep-22T11:02:55.8 & $14.8244\pm 0.0117$ \\
320 &  $  4499.61855\pm  0.00010$ &    2021-Oct-12T02:50:43.0 & $14.8236\pm 0.0117$ \\
321 &  $  4519.27709\pm  0.00010$ &    2021-Oct-31T18:39:00.6 & $14.8242\pm 0.0117$ \\
322 &  $  4538.93496\pm  0.00010$ &    2021-Nov-20T10:26:20.6 & $14.8257\pm 0.0117$ \\
\enddata
\end{deluxetable}

\newpage

\begin{deluxetable}{lrrr}
\tablecaption{Future primary eclipses for TIC 172900988, Family 3\label{family3primary}}
\tablewidth{0pt}
\tablehead{
\colhead{Cycle}   & 
\colhead{Time}    &
\colhead{Date} & 
\colhead{Duration}  \\
\colhead{}   & 
\colhead{(BJD -02,455,000)}    &
\colhead{(UT)} & 
\colhead{(hr)}  
}
\startdata
313 &  $  4232.18215\pm  0.00004$ &    2021-Jan-17T16:22:18.0 & $ 6.1570\pm 0.0044$ \\
314 &  $  4251.84096\pm  0.00004$ &    2021-Feb-06T08:10:58.8 & $ 6.1561\pm 0.0044$ \\
315 &  $  4271.49978\pm  0.00005$ &    2021-Feb-25T23:59:40.9 & $ 6.1565\pm 0.0044$ \\
316 &  $  4291.15765\pm  0.00005$ &    2021-Mar-17T15:47:01.2 & $ 6.1573\pm 0.0044$ \\
317 &  $  4310.81482\pm  0.00005$ &    2021-Apr-06T07:33:20.4 & $ 6.1576\pm 0.0044$ \\
318 &  $  4330.47215\pm  0.00005$ &    2021-Apr-25T23:19:54.0 & $ 6.1570\pm 0.0044$ \\
319 &  $  4350.13061\pm  0.00003$ &    2021-May-15T15:08:04.3 & $ 6.1560\pm 0.0044$ \\
320 &  $  4369.78988\pm  0.00004$ &    2021-Jun-04T06:57:25.2 & $ 6.1563\pm 0.0044$ \\
321 &  $  4389.44777\pm  0.00007$ &    2021-Jun-23T22:44:47.0 & $ 6.1575\pm 0.0044$ \\
322 &  $  4409.10452\pm  0.00005$ &    2021-Jul-13T14:30:30.2 & $ 6.1574\pm 0.0044$ \\
\enddata
\end{deluxetable}

\newpage

\begin{deluxetable}{lrrr}
\tablecaption{Future secondary eclipses for TIC 172900988, Family 3\label{family3secondary}}
\tablewidth{0pt}
\tablehead{
\colhead{Cycle}   & 
\colhead{Time}    &
\colhead{Date} & 
\colhead{Duration}  \\
\colhead{}   & 
\colhead{(BJD -02,455,000)}    &
\colhead{(UT)} & 
\colhead{(hr)}  
}
\startdata
313 &  $  4224.41118\pm  0.00008$ &    2021-Jan-09T21:52:05.9 & $14.8319\pm 0.0090$ \\
314 &  $  4244.06916\pm  0.00008$ &    2021-Jan-29T13:39:34.9 & $14.8339\pm 0.0090$ \\
315 &  $  4263.72605\pm  0.00009$ &    2021-Feb-18T05:25:30.4 & $14.8344\pm 0.0090$ \\
316 &  $  4283.38316\pm  0.00009$ &    2021-Mar-09T21:11:44.6 & $14.8334\pm 0.0090$ \\
317 &  $  4303.04111\pm  0.00009$ &    2021-Mar-29T12:59:12.1 & $14.8326\pm 0.0090$ \\
318 &  $  4322.69965\pm  0.00009$ &    2021-Apr-18T04:47:29.4 & $14.8333\pm 0.0090$ \\
319 &  $  4342.35782\pm  0.00009$ &    2021-May-07T20:35:15.2 & $14.8352\pm 0.0090$ \\
320 &  $  4362.01477\pm  0.00009$ &    2021-May-27T12:21:15.9 & $14.8360\pm 0.0090$ \\
321 &  $  4381.67157\pm  0.00009$ &    2021-Jun-16T04:07:03.7 & $14.8348\pm 0.0090$ \\
322 &  $  4401.33006\pm  0.00010$ &    2021-Jul-05T19:55:17.4 & $14.8342\pm 0.0090$ \\
\enddata
\end{deluxetable}

\begin{deluxetable}{lrrr}
\tablecaption{Future primary eclipses for TIC 172900988, Family 4\label{family4primary}}
\tablewidth{0pt}
\tablehead{
\colhead{Cycle}   & 
\colhead{Time}    &
\colhead{Date} & 
\colhead{Duration}  \\
\colhead{}   & 
\colhead{(BJD -02,455,000)}    &
\colhead{(UT)} & 
\colhead{(hr)}  
}
\startdata
313 &  $  4369.78957\pm  0.00004$ &    2021-Jun-04T06:56:58.7 & $ 6.1577\pm 0.0047$ \\
314 &  $  4389.44805\pm  0.00005$ &    2021-Jun-23T22:45:11.2 & $ 6.1589\pm 0.0047$ \\
315 &  $  4409.10486\pm  0.00005$ &    2021-Jul-13T14:30:59.5 & $ 6.1594\pm 0.0047$ \\
316 &  $  4428.76234\pm  0.00004$ &    2021-Aug-02T06:17:46.5 & $ 6.1583\pm 0.0047$ \\
317 &  $  4448.42129\pm  0.00005$ &    2021-Aug-21T22:06:39.0 & $ 6.1575\pm 0.0047$ \\
318 &  $  4468.08005\pm  0.00005$ &    2021-Sep-10T13:55:16.5 & $ 6.1580\pm 0.0047$ \\
319 &  $  4487.73781\pm  0.00005$ &    2021-Sep-30T05:42:26.5 & $ 6.1588\pm 0.0047$ \\
320 &  $  4507.39493\pm  0.00006$ &    2021-Oct-19T21:28:42.1 & $ 6.1590\pm 0.0047$ \\
321 &  $  4527.05233\pm  0.00005$ &    2021-Nov-08T13:15:21.6 & $ 6.1583\pm 0.0047$ \\
322 &  $  4546.71086\pm  0.00004$ &    2021-Nov-28T05:03:38.0 & $ 6.1573\pm 0.0047$ \\
\enddata
\end{deluxetable}

\newpage

\begin{deluxetable}{lrrr}
\tablecaption{Future secondary eclipses for TIC 172900988, Family 4\label{family4secondary}}
\tablewidth{0pt}
\tablehead{
\colhead{Cycle}   & 
\colhead{Time}    &
\colhead{Date} & 
\colhead{Duration}  \\
\colhead{}   & 
\colhead{(BJD -02,455,000)}    &
\colhead{(UT)} & 
\colhead{(hr)}  
}
\startdata
313 &  $  4362.01493\pm  0.00009$ &    2021-May-27T12:21:30.3 & $14.8303\pm 0.0097$ \\
314 &  $  4381.67153\pm  0.00010$ &    2021-Jun-16T04:07:00.0 & $14.8296\pm 0.0097$ \\
315 &  $  4401.32948\pm  0.00010$ &    2021-Jul-05T19:54:27.0 & $14.8281\pm 0.0097$ \\
316 &  $  4420.98848\pm  0.00009$ &    2021-Jul-25T11:43:24.7 & $14.8286\pm 0.0097$ \\
317 &  $  4440.64629\pm  0.00009$ &    2021-Aug-14T03:30:39.5 & $14.8302\pm 0.0097$ \\
318 &  $  4460.30310\pm  0.00010$ &    2021-Sep-02T19:16:28.1 & $14.8305\pm 0.0097$ \\
319 &  $  4479.96028\pm  0.00010$ &    2021-Sep-22T11:02:48.1 & $14.8295\pm 0.0097$ \\
320 &  $  4499.61833\pm  0.00010$ &    2021-Oct-12T02:50:23.9 & $14.8285\pm 0.0097$ \\
321 &  $  4519.27687\pm  0.00010$ &    2021-Oct-31T18:38:41.6 & $14.8289\pm 0.0097$ \\
322 &  $  4538.93495\pm  0.00010$ &    2021-Nov-20T10:26:19.8 & $14.8304\pm 0.0098$ \\
\enddata
\end{deluxetable}

\begin{deluxetable}{lrrr}
\tablecaption{Future primary eclipses for TIC 172900988, Family 5\label{family5primary}}
\tablewidth{0pt}
\tablehead{
\colhead{Cycle}   & 
\colhead{Time}    &
\colhead{Date} & 
\colhead{Duration}  \\
\colhead{}   & 
\colhead{(BJD -02,455,000)}    &
\colhead{(UT)} & 
\colhead{(hr)}  
}
\startdata
313 &  $  4369.78853\pm  0.00004$ &    2021-Jun-04T06:55:29.0 & $ 6.1528\pm 0.0047$ \\
314 &  $  4389.44775\pm  0.00004$ &    2021-Jun-23T22:44:45.6 & $ 6.1530\pm 0.0047$ \\
315 &  $  4409.10584\pm  0.00005$ &    2021-Jul-13T14:32:24.6 & $ 6.1541\pm 0.0047$ \\
316 &  $  4428.76281\pm  0.00004$ &    2021-Aug-02T06:18:27.0 & $ 6.1544\pm 0.0047$ \\
317 &  $  4448.42022\pm  0.00004$ &    2021-Aug-21T22:05:06.8 & $ 6.1535\pm 0.0047$ \\
318 &  $  4468.07879\pm  0.00004$ &    2021-Sep-10T13:53:27.8 & $ 6.1527\pm 0.0047$ \\
319 &  $  4487.73773\pm  0.00005$ &    2021-Sep-30T05:42:19.9 & $ 6.1528\pm 0.0047$ \\
320 &  $  4507.39587\pm  0.00005$ &    2021-Oct-19T21:30:03.2 & $ 6.1537\pm 0.0047$ \\
321 &  $  4527.05304\pm  0.00005$ &    2021-Nov-08T13:16:22.6 & $ 6.1542\pm 0.0047$ \\
322 &  $  4546.71021\pm  0.00005$ &    2021-Nov-28T05:02:42.1 & $ 6.1536\pm 0.0047$ \\
\enddata
\end{deluxetable}

\newpage

\begin{deluxetable}{lrrr}
\tablecaption{Future secondary eclipses for TIC 172900988, Family 5\label{family5secondary}}
\tablewidth{0pt}
\tablehead{
\colhead{Cycle}   & 
\colhead{Time}    &
\colhead{Date} & 
\colhead{Duration}  \\
\colhead{}   & 
\colhead{(BJD -02,455,000)}    &
\colhead{(UT)} & 
\colhead{(hr)}  
}
\startdata
313 &  $  4362.01555\pm  0.00009$ &    2021-May-27T12:22:23.2 & $14.8238\pm 0.0103$ \\
314 &  $  4381.67250\pm  0.00010$ &    2021-Jun-16T04:08:23.6 & $14.8247\pm 0.0103$ \\
315 &  $  4401.32930\pm  0.00010$ &    2021-Jul-05T19:54:11.5 & $14.8238\pm 0.0103$ \\
316 &  $  4420.98735\pm  0.00010$ &    2021-Jul-25T11:41:47.1 & $14.8225\pm 0.0103$ \\
317 &  $  4440.64611\pm  0.00010$ &    2021-Aug-14T03:30:23.6 & $14.8228\pm 0.0103$ \\
318 &  $  4460.30415\pm  0.00010$ &    2021-Sep-02T19:17:58.6 & $14.8243\pm 0.0103$ \\
319 &  $  4479.96117\pm  0.00010$ &    2021-Sep-22T11:04:04.8 & $14.8251\pm 0.0103$ \\
320 &  $  4499.61809\pm  0.00010$ &    2021-Oct-12T02:50:03.1 & $14.8243\pm 0.0103$ \\
321 &  $  4519.27589\pm  0.00010$ &    2021-Oct-31T18:37:16.7 & $14.8231\pm 0.0103$ \\
322 &  $  4538.93452\pm  0.00010$ &    2021-Nov-20T10:25:42.8 & $14.8230\pm 0.0103$ \\
\enddata
\end{deluxetable}

\begin{deluxetable}{lrrr}
\tablecaption{Future primary eclipses for TIC 172900988, Family 6\label{family6primary}}
\tablewidth{0pt}
\tablehead{
\colhead{Cycle}   & 
\colhead{Time}    &
\colhead{Date} & 
\colhead{Duration}  \\
\colhead{}   & 
\colhead{(BJD -02,455,000)}    &
\colhead{(UT)} & 
\colhead{(hr)}  
}
\startdata
313 &  $  4369.78821\pm  0.00004$ &    2021-Jun-04T06:55:01.5 & $ 6.1514\pm 0.0047$ \\
314 &  $  4389.44736\pm  0.00004$ &    2021-Jun-23T22:44:11.5 & $ 6.1509\pm 0.0048$ \\
315 &  $  4409.10606\pm  0.00005$ &    2021-Jul-13T14:32:43.2 & $ 6.1518\pm 0.0048$ \\
316 &  $  4428.76341\pm  0.00007$ &    2021-Aug-02T06:19:19.0 & $ 6.1526\pm 0.0048$ \\
317 &  $  4448.42044\pm  0.00005$ &    2021-Aug-21T22:05:25.9 & $ 6.1523\pm 0.0048$ \\
318 &  $  4468.07838\pm  0.00004$ &    2021-Sep-10T13:52:52.4 & $ 6.1513\pm 0.0048$ \\
319 &  $  4487.73727\pm  0.00005$ &    2021-Sep-30T05:41:40.5 & $ 6.1509\pm 0.0048$ \\
320 &  $  4507.39600\pm  0.00005$ &    2021-Oct-19T21:30:14.5 & $ 6.1514\pm 0.0048$ \\
321 &  $  4527.05368\pm  0.00006$ &    2021-Nov-08T13:17:17.8 & $ 6.1523\pm 0.0048$ \\
322 &  $  4546.71066\pm  0.00006$ &    2021-Nov-28T05:03:20.9 & $ 6.1524\pm 0.0048$ \\
\enddata
\end{deluxetable}

\begin{deluxetable}{lrrr}
\tablecaption{Future secondary eclipses for TIC 172900988, Family 6\label{family6secondary}}
\tablewidth{0pt}
\tablehead{
\colhead{Cycle}   & 
\colhead{Time}    &
\colhead{Date} & 
\colhead{Duration}  \\
\colhead{}   & 
\colhead{(BJD -02,455,000)}    &
\colhead{(UT)} & 
\colhead{(hr)}  
}
\startdata
313 &  $  4362.01554\pm  0.00009$ &    2021-May-27T12:22:22.9 & $14.8311\pm 0.0103$ \\
314 &  $  4381.67298\pm  0.00009$ &    2021-Jun-16T04:09:05.2 & $14.8328\pm 0.0103$ \\
315 &  $  4401.32963\pm  0.00010$ &    2021-Jul-05T19:54:40.3 & $14.8325\pm 0.0103$ \\
316 &  $  4420.98706\pm  0.00010$ &    2021-Jul-25T11:41:22.1 & $14.8311\pm 0.0103$ \\
317 &  $  4440.64561\pm  0.00010$ &    2021-Aug-14T03:29:40.8 & $14.8309\pm 0.0103$ \\
318 &  $  4460.30415\pm  0.00010$ &    2021-Sep-02T19:17:58.5 & $14.8325\pm 0.0103$ \\
319 &  $  4479.96169\pm  0.00010$ &    2021-Sep-22T11:04:49.9 & $14.8342\pm 0.0103$ \\
320 &  $  4499.61851\pm  0.00010$ &    2021-Oct-12T02:50:38.9 & $14.8342\pm 0.0103$ \\
321 &  $  4519.27571\pm  0.00010$ &    2021-Oct-31T18:37:01.3 & $14.8330\pm 0.0103$ \\
322 &  $  4538.93396\pm  0.00010$ &    2021-Nov-20T10:24:54.0 & $14.8323\pm 0.0103$ \\
\enddata
\end{deluxetable}

\newpage
\begin{deluxetable}{rrrr}
\rotate
\tablecolumns{4}
\tabletypesize{\footnotesize}
\tablecaption{Initial Dynamical Parameters,
Family 1 solution\tablenotemark{a}\label{Family1init}}
\tablewidth{0pt}
\tablehead{  
\colhead{parameter\tablenotemark{b}} & 
\colhead{orbit 1}   & 
\colhead{orbit 2}   & 
\colhead{} }
\startdata
Period  (days)&$ 1.96574580221424142E+01$ & $ 1.88789475383543561E+02$ & \cr 
$e\cos\omega$ &$ 1.56170869428264247E-01$ & $-7.84627636359548492E-02$ & \cr 
$e\sin\omega$ &$ 4.19679141794474964E-01$ & $-1.95706955564456352E-02$ & \cr 
$i$  (rad) &$ 1.56065838061355966E+00$ & $ 1.54077852290273420E+00$ & \cr 
$\Omega$  (rad) &$ 0.00000000000000000E+00$ & $-1.39615402591776010E-02$ & \cr 
$T_{\rm conj}$  (days)\tablenotemark{c} &$ 3.87817840712633324E+03$ & $ 3.97504901993195290E+03$ & \cr 
$a$  (AU) &$ 1.91922057111088368E-01$ & $ 8.67437940438783306E-01$ & \cr 
$e$     &$ 4.47794509250981132E-01$ & $ 8.08666643429462734E-02$ & \cr
\enddata
\tablenotetext{a}{Reference time =     -1750.00000,
integration step size = 0.05000 days}
\tablenotetext{b}{Jacobian instantaneous (Keplerian) elements}
\tablenotetext{c}{Times are relative to BJD 2,455,000.000}
\tablenotetext{d}{Barycentric Cartesian coordinates}
\end{deluxetable}

\newpage
\begin{deluxetable}{rrrr}
\rotate
\tablecolumns{4}
\tabletypesize{\footnotesize}
\tablecaption{Initial Dynamical Parameters, Family 2 
Solution\tablenotemark{a}\label{Family2init}}
\tablewidth{0pt}
\tablehead{  
\colhead{parameter\tablenotemark{b}} & 
\colhead{orbit 1}   & 
\colhead{orbit 2}   & 
\colhead{} }
\startdata
Period  (days)&$ 1.96581050348207640E+01$ & $ 1.90344409654595211E+02$ & \cr 
$e\cos\omega$ &$ 1.56095399121964296E-01$ & $-6.41839330129671676E-02$ & \cr 
$e\sin\omega$ &$ 4.19562643942974267E-01$ & $-3.41531319824112719E-03$ & \cr 
$i$  (rad) &$ 1.56069168597231789E+00$ & $ 1.53995695080737560E+00$ & \cr 
$\Omega$  (rad) &$ 0.00000000000000000E+00$ & $-1.23272607174403446E-02$ & \cr 
$T_{\rm conj}$  (days)\tablenotemark{c} &$ 3.87836379440340716E+03$ & $ 3.93178612576113301E+03$ & \cr 
$a$  (AU) &$ 1.91935253129229433E-01$ & $ 8.72243497642089749E-01$ & \cr 
$e$     &$ 4.47659006186029829E-01$ & $ 6.42747354817963384E-02$ & \cr
\enddata
\tablenotetext{a}{Reference time =     -1750.00000,
integration step size = 0.05000 days}
\tablenotetext{b}{Jacobian instantaneous (Keplerian) elements}
\tablenotetext{c}{Times are relative to BJD 2,455,000.000}
\tablenotetext{d}{Barycentric Cartesian coordinates}
\end{deluxetable}

\begin{deluxetable}{rrrr}
\rotate
\tablecolumns{4}
\tabletypesize{\footnotesize}
\tablecaption{Initial Dynamical Parameters, Family 3 
Solution\tablenotemark{a}\label{Family3init}}
\tablewidth{0pt}
\tablehead{  
\colhead{parameter\tablenotemark{b}} & 
\colhead{orbit 1}   & 
\colhead{orbit 2}   & 
\colhead{} }
\startdata
Period  (days)&$ 1.96581966864200162E+01$ & $ 1.93904791889272900E+02$ & \cr 
$e\cos\omega$ &$ 1.56121806427791304E-01$ & $-7.08444993703360865E-02$ & \cr 
$e\sin\omega$ &$ 4.19668697760405940E-01$ & $ 1.08423492641878030E-02$ & \cr 
$i$  (rad) &$ 1.58116590285353231E+00$ & $ 1.56931046735959434E+00$ & \cr 
$\Omega$  (rad) &$ 0.00000000000000000E+00$ & $-3.38391048322012472E-02$ & \cr 
$T_{\rm conj}$  (days)\tablenotemark{c} &$ 3.87839003218881726E+03$ & $ 3.87537072364502910E+03$ & \cr 
$a$  (AU) &$ 1.91945485746299049E-01$ & $ 8.83128528073461538E-01$ & \cr 
$e$     &$ 4.47767611962044176E-01$ & $ 7.16693772025415871E-02$ & \cr
\enddata
\tablenotetext{a}{Reference time =     -1750.00000,
integration step size = 0.05000 days}
\tablenotetext{b}{Jacobian instantaneous (Keplerian) elements}
\tablenotetext{c}{Times are relative to BJD 2,455,000.000}
\tablenotetext{d}{Barycentric Cartesian coordinates}
\end{deluxetable}

\newpage

\begin{deluxetable}{rrrr}
\rotate
\tablecolumns{4}
\tabletypesize{\footnotesize}
\tablecaption{Initial Dynamical Parameters, Family 4 
Solution\tablenotemark{a}\label{Family4init}}
\tablewidth{0pt}
\tablehead{  
\colhead{parameter\tablenotemark{b}} & 
\colhead{orbit 1}   & 
\colhead{orbit 2}   & 
\colhead{} }
\startdata
Period  (days)&$ 1.96566477698221753E+01$ & $ 1.98857206991129431E+02$ & \cr 
$e\cos\omega$ &$ 1.56289724178110490E-01$ & $-8.40521311420768802E-02$ & \cr 
$e\sin\omega$ &$ 4.19818036408871220E-01$ & $-2.78742314301126115E-02$ & \cr 
$i$  (rad) &$ 1.58093148741641842E+00$ & $ 1.60390271144704388E+00$ & \cr 
$\Omega$  (rad) &$ 0.00000000000000000E+00$ & $ 1.36027426644736976E-02$ & \cr 
$T_{\rm conj}$  (days)\tablenotemark{c} &$ 3.87794613658087519E+03$ & $ 3.95443034920364471E+03$ & \cr 
$a$  (AU) &$ 1.91930046466682352E-01$ & $ 8.98087983771095555E-01$ & \cr 
$e$     &$ 4.47966138874212660E-01$ & $ 8.85535630414952535E-02$ & \cr
\enddata
\tablenotetext{a}{Reference time =     -1750.00000,
integration step size = 0.05000 days}
\tablenotetext{b}{Jacobian instantaneous (Keplerian) elements}
\tablenotetext{c}{Times are relative to BJD 2,455,000.000}
\tablenotetext{d}{Barycentric Cartesian coordinates}
\end{deluxetable}

\begin{deluxetable}{rrrr}
\rotate
\tablecolumns{4}
\tabletypesize{\footnotesize}
\tablecaption{Initial Dynamical Parameters, Family 5 
Solution\tablenotemark{a}\label{Family5init}}
\tablewidth{0pt}
\tablehead{  
\colhead{parameter\tablenotemark{b}} & 
\colhead{orbit 1}   & 
\colhead{orbit 2}   & 
\colhead{} }
\startdata
Period  (days)&$ 1.96581697466411960E+01$ & $ 2.00453455776095382E+02$ & \cr 
$e\cos\omega$ &$ 1.56120265129208902E-01$ & $-2.53238980408201642E-02$ & \cr 
$e\sin\omega$ &$ 4.19765514968936770E-01$ & $ 1.03126288031170597E-02$ & \cr 
$i$  (rad) &$ 1.58078123663466807E+00$ & $ 1.60276140444680815E+00$ & \cr 
$\Omega$  (rad) &$ 0.00000000000000000E+00$ & $ 6.15744651978666815E-03$ & \cr 
$T_{\rm conj}$  (days)\tablenotemark{c} &$ 3.87838264397195871E+03$ & $ 3.93739746952321866E+03$ & \cr 
$a$  (AU) &$ 1.91925787812358140E-01$ & $ 9.02849883123379215E-01$ & \cr 
$e$     &$ 4.47857817550560366E-01$ & $ 2.73431915623053474E-02$ & \cr
\enddata
\tablenotetext{a}{Reference time =     -1750.00000,
integration step size = 0.05000 days}
\tablenotetext{b}{Jacobian instantaneous (Keplerian) elements}
\tablenotetext{c}{Times are relative to BJD 2,455,000.000}
\tablenotetext{d}{Barycentric Cartesian coordinates}
\end{deluxetable}

\begin{deluxetable}{rrrr}
\rotate
\tablecolumns{4}
\tabletypesize{\footnotesize}
\tablecaption{Initial Dynamical Parameters, Family 6 
Solution\tablenotemark{a}\label{Family6init}}
\tablewidth{0pt}
\tablehead{  
\colhead{parameter\tablenotemark{b}} & 
\colhead{orbit 1}   & 
\colhead{orbit 2}   & 
\colhead{} }
\startdata
Period  (days)&$ 1.96580278267864692E+01$ & $ 2.04044351487866408E+02$ & \cr 
$e\cos\omega$ &$ 1.56130598769293610E-01$ & $-2.03399939845695932E-02$ & \cr 
$e\sin\omega$ &$ 4.19636617208692664E-01$ & $ 1.32000037143129752E-02$ & \cr 
$i$  (rad) &$ 1.56039566396540219E+00$ & $ 1.58295631134958925E+00$ & \cr 
$\Omega$  (rad) &$ 0.00000000000000000E+00$ & $ 3.01046537868270951E-02$ & \cr 
$T_{\rm conj}$  (days)\tablenotemark{c} &$ 3.87834172008000996E+03$ & $ 3.93188579064755913E+03$ & \cr 
$a$  (AU) &$ 1.91938897677575687E-01$ & $ 9.13681045402537806E-01$ & \cr 
$e$     &$ 4.47740610593246102E-01$ & $ 2.42477927521248318E-02$ & \cr
\enddata
\tablenotetext{a}{Reference time =     -1750.00000,
integration step size = 0.05000 days}
\tablenotetext{b}{Jacobian instantaneous (Keplerian) elements}
\tablenotetext{c}{Times are relative to BJD 2,455,000.000}
\tablenotetext{d}{Barycentric Cartesian coordinates}
\end{deluxetable}

\end{document}